\documentclass[12pt]{article}

\usepackage{pstricks}
\usepackage{color}
\usepackage{pst-node}

\usepackage{epsf,amsfonts,hyperref}
\usepackage{nonfloat}
\catcode`\@=11
\def\marginnote#1{}

\newcount\hour
\newcount\minute
\newtoks\amorpm
\hour=\time\divide\hour by60 \minute=\time{\multiply\hour by60
\global\advance\minute by-\hour}
\edef\standardtime{{\ifnum\hour<12 \global\amorpm={am}%
        \else\global\amorpm={pm}\advance\hour by-12 \fi
        \ifnum\hour=0 \hour=12 \fi
        \number\hour:\ifnum\minute<10 0\fi\number\minute\the\amorpm}}
\edef\militarytime{\number\hour:\ifnum\minute<10 0\fi\number\minute}

\def\draftlabel#1{{\@bsphack\if@filesw {\let\thepage\relax
      \xdef\@gtempa{\write\@auxout{\string
          \newlabel{#1}{{\@currentlabel}{\thepage}}}}}\@gtempa \if@nobreak
    \ifvmode\nobreak\fi\fi\fi\@esphack} \gdef\@eqnlabel{#1}}
    \def\@eqnlabel{}
\def\@vacuum{}
\def\draftmarginnote#1{\marginpar{\raggedright\scriptsize\tt#1}}

\def\draft{
  \oddsidemargin -.5truein
  \def\@oddfoot{\footnotesize \sl preliminary draft \hfil
    \rm\thepage\hfil\sl\today\quad\militarytime}
  \let\@evenfoot\@oddfoot \overfullrule 3pt
    \let\label=\draftlabel
    \let\marginnote=\draftmarginnote
  \def\@eqnnum{(\theequation)\rlap{\kern\marginparsep\tt\@eqnlabel}%
    \global\let\@eqnlabel\@vacuum}

  }

\draft

\renewcommand{\theequation}{\thesection.\arabic{equation}}

\renewcommand{\appendix}[1]{
    \setcounter{equation}{0}
    \renewcommand{\thesection}{\Alph{section}}
    \section{Appendix: \protect\indent #1}
}

\newcommand\encadremath[1]{\vbox{\hrule\hbox{\vrule\kern8pt
\vbox{\kern8pt \hbox{$\displaystyle #1$}\kern8pt}
\kern8pt\vrule}\hrule}}
\def\enca#1{\vbox{\hrule\hbox{
\vrule\kern8pt\vbox{\kern8pt \hbox{$\displaystyle #1$}
\kern8pt} \kern8pt\vrule}\hrule}}

\newcommand\figureframex[3]{
\begin{figure}[bth]
\hrule\hbox{\vrule\kern8pt
\vbox{\kern8pt \vbox{
\begin{center}
{\mbox{\epsfxsize=#1.truecm\epsfbox{#2}}}
\end{center}
\caption{#3}
}\kern8pt}
\kern8pt\vrule}\hrule
\end{figure}
}
\newcommand\figureframey[3]{
\begin{figure}[bth]
\hrule\hbox{\vrule\kern8pt
\vbox{\kern8pt \vbox{
\begin{center}
{\mbox{\epsfysize=#1.truecm\epsfbox{#2}}}
\end{center}
\caption{#3}
}\kern8pt}
\kern8pt\vrule}\hrule
\end{figure}
}

\makeatletter
\@addtoreset{equation}{section}
\makeatother
\newtheorem{theorem}{Theorem}[section]
\newtheorem{conjecture}{Conjecture}[section]
\newtheorem{remark}{Remark}[section]
\newtheorem{proposition}{Proposition}[section]
\newtheorem{lemma}{Lemma}[section]
\newtheorem{corollary}{Corollary}[section]
\newtheorem{definition}{Definition}[section]
\def\br{\begin{remark}\rm\small}
\def\er{\end{remark}}
\def\bt{\begin{theorem}}
\def\et{\end{theorem}}
\def\bd{\begin{definition}}
\def\ed{\end{definition}}
\def\bp{\begin{proposition}}
\def\ep{\end{proposition}}
\def\bl{\begin{lemma}}
\def\el{\end{lemma}}
\def\bc{\begin{corollary}}
\def\ec{\end{corollary}}
\def\beaq{\begin{eqnarray}}
\def\eeaq{\end{eqnarray}}
\newcommand{\proof}[1]{{\noindent \bf proof:}\par
{#1} $\square$}

\newcommand{\beq}{\begin{equation}}
\newcommand{\eeq}{\end{equation}}
\newcommand{\bea}{\begin{eqnarray}}
\newcommand{\eea}{\end{eqnarray}}

\renewcommand{\and}{{\qquad {\rm and} \qquad}}

\newcommand{\tr}{{\,\rm tr}\:}

\newcommand{\td}[1]{{\widetilde{#1}}}
\newcommand{\ovl}[1]{{\overline{#1}}}

\renewcommand{\l}{\lambda}
\newcommand{\om}{\omega}

\newcommand{\ee}[1]{{{\rm e}^{#1}}}

\newcommand{\Pint}{{\int\kern -1.em -\kern-.25em}}

\renewcommand{\Re}{{\mathrm{Re}}}

\newcommand{\bcycle}{{\cal B}}
\newcommand{\acycle}{{\cal A}}

\newcommand{\genus}{{\mathfrak{g}}}

\newcommand{\sheet}[2]{{\stackrel{{#1}}{{#2}}}}

\renewcommand\l{\lambda}

\newcommand\Res{\mathop{{\rm Res}}}

\begin{document}

\sloppy


\pagestyle{empty}
\hfill ITEP/TH-34/10
\addtolength{\baselineskip}{0.20\baselineskip}
\begin{center}
\vspace{26pt} {\large \bf {Topological expansion of $\beta$-ensemble model
and quantum algebraic geometry in the sectorwise approach}}
\newline
\vspace{26pt}

{\sl L.\ O.\ Chekhov}\hspace*{0.05cm}${}^1$
,
{\sl B.\ Eynard}\hspace*{0.05cm}${}^2$
,
{\sl O.\ Marchal}\hspace*{0.05cm}${}^2$
\\
\vspace{6pt}
${}^1\,\,$ Steklov Mathematical Institute, ITEP and Laboratoire Poncelet,\\
Moscow, Russia\\
\vspace{1pt}
${}^2\,\,$ CEA, IPhT, F-91191 Gif-sur-Yvette, France, \\
CNRS, URA 2306, F-91191 Gif-sur-Yvette, France.\\
\end{center}

\vspace{20pt}
\begin{abstract}
We solve the loop equations of the $\beta$-ensemble
model analogously to the solution found for the
Hermitian matrices $\beta=1$. For $\beta=1$, the solution was
expressed using the algebraic
spectral curve of equation $y^2=U(x)$. For arbitrary $\beta$, the
spectral curve converts into a Schr\"odinger
equation $((\hbar\partial)^2-U(x))\psi(x)=0$ with $\hbar\propto
(\sqrt\beta-1/\sqrt\beta)/N$. This paper is similar to the sister
paper~I, in particular, all the main ingredients specific for the
algebraic solution of the problem remain the same, but here
we present the second approach to finding a solution of loop equations using sectorwise definition of
resolvents. Being technically more involved, it allows to define consistently
the $\bcycle$-cycle structure of the obtained quantum algebraic curve
(a D-module of the form $y^2-U(x)$, where $[y,x]=\hbar$)
and to construct explicitly the correlation functions and
the corresponding symplectic invariants ${\mathcal F}_h$, or the
terms of the free energy, in $1/N^2$-expansion at arbitrary $\hbar$. The set of
``flat'' coordinates comprises the potential times $t_k$ and the
occupation numbers $\widetilde{\epsilon}_\alpha$.
We define and investigate the properties of the ${\acycle}$- and ${\bcycle}$-cycles,
forms of 1st, 2nd and 3rd kind, and the Riemann bilinear identities. We use these
identities to find explicitly the singular part of $\mathcal F_0$ that depends exclusively
on $\widetilde{\epsilon}_\alpha$.
\end{abstract}

\tableofcontents

\vspace{26pt}
\pagestyle{plain}
\setcounter{page}{1}


\section{Introduction}

In the contemporary mathematical physics, one can often meet the notion of quantum surfaces, which appears in many different
aspects. Having no intension to describe all problems in which quantization of the very space--time coordinates takes place
(which pertains mainly to string or brane models) we however stress that the main feature of most, if not all, these models
is that the consideration is commonly restricted to simple geometries of sphere or torus. Observables in these theories are
not the coordinates, which cease to commute with each other and satisfy some postulated quantum algebras, but objects
related to representations of these algebras, because only these objects admit classical interpretation. In this paper, we
propose a new approach to the description of these so-called ``quantum surfaces,'' namely we begin with solutions of the
standard one-dimensional Schr\"odinger equation with a polynomial potential and construct a higher genus quantum surface
(which is the analogue of a classical hyperelliptic Riemann surface) for which we can define analogues of all the main
notions of algebraic geometry.

This paper is an ``alternative version'' of our paper~\cite{ChEyMarI} in which the notion of the
quantum algebraic geometry was introduced and which we refer to as paper~I in what follows.
In the both versions, the origin of quantum algebraic geometry is the same, the Schr\"odinger equation
$((\hbar\partial)^2-U(x))\psi(x)=0$. The principal difference is that in this, second version,
we use the sectorwise definition of all the quantities starting from the one-point resolvents,
i.e., we use different solutions of the Schr\"odinger equation to construct these
resolvents in different Stokes sectors of the complex plane.
This enables us to define in a rigorous way the integrations over ${\acycle}$- and ${\bcycle}$-cycles as well
as to present a self-consistent procedure for constructing the correlation functions and the symplectic invariants.

The correlation functions $W^{(h)}_n(x_1,\dots, x_n)$ and the symplectic invariants $F_h$
for any algebraic plane curve given by a polynomial equation
$$
{\cal E}(x,y)=\sum_{i,j} {\cal E}_{i,j}\, x^i y^j=0
$$
were defined in \cite{Eyn1loop,EOFg}.
The invariants ${\mathcal F}_h({\cal E})$ are defined in terms of algebraic
geometry quantities related to the Riemann surface of equation
${\cal E}(x,y)=0$. On the matrix model side, these invariants are
terms of the $1/N^2$- (the genus) expansion of the free energy
calculated in \cite{ChekEynFg} for the one-matrix model
and in \cite{CEO} for the two-matrix model.

We introduce the notion of a ``quantum curve" for which ${\cal E}(x,y)$ is a non-commutative polynomial of $x$ and $y$:
\beq\nonumber
{\cal E}(x,y)=\sum_{i,j} {\cal E}_{i,j}\, x^i\, y^j
\qquad , \quad
[y,x]=\hbar.
\eeq
The notion of quantum curve is also known as {\em D-modules}, i.e., a
quotient of the space of functions by ${\rm Ker}\, {\cal E}(x,y)$, where
$y=\hbar \partial/\partial x$.

Our construction is based on the functions $\psi(x)$ such that
\beq\nonumber
{\cal E}(x,\hbar \partial_x)\cdot \psi(x)=0
\eeq
and we show that one can consistently define
all the basic notion of algebraic geometry within this construction. Whereas some
objects, like branch points, become obsolete, we can define
cycles, forms, Bergman kernel, period matrix and the corresponding
Abel maps as well as other objects in a consistent way.

But, otherwise, it is striking to find that almost all
relationships of classical algebraic geometry remain unchanged
when $\hbar\neq 0$, for instance, the Riemann bilinear identity, the
modified Rauch variational formula, and the topological recursion defining
the correlation functions and the symplectic invariants.



The symplectic invariants ${\mathcal F}_h$ were first introduced for the solution
of loop equations arising in the 1-hermitian random matrix model
\cite{Eyn1loop, ChekEynFg}. They were later generalized to other
hermitian  multi--matrix models \cite{CEO, EPrats}.

The models that correspond to the quantum surface are the
{\em $\beta$-ensembles} classified by the exponent $\beta$.
The three Wigner ensembles (see
\cite{Mehtabook}, and we changed $\beta\to \beta/2$) correspond to
$\beta=1$ (hermitian case), $\beta=1/2$ (real symmetric case),
$\beta=2$ (real self-dual quaternion case), but we can easily define
a $\beta$-ensemble eigenvalue model for any real value of $\beta$ as the $N$-fold
integral of the form
$$
\int d\lambda_1\cdots d\lambda_N |\Delta(\lambda)|^{2\beta}e^{-N\sqrt{\beta}\sum_{j=1}^NV(\lambda_j)}
$$
($\Delta$ is the Vandermonde determinant).

\medskip

In \cite{ChekEynbeta}, the solution of
\cite{Eyn1loop} was generalized to the $\beta$-ensembles, but the
solution was presented as a double half-infinite sum for
$\beta=O(1)$ at large $N$,
\beq\nonumber
{\mathcal F}=\sum_{h,k=0}^\infty\, N^{2-2h-k}\,\,(\sqrt\beta-1/\sqrt\beta)^k {\mathcal F}_{h,k}.
\eeq
The coefficients ${\mathcal F}_{h,k}$ were computed in \cite{ChekEynbeta}.

In this paper, as in paper~I, we assume that
$\hbar=(\sqrt\beta-1/\sqrt\beta)/N$, so we perform an (infinite) resummation
in the above formula; the free-energy expansion then acquires the standard form,
\beq\nonumber
{\mathcal F} = \sum_{h=0}^\infty \, N^{2-2h}{\mathcal F}_h(\hbar).
\eeq
The ${\mathcal F}_{h,k}$'s of  \cite{ChekEynbeta} can be recovered by computing
the semi-classical small $\hbar$-expansion of ${\mathcal F}_h(\hbar)$. We demonstrate
that ${\mathcal F}_h(\hbar)$ is the natural
generalization of the symplectic invariants of \cite{EOFg} for a
``quantum spectral curve" ${\cal E}(x,y)$ with $[y,x]=\hbar$.

We define also analogues of the multi-point resolvents
$$
W_n(x_1,\dots,x_n)=N^{-n}\left\langle \sum_{j=1}^N(x_1-\lambda_j)^{-1}\cdots  \sum_{j=1}^N(x_n-\lambda_j)^{-1}\right\rangle_{conn},
$$
where we let angular brackets denote the averaging with the weight
$|\Delta(\lambda)|^{2\beta}e^{-N\sqrt{\beta}\sum_{j=1}^NV(\lambda_j)}$. These resolvents in turn admit the $1/N^2$-expansion,
$W_n(x_1,\dots,x_n)=\sum_{h=0}^\infty N^{2-2h-n}W_n^{(h)}(x_1,\dots,x_n)$, and we calculate all the terms $W_n^{(h)}$ using the
modified diagrammatic technique.

\medskip

The main tool applied for studying the $\beta$-eigenvalue model is the
loop equation method. We obtain loop equations from the invariance
of an integral under the special change of variables.
Loop equations for the $\beta$-eigenvalue model
can be found in \cite{Dum},~\cite{eynbeta}, and here we solve them
order by order in $1/N^2$, at fixed $\hbar$.

\medskip

Recently, models of this type got a new vim due to the conjecture by Alday, Gaiotto, and Tachikawa (AGT)~\cite{AGT}
relating Nekrasov's instanton function~\cite{Nekrasov} to conformal blocks of
the Liouville theory; these conformal blocks in turn can be described by
the matrix-like model (see~\cite{MMM},~\cite{EM}); the relation to the Nekrasov's
$\epsilon_{1,2}$ parameters is explicit: $\epsilon_1\epsilon_2\sim 1/N^2$ and $\epsilon_1/\epsilon_2\sim \beta$,
so using the approach in this paper, we can construct {\em nonperturbative} solutions of
Nekrasov's formulas in $\epsilon_1/\epsilon_2$. In this paper, we investigate only the case of
polynomial potentials, the generalization to the realistic logarithmic potentials appearing in the
AGT conjecture will follows.

\medskip

The structure of the paper is as follows: we collect the generalities on the Stokes phenomenon pertaining to
solutions of the Schr\"odinger equation in Sec.~\ref{secdef}. We describe our quantum Riemann surface in
Sec.~\ref{quantumRS} where we introduce $\acycle$- and $\bcycle$-cycles, filling fractions $\td{\epsilon}_i$, and the first-kind
functions (analogues of holomorphic and Krichever--Whitham meromorphic differentials) as well as the system
of flat coordinates and the
Riemann period matrix. In Sec.~\ref{kernels}, we introduce the recursion kernels and the second- and third-kind
(bi-)differentials. In Sec.~\ref{secdefWngFg}, we go beyond the leading approximation in $1/N^2$ and
construct correlation functions of all orders using the Feynman-like diagrammatic technique. We reveal the
origin of our recursion procedure in Sec.~\ref{variations}, where we develop in details the variations w.r.t. the
set of flat coordinates; the summary is in Sec.~\ref{summary}. In the next two (completely new as compared to paper~I) sections, we
investigate the link to the $\beta$-ensemble models (Sec.~\ref{secMM}) and construct on the base of
this analysis the free-energy terms (Sec.~\ref{secFree}). In the first three appendixes to the paper, we
present proofs of the three main theorems of Sec.~\ref{secdefWngFg} concerning properties of the
correlation functions whereas the fourth appendix contains the new formula expressing ${\mathcal F}_0$ through
the filling fractions $\td{\epsilon}_i$; in the matrix model approach,
the singular term has the structure $\frac12 \td{\epsilon}_i^2\log \td{\epsilon}_i$
whereas in the quantum geometry this term is proportional to $\int\log \Gamma(\td{\epsilon}_i)$, which is the first
actual example of calculations in the case of quantum Riemann surfaces.

\section{Schr\"odinger equation and resolvents}\label{secdef}

\subsection{Solutions of the Schr\"odinger equation}
We begin with the  Schr\"odinger equation
\beq\label{eqSchroedinger}
\hbar^2 \psi''(x) = U(x)\, \psi(x)
\eeq
with $U(x)$ being a polynomial of even degree $2d$ for which
we define the polynomial ``potential" $V(x)$ of degree $d+1$ to be
\beq\label{defV'}
V'(x) = 2\,(\sqrt{U})_+ = \sum_{k=0}^d t_{k+1}\,x^k
\eeq
We also define the polynomial of degree $d-1$,
\beq\label{defP}
P(x) = \frac{V'^2(x)}{4} - U(x) -\hbar \frac{V''(x)}{2}.
\eeq

Eventually, we define:
\beq\label{deft0}
t_0 = \mathop{{\rm lim}}_{x\to\infty}\, \frac{xP(x)}{V'(x)}
\eeq
In the matrix model language (see section \ref{secMM}), $t_1,\dots,t_{d+1}$ are called
the {\em times} associated to the potential $V(x)$, $t_0$ is the normalized
total number of eigenvalues (particles), or
the temperature, whereas the remaining coefficients of $P$ are defined by
introducing fixed ``filling fractions'' $\epsilon_i$ below.

\subsubsection{Stokes Sectors}

A function $\psi(x)$ that is a solution of the
Schr\"odinger equation exhibits the Stokes phenomenon, i.e.,
although $\psi(x)$ is an entire function, its asymptotics are
discontinuous near $\infty$ where it has an essential singularity.
Let $\theta_0= {\rm Arg}(t_{d+1})$
be the argument of the leading coefficient of the potential  $V(x)$.
We define the Stokes half-lines the asymptotic directions along which $\Re V(x)$ vanishes asymptotically,
$L_k = \left\{x\,\, / \,\, {\rm Arg}(x) =-\frac{\theta_0}{d+1} + \pi \,\frac{k+\frac{1}{2}}{d+1}\, \right\}$,
together with the corresponding Stokes sectors:
\beq
S_k =  \left\{{\rm Arg}(x) \in \left]-\frac{\theta_0}{d+1}+ \pi \,
\frac{k-\frac{1}{2}}{d+1},-\frac{\theta_0}{d+1}+ \pi\,\frac{k+\frac{1}{2}}{d+1}\right[\, \right\}
\eeq
i.e., $S_k$ is the sector between $L_{k-1}$ and $L_k$.

Notice that in even sectors we have asymptotically $\Re V(x)>0$ and in odd sectors we have $\Re V(x)<0$.

\begin{figure}[h]
\begin{center}
{\psset{unit=0.8}
\begin{pspicture}(-8,-4)(8,4)
\newcommand{\STOKES}{%
\pcline[linewidth=1pt,linestyle=dashed](0,0)(4,0)
\rput(3.5,.2){\tiny$\bullet$}
\rput(3.85,.2){\tiny$\bullet$}
\rput(3.6,-0.15){\tiny$\bullet$}
\rput(3.9,-.2){\tiny$\bullet$}
}
\newcommand{\PATTERN}[1]{%
{\psset{unit=#1}
\pscircle[linecolor=white,linewidth=0.5pt,fillstyle=solid,fillcolor=orange](0,0){4.0}
\pswedge[linecolor=orange,linewidth=1pt,fillstyle=solid,fillcolor=yellow](0,0){4.}{0}{45}
}
}
\rput(0,0){\PATTERN{1}}
\rput{90}(0,0){\STOKES}
\rput{135}(0,0){\STOKES}
\rput{180}(0,0){\STOKES}
\rput{225}(0,0){\STOKES}
\rput{270}(0,0){\STOKES}
\rput{315}(0,0){\STOKES}
\rput{45}(0,0){\pcline[linewidth=1pt,linestyle=dashed](0,0)(4,0)}
\pcline[linewidth=1pt,linestyle=dashed](0,0)(4,0)
\rput(0.2,0.5){$\bullet$}
\rput(.9,.2){$\bullet$}
\rput(.7,1){$\bullet$}
\rput(0.2,-1){$\bullet$}
\rput(1,0.3){\makebox(0,0)[lb]{\small\hbox{isolated}}}
\rput(1.5,-0.5){\makebox(0,0)[lb]{\small\hbox{zeros}}}
\rput(2.8,-2.8){\psarc[linewidth=1.5pt]{->}(0,0){5.6}{65}{85}}
\rput(0,4.){\psarc[linewidth=1.5pt]{<-}(0,0){5.6}{320}{340}}
\rput(5.5,2.2){\makebox(0,0)[lb]{\small\hbox{no zero concentration}}}
\end{pspicture}
}
\caption{Example of the Stokes sector partition and structure of zeros
for the Schr\"odinger equation solution $\psi(x)$ that decreases in the light-colored sector and
increases in all other sectors (the degree of the potential $V(x)$ is four).}
\end{center}
\label{fig:Stokes}
\end{figure}
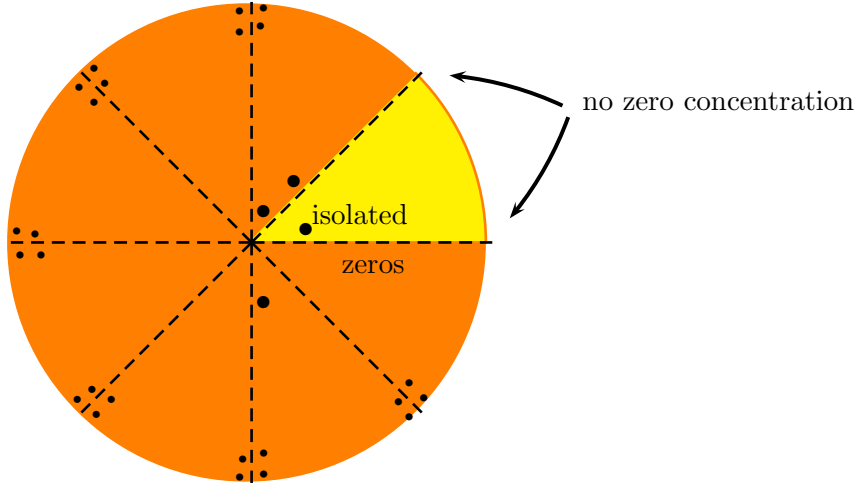

\subsubsection{The Stokes phenomenon. Decreasing solution}

From the study of the Schr\"odinger equation it is known that $\psi(x)$ is an entire
function having a large $x$ expansion in each sector $S_k$,
\beq
\label{psix}
\psi(x)  \mathop{{\sim}}_{S_k} e^{\pm\,{1\over
2\hbar}V(x)}\,x^{C_{k}}\,\, (A_k+\frac{B_k}{x}+\dots)
\eeq
and the
sign $\pm$, may jump discontinuously from one sector to another as
well as the numbers $A_k, B_k, C_k,\dots$  (and in general, all the
coefficients of the series in $\frac{1}{x^j}$ at infinity).\footnote{The corresponding series
is asymptotic, so we cannot continue it analytically to other sectors.}

In every sector $S_k$ there exists a unique solution that decreases
exponentially along each direction inside the sector. We now separate solutions in the
even and odd sectors and consider the set $\{\psi_\alpha(x)\}$ of solutions each of which
decreases in the corresponding even sector. We therefore introduce a {\em sectorwise} system of solutions to the
Schr\"odinger equation.

An important and useful result is the Stokes theorem, which claims
that if the asymptotics of $\psi(x)$ is exponentially small in some
sector, then the same asymptotic series expansion (\ref{psix}) is valid in the two adjacent sectors
(and therefore $\psi(x)$ is exponentially large in those two
sectors).

In the general case, (i.e., for a generic potential $U(x)$), the solution
$\psi_\alpha(x)$ decreases only in the sector $S_\alpha$, and is exponentially
large in all other sectors. But if the Schr\"odinger potential
$U(x)$ is non-generic, then there may exist several
sectors in which $\psi_\alpha(x)$ is exponentially small (which means that $\psi_{\alpha_1}(x)=\psi_{\alpha_2}(x)$
for some $\alpha_1\ne \alpha_2$).

In what follows, we mainly consider the general case, so in what follows we
assume all the functions $\psi_\alpha$ to be different if not stating the opposite.

The case studied in \cite{EynOM} was the most degenerate case in which one and the same
solution $\psi$ is exponentially small in $d+1$ sectors.

\subsubsection{Zeroes of $\psi$}
Every $\psi_\alpha(x)$ is an entire function
with an essential singularity at $\infty$, and with isolated zeroes
$s^{(\alpha)}_i$, $\psi_\alpha(s^{(\alpha)}_{i})=0$. The number of these zeros can be finite or
infinite. In the latter case, zeroes may only accumulate near $\infty$, and only along the Stokes
half-lines $L_j$ bordering the sectors (see
fig.\ref{fig:Stokes}). This accumulation of zeroes
along the half--line $L_j$ occurs if and only if $\psi_\alpha(x)$ is exponentially
large on both sides of the half-line. So, no accumulation of zeros of the function $\psi_\alpha(x)$
occurs along the lines that border the $\alpha$'s sector, and this function can therefore have only a finite
number of zeros inside the ``bigger'' sector when we join the $\alpha$'s sector with the adjacent parts of
the two neighboring sectors.

If $U(x)$ is generic, then each of $\psi_\alpha(x)$ has an infinite number of zeroes,
the zeroes accumulate at $\infty$ along all half-lines $L_j$
with $j\neq \alpha,\alpha-1$.

In paper~I, we define the genus of the Schr\"odinger equation to be related to the number of
half-lines of zeros accumulation of a selected function $\psi_0$. However, this definition is scheme-dependent,
and we can in principle obtain different genera for the very same
function $U(x)$. The clear understanding of this is still lacking; a possible explanation is that
we actually deal with different sections of an ambient infinite-genus quantum surface.

\subsubsection{Sheets}

In sector $S_\alpha$ we have the asymptotic behavior
\beq
\psi_\alpha(x)  \sim e^{-\hbar V(x)/2}\,x^{ t_0/ \hbar} (A_\alpha+\frac{B_\alpha}{x}+\dots),
\eeq
and the function $\psi_\alpha$ has the same asymptotic behavior in the two adjacent sectors.

We consider an $\alpha$'s sheet of the quantum Riemann surface to be the union of these three sectors with possible
analytic continuation into a finite domain of the complex plane. We consider only the sheets enumerated by even $\alpha$
and, in contrast to paper~I, introduce democracy of sheets: all of them will be equivalent in the approach of this paper.
Sheets obviously overlap; we have to choose boundaries (cuts) between them.

\subsection{Resolvent}

The first ingredient of our strategy is to define a resolvent similar to the one in matrix models.

\bd
We define the resolvent sectorwise:
\beq
\label{om}
\om(\sheet{\alpha}{x}) = \hbar \frac{\psi_\alpha'(x)}{\psi_\alpha(x)} +\frac{V'(x)}{2},\ \hbox{for}\ x\in S_\alpha.
\eeq
\ed

For a quantity defined sectorwise we indicate it by setting the sector index above the variable, as shown in (\ref{om}).

It follows from this definition that $\om(x)$ has simple poles at zeros of $\psi_\alpha$ in the corresponding
sector. The boundaries between sectors overlap, but in what follows we fix them in a more explicit form (see the partition
of the complex plane by ${\acycle}$-cycles).

A straightforward computation then gives
\beq
\om(\sheet{\alpha}{x}) \mathop{{\sim}}_{x\to {\infty_{\alpha},\infty_{\alpha\pm 1}}}
\frac{t_0}{x}  +O(1/x^2),
\eeq
that is, in each sheet the resolvent possesses asymptotic properties of a standard matrix-model resolvent.

An important property of any solution $\psi_\alpha$ is that
\beq
\Res_{s^{(\alpha)}_i} {1\over \psi_\alpha^2(x)} = 0
\eeq

The main property of $\om(\sheet{\alpha}{x})$ is that it satisfies the Ricatti equation.
We obtain
\bea
\label{eqBethe1}
V'(x)\om(\sheet{\alpha}{x}) - \om^2(\sheet{\alpha}{x}) - \hbar \om'(\sheet{\alpha}{x})
&=& {V'(x)^2\over 4} - \hbar^2{\psi_\alpha''(x)\over \psi_\alpha(x)} -\hbar {V''(x)\over 2} \cr
&=& {V'(x)^2\over 4} - U(x) -\hbar {V''(x)\over 2} \cr
&=& P(x),
\eea
with $P(x)$ being a polynomial of degree $d-1$ in $x$ and this polynomial is one and the same for all
sheets of the quantum Riemann surface introduced below.

\section{Quantum Riemann Surface}\label{quantumRS}

In this section we define the notions of $\acycle$- and $\bcycle$-cycles and the first kind differentials dual to them.

\subsection{The contour ${\mathcal C}_D$ and the set of ${\acycle}$- and $\bcycle$-cycles}

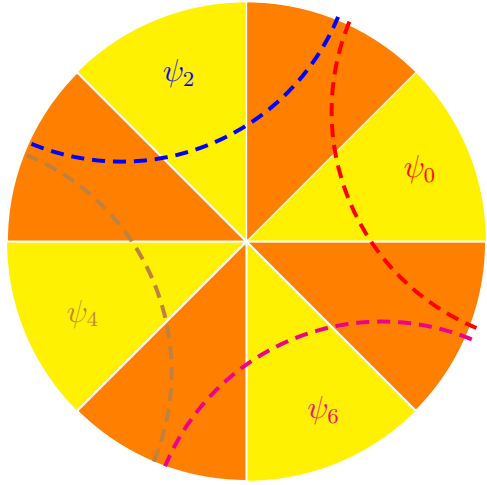
\begin{figure}[h]
\begin{center}
{\psset{unit=0.8}
\begin{pspicture}(-8,-5)(8,5)
\newcommand{\PATTERN}[1]{%
{\psset{unit=#1}
\pscircle[linecolor=white,linewidth=0.5pt,fillstyle=solid,fillcolor=yellow](0,0){4.0}
\pswedge[linecolor=white,linewidth=0.5pt,fillstyle=solid,fillcolor=orange](0,0){4.}{45}{90}
\rput{90}(0,0){\pswedge[linecolor=white,fillstyle=solid,fillcolor=orange](0,0){4.}{45}{90}}
\rput{180}(0,0){\pswedge[linecolor=white,fillstyle=solid,fillcolor=orange](0,0){4.}{45}{90}}
\rput{270}(0,0){\pswedge[linecolor=white,fillstyle=solid,fillcolor=orange](0,0){4.}{45}{90}}
}
}
\rput(0,0){\PATTERN{1}}
\rput{22.5}(0,0){\psarc[linecolor=red, linestyle=dashed, linewidth=1.5pt](5.66,0){3.9}{135}{225}}
\rput{112.5}(0,0){\psarc[linecolor=blue, linestyle=dashed, linewidth=1.5pt](5.66,0){3.9}{135}{225}}
\rput{202.5}(0,0){\psarc[linecolor=brown, linestyle=dashed, linewidth=1.5pt](5.66,0){3.9}{135}{225}}
\rput{292.5}(0,0){\psarc[linecolor=magenta, linestyle=dashed, linewidth=1.5pt](5.66,0){3.9}{135}{225}}
\rput(2.8,1.2){\makebox(0,0){\textcolor{red}{$\psi_0$}}}
\rput(-1.2,2.8){\makebox(0,0){\textcolor{blue}{$\psi_2$}}}
\rput(-2.8,-1.2){\makebox(0,0){\textcolor{brown}{$\psi_4$}}}
\rput(1.2,-2.8){\makebox(0,0){\textcolor{magenta}{$\psi_6$}}}
\end{pspicture}
}
\caption{The original integration contour ${\mathcal C}_D$.}
\end{center}
\label{fig:CD}
\end{figure}

In papers on matrix models (on an early stage, before coming to residues at the branch points), we have
the special contour of integration, ${\mathcal C}_D$, that encircles all the singularities of resolvents leaving apart all other
possible singular points.
The analogue of such a contour in our case is the union of $d+1$ contours, one per each sheet, that pairwise coincide in far asymptotic
domains of odd Stokes sectors and separate all the zeros of the function $\psi_\alpha$ from the infinity $\infty_\alpha$
(which is always possible because we have a finite number of zeros in each sheet). We have (see Fig.~\ref{fig:CD})
\beq
\label{contourCD}
\oint_{{\mathcal C}_D}f(x)dx\equiv \sum_{\alpha}\int_{\infty_{\alpha-1}}^{\infty_{\alpha+1}}f(\sheet{\alpha}{x})dx
\eeq
for {\em any} function $f(\sheet{\alpha}{x})$ that has no asymptotic zero concentration along the boundary lines of the sector $S_\alpha$.
Here and hereafter, we assume that $f(\sheet{\alpha}{x})$ may depend on a finite number of
derivatives of the function $\psi_\alpha(x)$, the symbol $f(\sheet{\alpha}{x})$ then indicates that we substitute the
solution $\psi_\alpha(x)$ as an argument.

\subsubsection{${\acycle}$- and $\bcycle$-cycles}

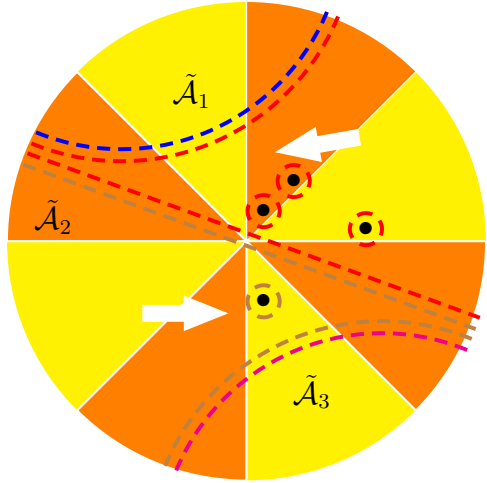
\begin{figure}[h]
\begin{center}
{\psset{unit=0.8}
\begin{pspicture}(-8,-5)(8,5)
\newcommand{\PATTERN}[1]{%
{\psset{unit=#1}
\pscircle[linecolor=white,linewidth=0.5pt,fillstyle=solid,fillcolor=yellow](0,0){4.0}
\pswedge[linecolor=white,linewidth=0.5pt,fillstyle=solid,fillcolor=orange](0,0){4.}{45}{90}
\rput{90}(0,0){\pswedge[linecolor=white,fillstyle=solid,fillcolor=orange](0,0){4.}{45}{90}}
\rput{180}(0,0){\pswedge[linecolor=white,fillstyle=solid,fillcolor=orange](0,0){4.}{45}{90}}
\rput{270}(0,0){\pswedge[linecolor=white,fillstyle=solid,fillcolor=orange](0,0){4.}{45}{90}}
}
}
\newcommand{\ARROW}[1]{%
{\psset{unit=#1}
\pspolygon[linecolor=white,linewidth=0.5pt,fillstyle=solid,fillcolor=white](0,0)(1,0.4)(1,-0.4)
\psframe[linecolor=white,linewidth=0.5pt,fillstyle=solid,fillcolor=white](0.8,-0.2)(2,0.2)
}
}
\rput(0,0){\PATTERN{1}}
\rput{112.5}(0,0){\psarc[linecolor=blue, linestyle=dashed, linewidth=1.5pt](5.66,0){3.7}{135}{225}}
\rput{112.5}(0,0){\psarc[linecolor=red, linestyle=dashed, linewidth=1.5pt](5.66,0){3.9}{135}{225}}
\rput{-20}(0,0){\pcline[linecolor=red, linestyle=dashed, linewidth=1.5pt](-4,0.1)(4,0.1)}
\rput{-20}(0,0){\pcline[linecolor=brown, linestyle=dashed, linewidth=1.5pt](-4,-0.1)(4,-0.1)}
\rput{292.5}(0,0){\psarc[linecolor=magenta, linestyle=dashed, linewidth=1.5pt](5.66,0){3.7}{135}{225}}
\rput{292.5}(0,0){\psarc[linecolor=brown, linestyle=dashed, linewidth=1.5pt](5.66,0){3.9}{135}{225}}
\rput{10}(0.5,1.5){\ARROW{0.7}}
\rput{180}(-0.5,-1.2){\ARROW{0.7}}
\rput(0.2,0.5){$\bullet$}
\pscircle[linecolor=red, linestyle=dashed, linewidth=1.5pt](0.2,0.5){.3}
\rput(1.9,.2){$\bullet$}
\pscircle[linecolor=red, linestyle=dashed, linewidth=1.5pt](1.9,0.2){.3}
\rput(.7,1){$\bullet$}
\pscircle[linecolor=red, linestyle=dashed, linewidth=1.5pt](0.7,1){.3}
\rput(0.2,-1){$\bullet$}
\pscircle[linecolor=brown, linestyle=dashed, linewidth=1.5pt](0.2,-1){.3}
\rput(-1,2.5){$\tilde {\acycle}_1$}
\rput(-3.3,0.4){$\tilde {\acycle}_2$}
\rput(1,-2.5){$\tilde {\acycle}_3$}
\end{pspicture}
}
\caption{Example of pushing the contour ${\mathcal C}_D$ from infinities
to the set of $\tilde {\acycle}$-cycles.}
\end{center}
\label{fig:An}
\end{figure}

We now deform the integration contour ${\mathcal C}_D$ pushing through the ``middle'' part of the complex plane and
taking the residues at the zeros $s_i^{(\alpha)}$ of the corresponding functions $\psi_\alpha$ as shown in Fig.~\ref{fig:An}.
On the way we might break some contours presenting them as the unions of newly introduced contours all of which are stretched between
different asymptotic directions.
As a result, we obtain a system of exactly $2d$ contours in which (leaving aside the residues at zeros $s_i^{(\alpha)}$)
all the contours are pairwise identified and represent edges of $d$ ``cuts''.
As the result, we obtain a complete system of $d$ cuts ${\widetilde {\acycle}}_i$, $i=1,\dots,d$,
that separate all the odd-numbered infinities\footnote{In what follows, we identify
an infinity ``point'' with the corresponding number with the related asymptotic direction.} and determining the corresponding
sheets of the quantum Riemann surface. If the functions $\psi_\alpha(x)$ coincide for some sheets, then we can identify these
sheets. Note that we definitely
have an arbitrariness in constructing this system of cuts; we can also arbitrarily
assign the residues inside the sheet to belong to
one of several contours bounding this sheet.

We call the cut separating two sheets a {\em cycle} ${\tilde {\acycle}}_\alpha$, and it is characterized by four indices:
$\alpha_+$ and $\alpha_-$ are indices of the sheets separated by this cut (they are even numbered in our classification);
$\td{\alpha}_+$ and $\td{\alpha}_-$ are indices of infinities that are asymptotic for this cut (they are odd numbered).

To each complete set $\{\widetilde{\acycle}_\alpha\}_{\alpha=1}^d$ of $\widetilde{\acycle}$-cycles we uniquely set into the
correspondence the set $\{\widetilde{\bcycle}_\alpha\}_{\alpha=1}^d$ of $\widetilde{\bcycle}$-cycles that go pairwise
between the even-numbered infinities ($\infty_{\alpha_+}$ and $\infty_{\alpha_-}$) such that the intersection index
$\widetilde{\acycle}_\alpha\circ \widetilde{\bcycle}_\beta=\delta_{\alpha,\beta}$.

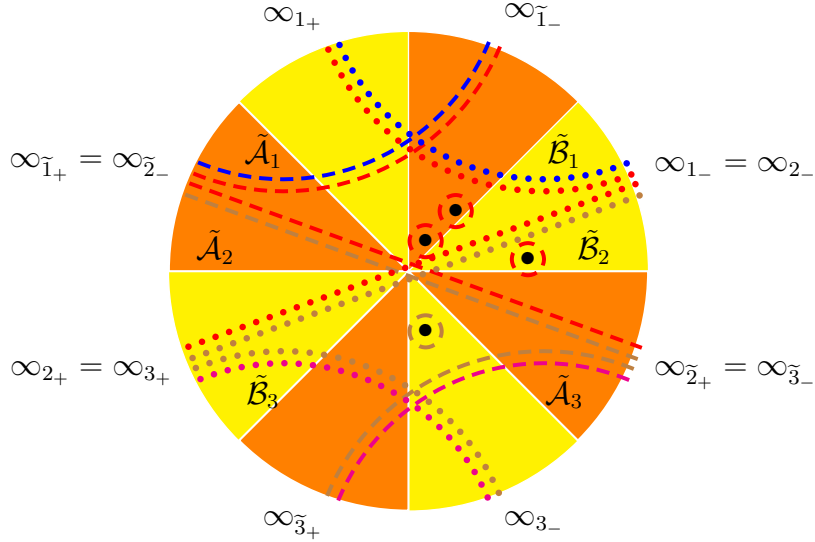
\begin{figure}[h]
\begin{center}
{\psset{unit=0.8}
\begin{pspicture}(-8,-6)(8,5)
\newcommand{\PATTERN}[1]{%
{\psset{unit=#1}
\pscircle[linecolor=white,linewidth=0.5pt,fillstyle=solid,fillcolor=yellow](0,0){4.0}
\pswedge[linecolor=white,linewidth=0.5pt,fillstyle=solid,fillcolor=orange](0,0){4.}{45}{90}
\rput{90}(0,0){\pswedge[linecolor=white,fillstyle=solid,fillcolor=orange](0,0){4.}{45}{90}}
\rput{180}(0,0){\pswedge[linecolor=white,fillstyle=solid,fillcolor=orange](0,0){4.}{45}{90}}
\rput{270}(0,0){\pswedge[linecolor=white,fillstyle=solid,fillcolor=orange](0,0){4.}{45}{90}}
}
}
\newcommand{\ARROW}[1]{%
{\psset{unit=#1}
\pspolygon[linecolor=white,linewidth=0.5pt,fillstyle=solid,fillcolor=white](0,0)(1,0.4)(1,-0.4)
\psframe[linecolor=white,linewidth=0.5pt,fillstyle=solid,fillcolor=white](0.8,-0.2)(2,0.2)
}
}
\rput(0,0){\PATTERN{1}}
\rput{112.5}(0,0){\psarc[linecolor=blue, linestyle=dashed, linewidth=1.5pt](5.66,0){3.7}{135}{225}}
\rput{112.5}(0,0){\psarc[linecolor=red, linestyle=dashed, linewidth=1.5pt](5.66,0){3.9}{135}{225}}
\rput{67.5}(0,0){\psarc[linecolor=blue, linestyle=dotted, linewidth=2.5pt](5.66,0){3.7}{135}{225}}
\rput{67.5}(0,0){\psarc[linecolor=red, linestyle=dotted, linewidth=2.5pt](5.66,0){3.9}{135}{225}}
\rput{-20}(0,0){\pcline[linecolor=red, linestyle=dashed, linewidth=1.5pt](-4,0.1)(4,0.1)}
\rput{-20}(0,0){\pcline[linecolor=brown, linestyle=dashed, linewidth=1.5pt](-4,-0.1)(4,-0.1)}
\rput{20}(0,0){\pcline[linecolor=red, linestyle=dotted, linewidth=2.5pt](-4,0.1)(4,0.1)}
\rput{20}(0,0){\pcline[linecolor=brown, linestyle=dotted, linewidth=2.5pt](-4,-0.1)(4,-0.1)}
\rput{292.5}(0,0){\psarc[linecolor=magenta, linestyle=dashed, linewidth=1.5pt](5.66,0){3.7}{135}{225}}
\rput{292.5}(0,0){\psarc[linecolor=brown, linestyle=dashed, linewidth=1.5pt](5.66,0){3.9}{135}{225}}
\rput{247.5}(0,0){\psarc[linecolor=magenta, linestyle=dotted, linewidth=2.5pt](5.66,0){3.7}{135}{225}}
\rput{247.5}(0,0){\psarc[linecolor=brown, linestyle=dotted, linewidth=2.5pt](5.66,0){3.9}{135}{225}}
\rput(0.2,0.5){$\bullet$}
\pscircle[linecolor=red, linestyle=dashed, linewidth=1.5pt](0.2,0.5){.3}
\rput(1.9,.2){$\bullet$}
\pscircle[linecolor=red, linestyle=dashed, linewidth=1.5pt](1.9,0.2){.3}
\rput(.7,1){$\bullet$}
\pscircle[linecolor=red, linestyle=dashed, linewidth=1.5pt](0.7,1){.3}
\rput(0.2,-1){$\bullet$}
\pscircle[linecolor=brown, linestyle=dashed, linewidth=1.5pt](0.2,-1){.3}
\rput(-2.5,2){$\tilde {\acycle}_1$}
\rput(-3.3,0.4){$\tilde {\acycle}_2$}
\rput(2.5,-2){$\tilde {\acycle}_3$}
\rput(2.5,2){$\tilde {\bcycle}_1$}
\rput(3,0.4){$\tilde {\bcycle}_2$}
\rput(-2.5,-2){$\tilde {\bcycle}_3$}
\rput[lb](1.5,4){$\infty_{\td{1}_-}$}
\rput[rb](-4,1.5){$\infty_{\td{1}_+}=\infty_{\td{2}_-}$}
\rput[lt](4,-1.5){$\infty_{\td{2}_+}=\infty_{\td{3}_-}$}
\rput[rt](-1.5,-4){$\infty_{\td{3}_+}$}
\rput[rb](-1.5,4){$\infty_{{1}_+}$}
\rput[lb](4,1.5){$\infty_{{1}_-}=\infty_{{2}_-}$}
\rput[rt](-4,-1.5){$\infty_{{2}_+}=\infty_{{3}_+}$}
\rput[lt](1.5,-4){$\infty_{{3}_-}$}
\end{pspicture}
}
\caption{The pattern of ${\tilde {\acycle}}$- (dashed lines) and ${\tilde {\bcycle}}$- (dotted lines) cycles for the
example in Fig.~\ref{fig:An}.}
\end{center}
\label{fig:AB}
\end{figure}

\bd\label{defabcycleint}
We define the integrals over the cycles ${\tilde {\acycle}}_\alpha$ and the conjugate cycle ${\tilde {\bcycle}}_\alpha$ to be
(see Fig.~\ref{fig:AB})
\beq
\label{defacycleint}
\oint_{{\tilde {\acycle}}_\alpha} \, f(x) dx \stackrel{{\rm def}}{=}\, \int_{\infty_{\td{\alpha}_-}}^{\infty_{\td{\alpha}_+}}\,
\bigr(f(\sheet{\alpha_+}{x})-f(\sheet{\alpha_-}{x}) \bigl)\, dx+\sum\mathop{res}_{s^{(\alpha_\pm)}_i(\alpha)}f(\sheet{\alpha_\pm}{x})
\eeq
and
\beq\label{defbcycleint}
\oint_{{\tilde {\bcycle}}_\alpha} \, f(x) dx \stackrel{{\rm def}}{=}\, \int_{\infty_{{\alpha}_-}}^{\infty_{{\alpha}_+}}\,
(f(\sheet{\alpha_+}{x})-f(\sheet{\alpha_-}{x}) )\, dx,
\eeq
where the residues in the first expression are taken at those zeros of $\psi_{\alpha_\pm}$ that are assigned to the
corresponding contour.
\ed

Because the prescription for the sheet assignment follows from the definitions (\ref{defacycleint}) and (\ref{defbcycleint})
of the cycle integrals, we omit the sheet labels in the corresponding integrands.

\br
Assignment of residues in the $\alpha$th sheet to the contours bounding this sheet is arbitrary; we have therefore
a (discrete) ambiguity in the definition (\ref{defacycleint}) of the ${\tilde {\acycle}}$-cycle integrals. However, the notion of
the integral over ${\mathcal C}_D$ is well defined and does not depend on the choice of ${\tilde {\acycle}}$-cycles.
\er

Obviously, $\oint_{{\mathcal C}_D}f(x)dx=\sum_{i=1}^d\oint_{{\tilde {\acycle}}_i}f(x)dx$.

We now introduce the ``genuine'' ${\acycle}$- and
${\bcycle}$-cycles, which are straightforward analogues of the set of
${\acycle}$- and ${\bcycle}$-cycles on a standard Riemann surface.
For this, we select one among the $\widetilde {\acycle}$-cycles, say,
the cycle $\widetilde {\acycle}_d$ and the conjugate cycle $\widetilde {\bcycle}_d$.
Then, we identify ${\acycle}_i={\widetilde {\acycle}}_i$
and ${\bcycle}_i={\widetilde {\bcycle}}_i-{\widetilde {\bcycle}}_d$ for
$i=1,\dots,d-1$ in the sense of Definition~\ref{defabcycleint}, that is
\bea
\label{abcycleint}
&&\oint_{{{\acycle}}_i} \, f(x) dx \stackrel{{\rm def}}{=}\,\oint_{{\tilde {\acycle}}_i} \, f(x) dx,\cr
&&\oint_{{{\bcycle}}_i} \, f(x) dx \stackrel{{\rm def}}{=}\,
\oint_{{\tilde {\bcycle}}_i} \, f(x) dx -\oint_{{\tilde {\bcycle}}_d} \, f(x) dx\ \ \hbox{for}\ \ i=1,\dots,d-1,
\eea
and we call the number $d-1=g$ of independent ${\acycle}$- and ${\bcycle}$- cycles the {\em genus} of the quantum Riemann surface.

The newly introduced ${\acycle}$- and ${\bcycle}$-cycles again satisfy the standard intersection formula,
\beq
{\acycle}_\alpha\cap {\bcycle}_\beta=\delta_{\alpha,\beta},
\eeq
and most of our construction features depend only on the
homology class of the paths ${\acycle}_\alpha$, ${\bcycle}_\alpha$ at the asymptotic infinities, but in the intermediate
considerations it is useful to choose a representant,
the intersection point $P_\alpha$,
\beq
{\acycle}_\alpha\cap {\bcycle}_\alpha=\{P_\alpha\}.
\eeq

\subsection{Filling fractions}

In random matrices, the notion of filling fractions is just the $\acycle$-cycle integrals of the resolvent.
Then, if the $\acycle$-cycles are chosen to lie in the physical sheet (which is possible, say, in the hyperelliptic case),
the discontinuity of the resolvent along the corresponding cuts determines the eigenvalue density and the $\acycle$-cycle integrals
determine the portions of eigenvalues lying on the corresponding interval of eigenvalue distribution. They are called
therefore the filling fractions.

In the case of quantum surface, we define the ``filling fractions'' $\widetilde{\epsilon}_\alpha$ to be
\beq
\label{deffilling}
\widetilde{\epsilon}_\alpha
= {1\over 2i\pi}\,\oint_{\td{{\acycle}}_\alpha} \om(x)dx \stackrel{{\rm def}}{=} \int_{\infty_{\td{\alpha}_-}}^{\infty_{\td{\alpha}_+}}
\bigl(\om(\sheet{\alpha_+}{x})-\om(\sheet{\alpha_-}{x})\bigr)dx,\quad \alpha=1,\dots,d.
\eeq
Note that this definition depends on where we place the contours and (in the case where a sheet is bounded by more than one
$\td {\acycle}$-cycle) we also have a freedom to assign residues inside the sheet to different $\td {\acycle}$-cycles, so
the filling fractions are defined up to integers times $\hbar$.

For the difference, we have
$$
\om(\sheet{\alpha_+}{x})-\om(\sheet{\alpha_-}{x})=\frac{w_{\alpha_+,\alpha_+}}{\psi_{\alpha_+}(x)\psi_{\alpha_-}(x)},
$$
where $w_{\alpha_+,\alpha_+}=\psi'_{\alpha_+}\psi_{\alpha_-}-\psi'_{\alpha_-}\psi_{\alpha_+}$ is the Wronskian
of the two solutions. Therefore, this difference decreases exponentially in sectors where the both solutions
$\psi_{\alpha_+}$ and $\psi_{\alpha_-}$ increase, and this is why we identify the asymptotic
domains of the $\acycle$-cycle integrals with ``branch points.''

We have
\beq
\sum_{\alpha=1}^d \widetilde{\epsilon}_\alpha = t_0,
\eeq
which just follows from that the sum of integrals over $\td {\acycle}$-cycles is equivalent to evaluating the integral over ${\mathcal C}_D$.
This also means that we should take as independent variables only
$d-1=g$ of the variables $\td{\epsilon}_\alpha$ if we consider $t_0$ to be an independent variable,
and we naturally choose these $g$ variables $\epsilon_\alpha$ to be filling
fractions corresponding to the cycles $\acycle_\alpha$ of the quantum Riemann surface.

\br
In the case $g=-1$ in~\cite{ChEyMarI}, the only filling fraction is
$\widetilde{\epsilon}_d=t_0$, and it is given by the (finite) sum of residues of the function $\om$ at the zeros $s_i$:
$$
\widetilde{\epsilon}_d=t_0=\sum_i \Res_{s_i}\om = \hbar\,\#\{s_i\},
$$
so $t_0$ is discrete in this case.
For $g\ge 0$, the variables
$\widetilde{\epsilon}_\alpha$, $\alpha=1,\dots,g$, and $t_0$ may take arbitrary, not necessarily integer, values.
\er

\subsection{First kind functions}

After defining the cycles, another important step is to define the
equivalent of the first, second and third kind differentials. We begin
with the definition of the first-kind differentials.

Let $h_k$, $k=1,\dots,d-1$, be a basis in the complex vector
space of polynomials of degree $\leq d-2$.

We introduce the functions
\beq
v_k(\sheet{\alpha}{x}) = {1\over \hbar\, \psi_\alpha^2(x)}\,\int_{\infty_\alpha}^x h_k(x')\,\psi_\alpha^2(x')\,dx'.
\label{VK}
\eeq
We use the same polynomial $h_k(x')$ for all the sheets of the Riemann surface.

Note that because every $\psi_\alpha(x)$ is a solution to the Schr\"odinger
equation, $v_{k}(\sheet{\alpha}{x})$ has double poles
with no residue at the $s^{(\alpha)}_{j}$ (at the zeroes of $\psi_\alpha$), and behaves
like $O(1/x^2)$ in the sector $S_\alpha$ and inside all the sectors
in which $\psi_\alpha$ is exponentially large (if the polynomial $h_k(x')$ has power less than
$d-2$). Therefore, the following integrals are well defined in the general case:
\beq
I_{k,\alpha} = \oint_{\acycle_\alpha} v_k(x)\,dx
\quad \alpha=1,\dots,g, \, k=1,\dots,d-1.
\eeq
If the matrix $I_{k,\alpha}$ with $k,\alpha=1,\dots,d-1$ has the full rank (which we assume in what follows),
then it is possible to choose the canonically normalized basis of $h_k$ such that
\beq
\label{eqdualvkacycle}
I_{k,\alpha}=\delta_{k,\alpha}.
\eeq


The functions $v_k(x) \, k=1,\dots,g$ are therefore the
natural analogues of canonically normalized holomorphic forms (1st kind differentials).

We now extend this notion to the {\it meromorphic} (Whitham--Krichever)~\cite{Krich} differentials. For this, let us
consider the following basis $h_k$, $k=1,\dots$, in the space of polynomials of arbitrary order: the first $d-1$ elements of this basis
are the original polynomials $h_k$ each of which has degree not higher than $d-2$, each polynomial $h_k$ with $k>d-1$ has degree
exactly $k-1$ and must be chosen on the following grounds.

Define the functions $v_k(\sheet{\alpha}{x})$ with $k>d-1$ exactly as in (\ref{VK}) with $h_k$ being now a polynomial of
arbitrary (but fixed) degree $k-1$. The coefficients of $h_k$ with $k\ge d-1$ are unambiguously fixed by the normalization
conditions:
\begin{itemize}
\item (the residue condition)
\beq
\oint_{{\mathcal C}_D}x^{-l}\,v_k(x) dx=\delta_{l,k-d}, \quad l=0,1,\dots, \ k\ge d-1.
\label{residue}
\eeq
\item (the normalizing condition)
\beq
\oint_{{\acycle}_\alpha}v_k(x) dx=0, \ \alpha=1,\dots,d-1, \ k=d,\dots.
\label{norm}
\eeq
\end{itemize}

\br
Although the functions $v_k(\sheet{\alpha}x)$ generally increase as $x^{k-d}$ as $x\to\infty$, the integral
(\ref{residue}) as well as the normalizing condition (\ref{norm}) are well defined for any finite $l$ and $k$.
This is because the difference $v_k(\sheet{\alpha_+}x)-v_k(\sheet{\alpha_-}x)$ is exponentially small
as $x\to\infty_{{\tilde\alpha}_\pm}$ for any $k$, and we can integrate it along ${\mathcal C}_D$ weighted by any polynomially growing
function. The integral (\ref{residue}) is therefore a natural analogue of the residue at infinity of order $l+1$.
\er

%
%

\subsection{Riemann matrix of periods}

An interesting quantity in standard algebraic geometry is the
Riemann matrix of periods provided by integrals of the
holomorphic differentials over $\bcycle$-cycles. So,
an analogous ``quantum'' Riemann period matrix $\tau_{i,j}$, $i,j=1,\ldots,g$ is
\beq
\tau_{\alpha,i} \stackrel{{\rm def}}{=} \oint_{\bcycle_\alpha} v_i(x) dx.
\eeq

Note that this definition makes sense since $v_i(x)$ ($i=1,\dots, g$) behaves as $O(1/x^2)$ in the sectors asymptotic
for the $\bcycle$-cycles. And because the residues of $v_i(\sheet{\alpha}{x})$ vanish at all zeros $s_j^{(\alpha)}$,
these integrals depend only on the homology class of $\bcycle$-cycles.

Like for the classical Riemann matrix of periods we have the following property:

\begin{theorem}
The period matrix $\tau$ is symmetric: $\tau_{i,j}=\tau_{j,i}$.
\end{theorem}

\proof{
This result follows from Theorem \ref{thBergmanABcycles} below, since:
$$
\oint_{\mathcal{B}_{\beta}} dx \oint_{\mathcal{B}_{\alpha}}B(x,z)dz=2i\pi \oint_{\mathcal{B}_{\beta}} dx v_{\alpha}(x)=2i\pi \tau_{\beta,\alpha}
$$
and from the symmetry theorem \ref{thBergmanSymmetry} for the Bergman kernel, $B(\sheet{\alpha}{x},\sheet{\beta}{z})=
B(\sheet{\beta}{z},\sheet{\alpha}{x})$.
}

\section{Recursion kernels}\label{kernels}

One of the key geometric objects in \cite{EynOM} and in \cite{EOFg}, is the ``recursion kernel" $K(x,z)$.
It was used in the context of matrix models~\cite{CEO} for constructing a solution of loop equations.
We use its analogue below for constructing the 3rd and 2nd kind differentials.

\subsection{The recursion kernel $K$}

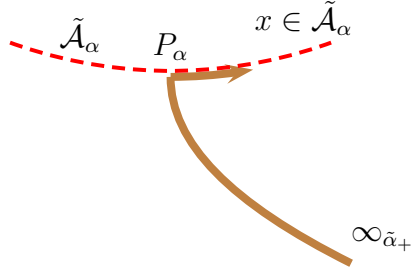
\begin{figure}[h]
\begin{center}
{\psset{unit=0.8}
\begin{pspicture}(-8,-3)(8,3)
\psbezier[linecolor=brown, linewidth=3pt](0,0.1)(0,-1)(1,-2)(3,-3)
\psarc[linecolor=brown, linewidth=3pt]{->}(0,8){7.9}{270}{280}
\psarc[linecolor=red, linestyle=dashed, linewidth=1.5pt](0,8){7.8}{250}{290}
\rput[cb](0,0.6){$P_\alpha$}
\rput[lb](1.4,0.8){$x\in\tilde {\acycle}_\alpha$}
\rput[lb](3,-2.8){$\infty_{\tilde \alpha_+}$}
\rput[cb](-1.5,0.8){$\tilde {\acycle}_\alpha$}
\end{pspicture}
}
\caption{The path of integration w.r.t. the variable $x'$ in the expression for the recursion kernel
$K(\sheet{\alpha}{x},y)$.}
\end{center}
\label{fig:path}
\end{figure}

First we define the kernel
\beq
\label{hatK}
\widehat K(\sheet{\alpha}{x},z) = {1\over \hbar\psi_\alpha^2(x)}\, \int_{\infty_\alpha}^x \psi_\alpha^2(x')\,{dx'\over x'-z}
\eeq
and for each $\alpha=1,\dots,g$, we define
\beq
\label{Cfactor}
\hbar C_\alpha(z)
=\oint_{\acycle_\alpha}\widehat K(x,z)\equiv \int_{\infty_{\td\alpha_-}}^{\infty_{\td\alpha_+}}
\bigl(K(\sheet{\alpha_+}{x},z)-K(\sheet{\alpha_-}{x},z)\bigr).
\eeq
In these expressions, we must also specify the contours of integration w.r.t. the variable $x'$
from the infinities $\infty_{\alpha_\pm}$ to the point $x$ lying on the cycle $\acycle_\alpha$.
We assume that these contours go first from the corresponding infinity along the part
of the adjoint cycle $\bcycle_\alpha$ that lies in the sheet $\alpha_\pm$ until it reaches the intersection point $P_\alpha$;
after this point, we integrate along the cycle $\acycle_\alpha$ towards the final point $x$ (see Fig.~\ref{fig:path})

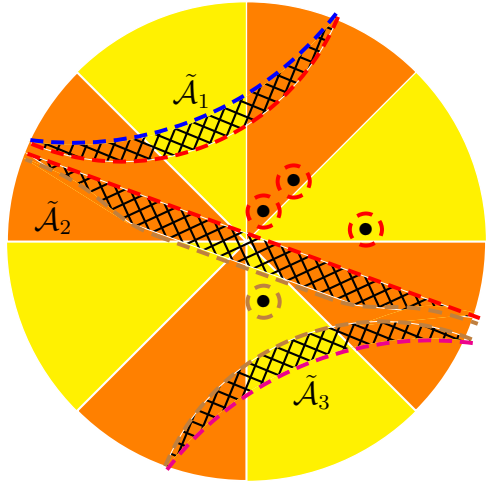
\begin{figure}[h]
\begin{center}
{\psset{unit=0.8}
\begin{pspicture}(-8,-5)(8,5)
\newcommand{\PATTERN}[1]{%
{\psset{unit=#1}
\pscircle[linecolor=white,linewidth=0.5pt,fillstyle=solid,fillcolor=yellow](0,0){4.0}
\pswedge[linecolor=white,linewidth=0.5pt,fillstyle=solid,fillcolor=orange](0,0){4.}{45}{90}
\rput{90}(0,0){\pswedge[linecolor=white,fillstyle=solid,fillcolor=orange](0,0){4.}{45}{90}}
\rput{180}(0,0){\pswedge[linecolor=white,fillstyle=solid,fillcolor=orange](0,0){4.}{45}{90}}
\rput{270}(0,0){\pswedge[linecolor=white,fillstyle=solid,fillcolor=orange](0,0){4.}{45}{90}}
}
}
\rput(0,0){\PATTERN{1}}
\rput{-20}(0,0){\pspolygon[linecolor=white,linewidth=0pt,fillstyle=crosshatch](-4,0.1)(-2,-0.4)(2,-0.4)(4,0.1)}
\psclip{\psbezier*[linecolor=white](-3.8,1.4)(-2.85,1)(-2.85,1)(-2.05,0.3)}
\rput(0,0){\PATTERN{1}}
\endpsclip
\psclip{\psbezier*[linecolor=white](3.8,-1.2)(2.85,-1)(2.85,-1)(1.75,-1.4)}
\rput(0,0){\PATTERN{1}}
\endpsclip
\rput{112.5}(0,0){\pswedge[linecolor=white,linewidth=0pt,fillstyle=crosshatch](5.66,0){3.9}{135}{225}}
\rput{112.5}(0,0){\pswedge[linecolor=white,linewidth=0pt,fillstyle=solid,fillcolor=white](7.73,0){5.5}{150}{210}}
\psclip{\pswedge*[linecolor=white,linewidth=0pt](-2.95,7.11){5.5}{260}{325}}
\rput(0,0){\PATTERN{1}}
\endpsclip
\rput{292.5}(0,0){\pswedge[linecolor=white,linewidth=0pt,fillstyle=crosshatch](5.66,0){3.9}{135}{225}}
\rput{292.5}(0,0){\pswedge[linecolor=white,linewidth=0pt,fillstyle=solid,fillcolor=white](7.73,0){5.5}{150}{210}}
\psclip{\pswedge*[linecolor=white,linewidth=0pt](2.95,-7.11){5.5}{80}{145}}
\rput(0,0){\PATTERN{1}}
\endpsclip
\rput{112.5}(0,0){\psarc[linecolor=blue, linestyle=dashed, linewidth=1.5pt](7.73,0){5.5}{150}{210}}
\rput{112.5}(0,0){\psarc[linecolor=red, linestyle=dashed, linewidth=1.5pt](5.66,0){3.9}{135}{225}}
\rput{-20}(0,0){\pcline[linecolor=red, linestyle=dashed, linewidth=1.5pt](-4,0.1)(4,0.1)}
\rput{-20}(0,0){\psbezier[linecolor=brown, linestyle=dashed, linewidth=1.5pt](-4,0)(-2.5,0)(-2.5,-0.45)(-1.5,-0.45)}
\rput{-20}(0,0){\psbezier[linecolor=brown, linestyle=dashed, linewidth=1.5pt](4,0)(2.5,0)(2.5,-0.45)(1.5,-0.45)}
\rput{-20}(0,0){\pcline[linecolor=brown, linestyle=dashed, linewidth=1.5pt](-1.5,-0.45)(1.5,-0.45)}
\rput{292.5}(0,0){\psarc[linecolor=magenta, linestyle=dashed, linewidth=1.5pt](7.73,0){5.5}{150}{210}}
\rput{292.5}(0,0){\psarc[linecolor=brown, linestyle=dashed, linewidth=1.5pt](5.66,0){3.9}{135}{225}}
\rput(0.2,0.5){$\bullet$}
\pscircle[linecolor=red, linestyle=dashed, linewidth=1.5pt](0.2,0.5){.3}
\rput(1.9,.2){$\bullet$}
\pscircle[linecolor=red, linestyle=dashed, linewidth=1.5pt](1.9,0.2){.3}
\rput(.7,1){$\bullet$}
\pscircle[linecolor=red, linestyle=dashed, linewidth=1.5pt](0.7,1){.3}
\rput(0.2,-1){$\bullet$}
\pscircle[linecolor=brown, linestyle=dashed, linewidth=1.5pt](0.2,-1){.3}
\rput(-1,2.5){$\tilde {\acycle}_1$}
\rput(-3.3,0.4){$\tilde {\acycle}_2$}
\rput(1,-2.5){$\tilde {\acycle}_3$}
\end{pspicture}
}
\caption{The domain of variable $z$ (crosshatched) in (\ref{hatK}) (we
slightly deform the $\td \acycle$-cycle integrals).}
\end{center}
\label{fig:Zdomain}
\end{figure}

To obtain the domain of the function $\widehat K(\sheet{\alpha}{x},z)$, we slightly deform the contours of integration over
edges of the $\td \acycle$-cycles as shown in Fig.~\ref{fig:Zdomain}; then, for the variable $z$ lying in the domain that is ``inner''
w.r.t. integrations from infinities for all the functions $\psi_{\gamma_{\pm}}$, that is, for the domain that is separated from all
the infinities $\infty_{\gamma_{\pm}}$ by the drawn apart edges of the  $\td \acycle$-cycles,
the kernel $\widehat K(\sheet{\alpha}{x},z)$ is well defined (and it develops
logarithmic cuts if we push the variable $z$ through the boundary of the sheet $S_\alpha$).

We now need to describe analytic properties of these functions.

For a fixed $x$, the kernel $\widehat K(\sheet{\alpha}{x},z)$
is defined for $z$ in the cross-hatched domain in Fig.~\ref{fig:AB}.

Taking an integration path between $\infty_\alpha$ and $x$ we obtain that $\widehat
K(\sheet{\alpha}{x},z)$ is defined for $z$ outside this path. Across the path
$]\infty_\alpha,x]$, $\widehat K(\sheet{\alpha}{x},z)$ has a discontinuity w.r.t. the argument $z$:
\beq
\delta_z \widehat K(\sheet{\alpha}{x},z) = {2i\pi\over \hbar}\,\,{\psi_\alpha^2(z)\over \psi_\alpha^2(x)}.
\eeq
The similar statement is true for $C_\alpha(z)$: when $z$ cross the line of the cycle $\acycle_\beta$,
we have
\beq
\label{discontC}
\delta_z C_\beta(z) = {2i\pi\,\,\psi_{\beta_{\pm}}^2(z)\over \hbar}\,\,\int_z^{\infty_{\td\beta_{\pm}}}
{dx''\over \psi_{\beta_{\pm}}^2(x'')}.
\eeq



We now define the recursion kernel $K(\sheet{\alpha}{x},z)$, which is the main ingredient of our construction.
\bd\label{kernelK}
The recursion kernel  $K(\sheet{\alpha}{x},z)$ reads
\beq
K(\sheet{\alpha}{x},z) = \widehat K(\sheet{\alpha}{x},z) - \sum_{j=1}^{d-1} v_j(\sheet{\alpha}{x})C_j(z).
\eeq
It is defined for $z$ in the cross-hatched domain in Fig.~\ref{fig:Zdomain}.
\ed

\bt\label{thlargeKx}
The kernel $K$ has the following properties:
\begin{itemize}
\item For a given $z$,
\beq
K(\sheet{\alpha}{x},z) \mathop{\sim}\, O(x^{-2})
\eeq
when $x\to \infty$ in all sectors (if the function $\psi_\alpha(x)$ increases in all
the sectors except $S_\alpha$).
\item The normalization condition reads
\beq
\label{acycleK}
\oint_{\acycle_j}K({x},z)dx=0,\quad j=1,\dots,d-1.
\eeq
\item $K(\sheet{\alpha}{x},z)$ has double poles with zero residues at the zeros
$s_j^{(\alpha)}$ of $\psi_\alpha$.
\end{itemize}
\et


\bt\label{thKlargez}
We have in all sectors at infinity :
\beq
K(\sheet{\alpha}x,z) \mathop{\sim}_{z\to\infty}\, O(z^{-d}).
\eeq
More precisely we have:
\beq
K(\sheet{\alpha}x,z) \sim -\,\sum_{k=d-1}^\infty {K_k(\sheet{\alpha}x)\over z^{k+1}}
\eeq
with
\beq
\widehat K_k(\sheet{\alpha}x) = {1\over \hbar \psi_\alpha^2(x)}\,\int^x_{\infty_\alpha} x'^k\,\psi_\alpha^2(x')\, dx',
\eeq
and
\beq
K_k(\sheet\alpha{x}) = \widehat K_k(\sheet\alpha{x}) - \sum_{j=1}^{g}\, v_j(\sheet{\alpha}x)\, \oint_{\acycle_j} \widehat K_k(x')\,dx'.
\eeq
\et

\proof{
We can expand $\widehat K(\sheet{\alpha}x,z)$ as
\beq
\widehat K(\sheet{\alpha}x,z) \sim -\sum_{k=0}^\infty {\widehat K_k(\sheet{\alpha}x)\over z^{k+1}}
\eeq
where
\beq
\widehat K_k(\sheet{\alpha}x) = {1\over \hbar \psi_\alpha^2(x)}\,\int^x_{\infty_\alpha} x'^k\,\psi_\alpha^2(x')\, dx',
\eeq
and therefore
\beq
K_k(\sheet{\alpha}x) = \widehat K_k(\sheet{\alpha}x) -
\sum_{\alpha=1}^{g}\, v_\alpha(\sheet{\alpha}x)\, \oint_{\acycle_\alpha} \widehat K_k(x')\,dx'.
\eeq
For $k\leq d-2$, \ ${(x')}^k$ can be presented as a linear combinations of $h_j(x')$,
\beq
x'^k = \sum_{\beta=1}^{d-1}\, b_{k,\beta}\, h_\beta(x'),
\eeq
and from the normalization condition, we immediately obtain that
\beq
\oint_{\acycle_\alpha} \widehat K_k(x')\,dx' = b_{k,\alpha},
\eeq
and therefore $K_k(x)=0$ for $k\leq d-2$, which implies that
\beq
K(x,z) = O(z^{-d}).
\eeq
}



%



%

%


\subsection{Third kind differential: the kernel $G(\sheet{\alpha}x,\sheet{\beta}z)$}

The second important kernel to define is the equivalent of the third kind differential.

\bd The kernel $G(\sheet{\alpha}x,\sheet{\beta}z)$ is
\beq\label{eqdefG}
G(\sheet{\alpha}x,\sheet{\beta}z) =
- \hbar\,\psi_\beta^2(z)\, \partial_z\, {K(\sheet{\alpha}x,z)\over \psi_\beta^2(z)}=2\hbar\frac{\psi_\beta'(z)}{\psi_\beta(z)}
K(\sheet{\alpha}x,z)-\hbar \partial_z\,K(\sheet{\alpha}x,z)
\eeq
\ed

Integration by parts gives
\bea
G(\sheet{\alpha}x,\sheet{\beta}z)
&=& -  {1\over x-z}  + {2\over \psi_\alpha^2(x)}\,\int^x_{\infty_\alpha} {dx'\over x'-z}\, \psi_\alpha^2(x')
\left( {\psi_\alpha'(x')\over \psi_\alpha(x')} - {\psi_\beta'(z)\over \psi_\beta(z)}  \right) \cr
&& - \hbar \sum_{j=1}^{d-1} v_{j}(\sheet{\alpha}x)\,  \psi_\beta^2(z) \partial_z\, {C_{j}(z)\over \psi_\beta^2(z)}.
\eea
In what follows we often are in the situation when we take two integration contours, ${\mathcal C}_{D_x}$ and  ${\mathcal C}_{D_z}$,
and must interchange the order of integration (or the order in which these two contours intersect the $\bcycle$-cycles).
From the definition of the $\acycle$-cycles, it is then obvious that we must interchange the variables $x$ and $z$
within the same sector, so we need permutation relations for $G(\sheet{\alpha}x,\sheet{\beta}z)$
with $\alpha=\beta$. Then, as $x\to z$, we find that $G(\sheet{\alpha}x,\sheet{\alpha}z) \sim {1\over z-x}$, i.e.,
there is a simple pole with the unit residue at $z=x$. Because the combination
${1 \over x'-z}\left( {\psi_\alpha'(x')\over \psi_\alpha(x')} -
{\psi_\alpha'(z)\over \psi_\alpha(z)}\right)$ is regular at $x'=z$,
interchanging the order of integration over  ${\mathcal C}_{D_x}$ and  ${\mathcal C}_{D_z}$ then
just gives the residue at $z=x$; no logarithmic cut takes place.

\bt\label{thdiscG}

$G(\sheet{\alpha}x,\sheet{\beta}z)$ is an analytical function of $x$ with a simple pole at
$x=z$ with residue $-1$ for $\alpha=\beta$ and with double poles at the $s^{(\alpha)}_{j}$'s (zeros of
$\psi_\alpha(x)$) with vanishing residues, and possibly an essential
singularity at $\infty$.

$G(\sheet{\alpha}x,\sheet{\beta}z)$ is an analytical function of $z$, with a simple pole at
$z=x$ with residue $+1$ for $\alpha=\beta$, with simple poles at $z=s^{(\beta)}_{j}$, and with a
discontinuity across $\acycle_\gamma$-cycles with
$\gamma=1,\dots,g$ (this discontinuity has opposite signs depending on which line of the cycle $\acycle_\gamma$,
$\gamma_+$ or $\gamma_-$, we cross; no discontinuity takes place when crossing the last cycle $\td\acycle_n$):
\beq
\delta_z G(\sheet{\alpha}x,\sheet{\beta_{\pm}}z) = \mp 2i\pi \, v_\beta(\sheet{\alpha}x).
\eeq
We also have
\beq
\oint_{\acycle_\alpha} G(x,\sheet{\beta}z)\,dx = 0.
\eeq
\et
\proof{All the discontinuities of $K(\sheet{\alpha}x,z)$ except the one for $C_j(z)$ (\ref{discontC})
are proportional to $\psi_\alpha^2(z)$ and vanish in (\ref{eqdefG}). The discontinuity of
$C_j(z)$ gives $\mp 2\pi i$, and thus, the discontinuity of $G(\sheet{\alpha}x,\sheet{\beta}z)$ is
$\delta_z G(\sheet{\alpha}x,\sheet{\beta_\pm}z) = \mp 2i\pi \, v_\beta(\sheet{\alpha}x)$.

Since $K(\sheet{\alpha}x,z)$ is regular when $z=s^{(\beta)}_{j}$, then it is clear that
$G(\sheet{\alpha}x,\sheet{\beta}z)$ has simple poles at $z=s^{(\beta)}_{j}$, with residue $-2\hbar
K(x,s^{(\beta)}_{j})$.

In the variable $x$, \ $K(\sheet{\alpha}x,z)$ has double poles at
$x=s^{(\alpha)}_{j}$ without residue, and this property holds for $G(\sheet{\alpha}x,\sheet{\beta}z)$ as well.

The vanishing $\acycle$-integral property follows immediately from (\ref{acycleK}).
}

\bt\label{thGlargexz}
\beq
G(\sheet{\alpha}x,\sheet{\beta}z)=O(1/x^2)
\eeq
when $x\to \infty_\alpha$ for any $\alpha$.

At large $z$ in the sector $S_\gamma$ we have
\beq
\mathop{{\lim}}_{z\to \infty_\gamma}\,\,  G(\sheet{\alpha}x,\sheet{\beta}z) = G(\sheet{\alpha}x,\infty_\beta)
=  \eta_{\gamma,\beta}\,\,t_{d+1}\,K_{d-1}(\sheet{\alpha}x)
\eeq
where $\eta_{\gamma,\beta}=\pm 1$ depending on the asymptotic behavior
of $\psi_\beta\sim e^{\pm V/2\hbar}$ in the sheet $S_\gamma$.
\et

\proof{The large $x$ behavior follows from theorem \ref{thlargeKx}.
The large $z$ behavior is given by theorem \ref{thKlargez}, i.e.
$G(\sheet{\alpha}x,\sheet{\beta}z)\sim \pm V'(z) K(\sheet{\alpha}x,z) \sim \pm\, t_{d+1} K_{d-1}(\sheet{\alpha}x)$.
The sign depends on the behavior of the solution in this sector.
}




\subsection{The Bergman kernel $B(\sheet{\alpha}x,\sheet{\beta}z)$}

In classical algebraic geometry, the Bergman kernel is the
fundamental second kind bi-differential, it is the derivative of the
3rd kind differential. Using the same definition as in
\cite{EynOM}, we define:
\beq
B(\sheet{\alpha}x,\sheet{\beta}z) = -{1\over 2}\, \partial_z\, G(\sheet{\alpha}x,\sheet{\beta}z).
\eeq
We call the kernel $B$ the ``quantum" Bergman kernel.

\bt
$B(\sheet{\alpha}x,\sheet{\beta}z)$ is an analytical function of $x$.
When $\alpha=\beta$, it has a double pole at
$x=z$ in the both variables $x$ and $z$
with no residue, and it has double poles in $x$ and in $z$ at the respective
zeros $s^{(\alpha)}_{j}$ and $s^{(\beta)}_{j}$ with
vanishing residues, and possibly an essential singularity at
$\infty$. The discontinuity across $\acycle$-cycles that was present in the kernel $G$ disappears upon
differentiation, so $B(\sheet{\alpha}x,\sheet{\beta}z)$ is defined
analytically in the whole complex plane.
\et

\proof{
These properties follow from those of $G(\sheet{\alpha}x,\sheet{\beta}z)$ of theorem
\ref{thdiscG}. In particular, the
only discontinuity of $G(\sheet{\alpha}x,\sheet{\beta}z)$ is along the $\acycle$-cycles, and
it is independent of $z$, therefore $B(\sheet{\alpha}x,\sheet{\beta}z)$ has no discontinuity there.
}

\subsubsection{Properties of the Bergman kernel}

\bt\label{thBlargexz}
\beq
B(\sheet{\alpha}x,\sheet{\beta}z) = O(1/x^2)
\eeq
when $x\to \infty$ in all sectors, and
\beq
B(\sheet{\alpha}x,\sheet{\beta}z) = O(1/z^2)
\eeq
when $z\to \infty$ in all sectors.
\et

\proof{Follows from the large $x$ and $z$ behaviors of $G(\sheet{\alpha}x,\sheet{\beta}z)$.}

\bt\label{thloopeqB}
The kernel $B$ satisfies the loop equations:
\beq\label{loopeqBx}
\left(2{\psi_\alpha'(x)\over \psi_\alpha(x)}+\partial_x\right)\,\left(B(\sheet{\alpha}x,\sheet{\beta}z)-{1\over 2(x-z)^2}\right)
+ \partial_z\,{{\psi_\alpha'(x)\over \psi_\alpha(x)}-{\psi_\beta'(z)\over \psi_\beta(z)}\over x-z} = P_2^{(0)}(x,\sheet{\beta}z),
\eeq
where $P_2^{(0)}(x,\sheet{\beta}z)$ is a polynomial in $x$ of degree at most $d-2$, and
\beq\label{loopeqBz}
\left(2{\psi_\beta'(z)\over \psi_\beta(z)}+\partial_z\right)\,\left(B(\sheet{\alpha}x,\sheet{\beta}z)-{1\over 2(x-z)^2}\right)
+ \partial_x\,{{\psi_\alpha'(x)\over \psi_\alpha(x)}-{\psi_\beta'(z)\over \psi_\beta(z)}\over x-z} =\td{P}_2^{(0)}(\sheet{\alpha}x,z)
\eeq
where  $\td{P}_2^{(0)}(\sheet{\alpha}x,z)$ is a polynomial in $z$ of degree at most $d-2$.
\et

\proof{We begin with proving the first loop equation for $B(\sheet{\alpha}x,\sheet{\beta}z)$.
We define:
\beq
\widehat B(\sheet{\alpha}x,\sheet{\beta}z) = {1\over 2}\,\partial_z\,
\left(2{\psi_\beta'(z)\over \psi_\beta(z)}-\partial_z\right)\, \widehat K(\sheet{\alpha}x,z)
\eeq
that is, we have
\beq
B(\sheet{\alpha}x,\sheet{\beta}z) = \widehat B(\sheet{\alpha}x,\sheet{\beta}z)
- \sum_{j=1}^{d-1} v_j(\sheet{\alpha}x)\, \oint_{\acycle_j} \widehat B(x'',\sheet{\beta}z)dx''.
\eeq

Since
$(2{\psi_\alpha'(x)\over \psi_\alpha(x)}+\partial_x)\, v_j(\sheet{\alpha}x) = h_j(x)$
is itself a polynomial of degree $\leq d-2$, it suffices to prove Eq. (\ref{loopeqBx}) for $\widehat B(\sheet{\alpha}x,\sheet{\beta}z)$.

We have
\bea
\left(2{\psi_\alpha'(x)\over \psi_\alpha(x)}+\partial_x\right)\,\widehat B(\sheet{\alpha}x,\sheet{\beta}z)
&=& {1\over 2}\,\partial_z\,\left(2{\psi_\beta'(z)\over \psi_\beta(z)}-\partial_z\right)\, {1\over x-z}\,  \cr
&=& -{1\over (x-z)^3} + \partial_z\,{\psi'_\beta(z)\over \psi_\beta(z)(x-z)}
\eea
and therefore:
\beq
\left(2{\psi_\alpha'(x)\over \psi_\alpha(x)}+\partial_x\right)\,\left(\widehat B(\sheet{\alpha}x,\sheet{\beta}z)-{1\over 2(x-z)^2}\right) +
\partial_z\left({{\psi_\alpha'(x)\over \psi_\alpha(x)}-{\psi_\beta'(z)\over \psi_\beta(z)}\over x-z}\right)= 0
\eeq
This proves Eq. (\ref{loopeqBx}) with
\beq
P_2^{(0)}(x,\sheet{\beta}z) = - \sum_{j=1}^{\genus} h_j(x)\, \oint_{\acycle_j} \widehat B(x'',\sheet{\beta}z)dx''.
\eeq

\medskip

We now prove the second loop equation for $B(\sheet{\alpha}x,\sheet{\beta}z)$. We have
\beq
\left(2{\psi_\beta'(z)\over \psi_\beta(z)}+\partial_z\right)\,\widehat B(\sheet{\alpha}x,\sheet{\beta}z)
= {1\over 2} \left(2{\psi_\beta'(z)\over \psi_\beta(z)}+\partial_z\right)\,\partial_z\,
\left(2{\psi_\beta'(z)\over \psi_\beta(z)}-\partial_z\right)\,\widehat K(\sheet{\alpha}x,z),
\eeq
where the operator $\widehat U(z)\equiv {1\over 2} \left(2{\psi_\beta'(z)\over \psi_\beta(z)}+\partial_z\right)\,\partial_z\,
\left(2{\psi_\beta'(z)\over \psi_\beta(z)}-\partial_z\right)$ reads
\beq
\widehat U(z) = -{1\over 2}\,\partial_z^3 + \frac{2}{\hbar^2} U(z) \partial_z + \frac{1}{\hbar^2}U'(z)
\label{KdVoperator}
\eeq
and is therefore independent on the solution $\psi_\beta(z)$ we started with. This is
the Gelfand--Dikii operator \cite{GD}. We then have
\beq
\left(2{\psi_\beta'(z)\over \psi_\beta(z)}+\partial_z\right)\,\widehat B(\sheet{\alpha}x,\sheet{\beta}z)
= {1\over \psi_\alpha^2(x)}\int_{\infty_\alpha}^x\, \psi_\alpha^2(x')\,dx'
\left(-{3\over (x'-z)^4} + {2U(z)\over (x'-z)^2} + {U'(z)\over x'-z} \right)
\eeq
We integrate the first term by parts three times introducing $Y_\alpha(x)=\psi_\alpha'(x)/\psi_\alpha(x)$ (and
exploiting that $Y_\alpha'+Y_\alpha^2=U$). The result reads
\bea
&& \left(2{\psi_\beta'(z)\over \psi_\beta(z)}+\partial_z\right)\,\widehat B(\sheet{\alpha}x,\sheet{\beta}z)
={1\over (x-z)^3} - {\partial \over \partial x}\, {Y_\alpha(x)\over x-z}+\cr
&{}&\qquad\qquad + {1\over \psi_\alpha^2(x)}\,\int_{\infty_\alpha}^x \psi_\alpha^2(x')\,dx'
\left( 2{U(z)-U(x')\over (x'-z)^2} + {U'(z)+U'(x')\over x'-z} \right)
\eea
This implies that
\bea
\label{A10}
&& \left(2{\psi_\beta'(z)\over \psi_\beta(z)}+\partial_z\right)\,\Big(\widehat B(\sheet{\alpha}x,\sheet{\beta}z)-{1\over 2(x-z)^2}\Big)
+ {\partial \over \partial x}\, {Y_\alpha(x)-Y_\beta(z)\over x-z}  \cr
&=& {1\over \psi_\alpha^2(x)}\,\int_{\infty_\alpha}^x\, \psi_\alpha^2(x')\,dx'\,
\Big( 2{U(z)-U(x')\over (x'-z)^2} + {U'(z)+U'(x')\over x'-z} \Big),
\eea
and the obtained expression is obviously a polynomial in $z$. The expression in the brackets in (\ref{A10}) is a
skew-symmetric polynomial in $x'$ and $z$ of degree not higher than $2d-2$. Moreover, all the terms with
$(x')^k$ with $k\le d-2$ become upon integration linear combinations of $v_j(\sheet{\alpha}x)$ and vanish identically
when we apply the projection to the subspace of zero $\acycle$-cycle integrals. So, the minimal power of $x'$ that contributes to the
answer is $(x')^{d-1}$. But there is no term $(x')^{d-1}z^{d-1}$ in the brackets because it contradicts the skew-symmetricity.
The first nonzero term that might contribute is proportional to $(x')^{d-1}z^{d-2}$, which obviously means
that the obtained polynomial $\td{P}_2^{(0)}(\sheet{\alpha}x,z)$ has the maximum degree at most $d-2$ in $z$.
}

\bt\label{thBergmanABcycles}
We have for every $\alpha=1,\dots,g$:
\beq
\oint_{\acycle_i} B(x,\sheet{\beta}z)\,dx = 0,
\qquad
\oint_{\acycle_j} B(\sheet{\alpha}x,z)\,dz = 0
\eeq
and
\beq
\label{bcycle}
\oint_{\bcycle_j} B(\sheet{\alpha}x,z)\,dz = 2i\pi v_j(\sheet{\alpha}x).
\eeq
\et

\proof{
The vanishing of $\acycle$-cycle integrals in the $x$ variable is by
construction. For the $z$ variable, we have
\bea
\oint_{\acycle_\beta} B(\sheet{\alpha}x,z)dz &=&
\int_{\infty_{\td\beta_-}}^{\infty_{\td\beta_+}}  (B(\sheet{\alpha}x,\sheet{\beta_+}z
-B(\sheet{\alpha}x,\sheet{\beta_-}z)dz \cr
&=& -{1\over 2}\,\left(G(\sheet{\alpha}x,\sheet{\beta_+}{\infty_{\td\beta_+}})-G(\sheet{\alpha}x,\sheet{\beta_-}{\infty_{\td\beta_+}})
-G(\sheet{\alpha}x,\sheet{\beta_+}{\infty_{\td\beta_-}})+G(\sheet{\alpha}x,\sheet{\beta_-}{\infty_{\td\beta_-}})\right),
\eea
and the asymptotic conditions in all four cases for the function $G$ in the second line are the same,
so from theorem \ref{thGlargexz} we conclude that the result is zero.

We begin with the integral over a cycle $\td\bcycle_\beta$:
\bea
\oint_{\td\bcycle_\beta}B(\sheet{\alpha}x,y)dy&=&\hbox{jump of $G(\sheet{\alpha}x,y)$ on $\td\acycle_\beta$}\cr
&&-{1\over 2}\,\left(G(\sheet{\alpha}x,\sheet{\beta_+}{\infty_{\beta_+}})-G(\sheet{\alpha}x,\sheet{\beta_-}{\infty_{\beta_+}})
-G(\sheet{\alpha}x,\sheet{\beta_+}{\infty_{\beta_-}})+G(\sheet{\alpha}x,\sheet{\beta_-}{\infty_{\beta_-}})\right)\cr
&=&2\pi i(1-\delta_{\beta,d})v_\beta(\sheet{\alpha}x)+2K_{d-1}(\sheet{\alpha}x),
\eea
where we have used again the asymptotic conditions from Theorem \ref{thGlargexz}. This formula implies that if we
perform the integration $\oint_{\bcycle_j}$ for $j=1,\dots,d-1$, which is the difference of integrals over the cycles
$\td\bcycle_j$ and $\td\bcycle_d$, we obtain the formula (\ref{bcycle}).
}

\medskip

The key property of the Bergman kernel is provided by the following theorem.
\bt
\label{thBergmanSymmetry}
$B(\sheet{\alpha}x,\sheet{\beta}z)$ is symmetric,
\beq
B(\sheet{\alpha}x,\sheet{\beta}z) = B(\sheet{\beta}z,\sheet{\alpha}x)
\eeq
\et

\proof{
The proof uses that $B(\sheet{\alpha}x,\sheet{\beta}z)$ satisfies the loop equation in the two variables. We have:
\bea
&& \left(2{\psi'_\beta(z)\over \psi_\beta(z)}+\partial_z\right)\,\left(2{\psi_\alpha'(x)\over \psi_\alpha(x)}+\partial_x\right)\,
\left(B(\sheet{\alpha}x,\sheet{\beta}z)-{1\over 2(x-z)^2}\right) \cr
&=& \left(2{\psi_\beta'(z)\over \psi_\beta(z)}+\partial_z\right)\,\Big( P_2^{(0)}(x,\sheet{\beta}z)
- \partial_z\,{{\psi_\alpha'(x)\over \psi_\alpha(x)}-{\psi'_\beta(z)\over \psi_\beta(z)}\over x-z} \Big) \cr
&=& \left(2{\psi_\alpha'(x)\over \psi_\alpha(x)}+\partial_x\right)\,\Big( \td{P}_2^{(0)}(\sheet{\alpha}x,z)
- \partial_x\,{{\psi_\alpha'(x)\over \psi_\alpha(x)}-{\psi_\beta'(z)\over \psi_\beta(z)}\over x-z} \Big)
\eea
We then have
\bea
&& \left(2{\psi_\beta'(z)\over \psi_\beta(z)}+\partial_z\right)\,
P_2^{(0)}(x,\sheet{\beta}z) -\left(2{\psi_\alpha'(x)\over \psi_\alpha(x)}+\partial_x\right)\, \td{P}_2^{(0)}(\sheet{\alpha}x,z) \cr
&=&  \left(2{\psi_\beta'(z)\over \psi_\beta(z)}+\partial_z\right)\partial_z\,
{{\psi_\alpha'(x)\over \psi_\alpha(x)}-{\psi_\beta'(z)\over \psi_\beta(z)}\over x-z}
- \left(2{\psi_\alpha'(x)\over \psi_\alpha(x)}+\partial_x\right) \partial_x\,
{{\psi_\alpha'(x)\over \psi_\alpha(x)}-{\psi_\beta'(z)\over \psi_\beta(z)}\over x-z}  \cr
&=& 2 {U(x)-U(z)\over (x-z)^2} - {U'(x)+U'(z)\over x-z},
\eea
so that
\bea
\label{definitionofR(x,z)}
&& (x-z)^2\left(2{\psi_\beta'(z)\over \psi_\beta(z)}+\partial_z\right) P_2^{(0)}(x,\sheet{\beta}z) +2U(z)+(x-z)U'(z) \cr
&=& (x-z)^2\left(2{\psi_\alpha'(x)\over \psi_\alpha(x)}+\partial_x\right) \td{P}_2^{(0)}(\sheet{\alpha}x,z) +2U(x)+(z-x)U'(x) \cr
&\stackrel{{\rm def}}{=}& R(x,z).
\eea
Here, the first line is a polynomial in $x$, whereas the second line
is in turn a polynomial in $z$. Therefore, $R(x,z)$ is a polynomial in
the both variables, of degree at most $d$ in each variable. Moreover, we
must have:
\beq
R(x,x) = 2U(x)
\eeq
Therefore we must have:
\beq
R(x,z)
= {\frac{1}{\hbar^2}}\left({1\over {2}}\,V'(x)V'(z) -\hbar\,{V'(x) -V'(z)\over x-z} - P(x)-P(z) \right) + (x-z)^2 \td{R}(x,z)
\eeq
where $\td{R}(x,z)$ is a polynomial in the both variables of degree at most $d-2$ in each variable.

Putting this back into (\ref{definitionofR(x,z)}) and using the symmetry $x \leftrightarrow z$, we obtain
\beq \label{symm}
 \left(2{\psi_\beta'(z)\over \psi_\beta(z)}+\partial_z\right) (P_2^{(0)}(x,\sheet{\beta}z)-\td{P}_2^{(0)}(\sheet{\beta}z,x))
=  \td{R}(x,z)-\td{R}(z,x)
\eeq
Then, we can decompose the r.h.s into the basis $h_i(x)h_j(z)$,
\beq \td{R}(x,z)-\td{R}(z,x)=\sum_{i,j=1}^{d-1} (\td R_{i,j}-\td R_{j,i}) h_i(x)h_j(z)
\eeq
Applying the integral operator
\bea
\label{intop}
f(z)\mapsto\frac{1}{\psi^2_\beta(z)}\int_{\infty_\beta}^z\psi^2_\beta(z')f(z')dz'
\eea
to the differential equation (\ref{symm}), we obtain
\beq
P_2^{(0)}(x,\sheet{\beta}z)-\td{P}_2^{(0)}(\sheet{\alpha}z,x)
= \sum_{i,j=1}^{d-1} (\td R_{i,j}-\td R_{j,i}) h_i(x)\, v_j(\sheet{\beta}z) + A_1(x)
\eeq
where $A_1(x)$ is some integration constant.

Then using the loop equations (\ref{thloopeqB}) we find by substraction that:
\beq
\left(2\frac{\psi_\alpha'(x)}{\psi(x)}+\partial_x\right)\left(B(\sheet{\alpha}x,\sheet{\beta}z)-
B(\sheet{\beta}z,\sheet{\alpha}x)\right)=P_2^{(0)}(x,\sheet{\beta}z)-\td{P}_2^{(0)}(\sheet{\beta}z,x)
\eeq
and applying the integral operator (\ref{intop}) w.r.t. the variable $x$ in the sheet $S_\alpha$, we obtain
\beq
B(\sheet{\alpha}x,\sheet{\beta}z)-B(\sheet{\beta}z,\sheet{\alpha}x)
= \sum_{i,j=1}^{d-1} (\td R_{i,j}-\td R_{j,i}) v_i(\sheet{\alpha}x)\, v_j(\sheet{\beta}z) + A(x)+\td A(z),
\eeq
with $A(x)$ and $\td A(z)$ the integration constants.

Next, the large $x$ and large $z$ behavior of $B$ imply that $A(x)=\td A(z)=0$, and therefore
\beq \label{difference}
B(\sheet{\alpha}x,\sheet{\beta}z)-B(\sheet{\beta}z,\sheet{\alpha}x)
= \sum_{i,j} (\td{R}_{i,j}-\td{R}_{j,i})\, {v}_{i}(\sheet{\alpha}x){v}_{j}(\sheet{\beta}z).
\eeq

Then, using theorem \ref{thBergmanABcycles}
\beq
\oint_{\acycle_i} B(x,\sheet{\beta}z)dx=\oint_{\acycle_j} B(\sheet{\alpha}x,z)dz=0\ \forall \ i,j,
\eeq
we obtain that
\beq
\td{R}_{i,j}=\td{R}_{j,i}\ \forall i,j,
\eeq
which completes the proof of the symmetricity of the Bergman kernel.
}

We see therefore that our ``quantum Bergman kernel'' enjoys all the features of the standard
Bergman kernel associated with a Riemann surface: it is symmetric, has no discontinuities, and
possesses the double pole with no residue at the coinciding arguments (which corresponds in our
case to the coinciding arguments on the {\em same} sheet $S_k$).
Using all these kernels we can then generalize the recursion of \cite{Eyn1loop, CEO, EynOM} defining the
correlation functions (see the next section).

\subsection{Meromorphic forms and the Riemann bilinear identity}

\bd
A meromorphic form ${\cal R}(\sheet{\alpha}x)$ reads
\beq
\label{form-r}
{\cal R}(\sheet{\alpha}x) = {1\over \hbar\psi_\alpha^2(x)}\,\int_{\infty_\alpha}^x\,\, r(x')\,\psi_\alpha^2(x')\,dx'
\eeq
where $r(x)$ is a rational function of $x$, which behaves at most like $O(x^{d-2})$ at large $x$ and whose poles $r_i$ are such that
\beq
\Res_{x\to r_i}\, \psi_\alpha^2(x)\,r(x)=0.
\eeq
\ed

The holomorphic forms $v_j(\sheet{\alpha}x)$, the kernels $G(\sheet{\alpha}x,\sheet{\beta}z)$ and
$B(\sheet{\alpha}x,\sheet{\beta}z)$ are meromorphic forms of $x$.

\medskip


A meromorphic form ${\cal R}(\sheet{\alpha}x)$ defined by (\ref{form-r})
has poles at $x=r_i$, the poles of
$r(x)$, with degree one less than that of $r(x)$, it has
double poles at the $s^{(\alpha)}_i$ with vanishing residues, and it behaves like
$O(x^{-2})$ in all sectors (having also
an accumulation of poles along the half-lines $L_i$ of accumulations of zeroes of $\psi_\alpha$).
Note that the integrals $\oint_{\acycle_\alpha}\, {\cal R}(x)dx$ are well defined.

We then have the following theorem.

\bt {\bf Riemann bilinear identity}

Consider a meromorphic form ${\cal R}(\sheet{\alpha}z)$.
For $z$ in the sector $S_\alpha$ (outside the crosshatched domain in Fig.~\ref{fig:Zdomain}),
we obtain the representation formula
\bea
{\cal R}(\sheet{\alpha}z)
&=&  -\sum_{\beta}\sum_{r_i\in S_{\beta}} \Res_{r_i} G(\sheet{\alpha}z,\sheet{\beta}y){\cal R}(\sheet{\beta}y)dy
-\sum_\beta\sum_{s_k^{(\beta)}\in S_\beta} \Res_{s_k^{(\beta)}} G(\sheet{\alpha}z,\sheet{\beta}y){\cal R}(\sheet{\beta}y)dy \cr
&&\quad +  \sum_{j=1}^{\genus}\, v_j(\sheet{\alpha}z)\oint_{\acycle_j} {\cal R}(y)\,dy.
\label{holomor}
\eea
\et

\proof{
We begin with the integral over ${\mathcal C}_D$ (Fig.~\ref{fig:CD}) of $G(\sheet{\alpha}z,y){\mathcal R}(y)dy$:
\beq
\int_{{\mathcal C}_D}G(\sheet{\alpha}z,y){\mathcal R}(y)dy=\sum_\beta\int_{\infty_{\beta-1}}^{\infty_{\beta+1}}
G(\sheet{\alpha}z,\sheet{\beta}y){\mathcal R}(\sheet{\beta}y)dy.
\eeq
This integral is identically zero because of asymptotic conditions $G(\sheet{\alpha}z,\sheet{\beta}y)\to 1/y$ as $y\to\infty_\beta$
and ${\mathcal R}(\sheet{\beta}y)\sim 1/y^2$ as $y\to\infty_\beta$ and by the fact that no accumulation of zeros occurs
for the function  ${\mathcal R}(\sheet{\beta}y)$ on the boundaries between the sectors $S_\beta$ and $S_{\beta\pm1}$. We then push the
integration contours through the complex plane towards the $\td\acycle$-cycles as in Fig.~\ref{fig:An}; the residues at the
points $r_i$ and $s_k^{(\beta)}$ give the first line of (\ref{holomor}), the residue at the point $x=y$ in the sector $S_\alpha$ gives
the left-hand side because we have from Theorem~\ref{thdiscG} that $G(\sheet{\alpha}z,\sheet{\alpha}y)=1/(z-y)+\hbox{reg.}$; it remains
to consider the integrals along the $\td\acycle$-cycles. For this, note that our integrations over $y$ are outer w.r.t.
the integrations over the variable $x$ along $\acycle$-cycles in the formula (\ref{Cfactor}) for the factors $C_j$,
and when we push the integration over $y$ through that for $x$, we have the discontinuity of  $G(\sheet{\alpha}z,\sheet{\beta_{\pm}}y)$,
which is equal to $2\pi i v_j(\sheet{\alpha}z)$; no such discontinuity occurs for the cycle $\td\acycle_d$.
All these discontinuities are independent of $y$, so the contour
integral of the product factorizes for each cycle ${\acycle}_j$ and gives the second line of (\ref{holomor}).

To evaluate the remaining integrals inside the crosshatched
domain in Fig.~\ref{fig:Zdomain} recall that $G(\sheet{\alpha}x,\sheet{\beta}y)=\psi_\beta^2(y)\,\partial_y\,
\left(K(\sheet{\alpha}x,y)/\psi_\beta^2(y)\right)$, so integrating by parts we obtain
\beq
\int_{\infty_{\td\beta_-}}^{\infty_{\td\beta_+}} \Bigl(G(\sheet{\alpha}x,\sheet{\beta_+}z){\cal R}(\sheet{\beta_+}z)
-G(\sheet{\alpha}x,\sheet{\beta_-}z){\cal R}(\sheet{\beta_-}z)\Bigr)dz
=
-\int_{\infty_{\td\beta_-}}^{\infty_{\td\beta_+}} \bigl(K(\sheet{\alpha}x,z)r(z)
-K(\sheet{\alpha}x,z)r(z)\bigr)dz=0,
\eeq
and these contributions vanish for all the cycles $\acycle_\beta$.
}

\section{Correlation functions. Diagrammatic representation}
\label{secdefWngFg}

In this section, we define the sectorwise defined versions of the quantum
correlation functions from paper~I (deformations of ``classical'' correlation functions
introduced in \cite{Eyn1loop, CEO, EOFg}). Our
definitions are inspired from (non-Hermitian) eigenvalue
models (see section \ref{secMM}), but they are valid as well in a general setting
of an arbitrary Schr\"odinger equation.

\subsection{Definition and properties of correlation functions}

\bd\label{defWng}
We define the functions $W_n^{(h)}(\sheet{\alpha_1}{x}_1,\dots,\sheet{\alpha_n}{x}_n)$ called
the {\em $n$-point correlation function of ``genus'' $h$} by the
recursion
\beq
W_1^{(0)}(\sheet{\alpha}x) = \omega(\sheet{\alpha}x),
\qquad
W_2^{(0)}(\sheet{\alpha_1}{x}_1,\sheet{\alpha_2}{x}_2)=B(\sheet{\alpha_1}{x}_1,\sheet{\alpha_2}{x}_2)
\eeq
\beq\label{mainrecformula}
W^{(h)}_{n+1}(\sheet{\alpha_0}{x}_0,J)
=   \oint_{{\cal C}_{D_x}} dx\,  K(\sheet{\alpha_0}{x}_0,x) \Big( \ovl{W}_{n+2}^{(h-1)}(x,x,J)
+ \sum_{r=0}^h\sum'_{I\subset J} {W}_{|I|+1}^{(r)}(x,I) {W}_{n-|I|+1}^{(h-r)}(x,J/I) \Big)
\eeq
where $J$ and $I$ are the collective notation for the variables ($J=\{
x_{1},\dots,x_{n} \}$), the symbol $\sum\sum'$ means that we exclude
the terms $(r=0,I=\emptyset)$,  $(r=0,I=\{x_i\})$,  $(r=h,I=J/\{x_i\})$, and $(r=h,I=J)$,
the integration over the contour ${\cal C}_{D_x}$
is defined in (\ref{contourCD}), and where
\beq
\label{normalization-W}
\ovl{W}_{n}^{(h)}(\sheet{\alpha_1}{x}_1,\dots,\sheet{\alpha_n}{x}_n) =
W_{n}^{(h)}(\sheet{\alpha_1}{x}_1,\dots,\sheet{\alpha_n}{x}_n) - {\delta_{n,2}\delta_{h,0}\delta_{\alpha_1,\alpha_2}\over 2(x_1-x_2)^2}.
\eeq
Here the point $x_0$ is outside the integration contour ${\cal C}_{D_x}$ for the variable $x$ and
all the $x_i$ are outside the $\acycle$-cycles of the projection integrals.
\ed

To shorten equations we introduce the notation
\beq
\label{U}
U_n^{(h)}(\sheet{\alpha}x,J)=\ovl{W}_{n+2}^{(h-1)}(\sheet{\alpha}x,\sheet{\alpha}x,J)
+ \sum_{r=0}^{h}\sum'_{I\subset J} \ovl{W}_{|I|+1}^{(r)}(\sheet{\alpha}x,I) \ovl{W}_{n-|I|+1}^{(h-r)}(\sheet{\alpha}x,J/I).
\eeq



The main property of \ref{defWng} is that these quantities solve the loop
equations in the $1/N^2$-expansion. We also prove the following properties:

\bt\label{thpolessiWng}
Each  $W_n^{(h)}(\sheet{\alpha_1}{x}_1,\dots,\sheet{\alpha_n}{x}_n)$
with $2-2h-n<0$ is an analytical functions of all its arguments with poles only when $x_i\to s^{(\alpha_i)}_{j}$.
It vanishes at least as $O\left(1/{x_i^2}\right)$ when $x_i\to\infty_{\alpha_i}$ and
has no discontinuities across $\acycle$-cycles.
\et

\bc
\label{corresidue}
We have that
$$
\int_{{\cal C}_{D_x}}W_{n+1}^{(h)}(x,J)dx=t_0\delta_{n,0}\delta_{h,0}.
$$
\ec

\proof{
We proceed by recursion on $2h+n$. The analyticity is obvious;
the theorem is true for $W_2^{(0)}$.
Assume it is true up to $2g+n$, we shall prove it for $W_{n+1}^{(g)}(x_0,x_1,\dots,x_n)$.
To prove the asymptotic behavior, note that the definition~\ref{kernelK} implies, first, that
the term $\int_{{\cal C}_{D_y}}\widehat K(\sheet{\alpha}x,y)U_n^{(h)}(y,J)dy$ is of order of
$1/(\psi_\alpha^2(x))\int_{\infty_\alpha}^x \psi_\alpha^2(x')dx'/(x')^2\sim x^{-d-2}$,
so the leading contribution comes from the terms proportional to $v_j(\sheet{\alpha}x)\sim x^{-2}$,
which completes the proof.
}

We also have the following simple lemma that follows from the corollary \ref{corresidue} and from the
normalization conditions for the kernel $K(\sheet{\alpha}x,y)$.

\bl For all $(n,h)\neq (0,0)$ we have
\beq
\oint_{{\td\acycle}_\alpha} W_{n+1}^{(h)}(x,J) dx=0.
\eeq
\el

Now comes first of the main theorems.

\bt\label{thWngPng}
For $2-2h-n<0$, the $W_n^{(h)}$ satisfy the loop equation.
This means that the quantity
\bea\label{loopeqPng}
P_{n+1}^{(h)}(x;\sheet{\alpha_1}{x}_1,\dots,\sheet{\alpha_n}{x}_n)
 &=&
\hbar\left(2\frac{\psi_\alpha'(x)}{\psi_\alpha(x)}+\partial_{x}\right)
\overline{W}_{n+1}^{(h)}(\sheet{\alpha}{x},\sheet{\alpha_1}{x}_1,\dots,\sheet{\alpha_n}{x}_n)\cr
&& + \sum_{r=0}^h\sum'_{I\subset J} \ovl{W}_{|I|+1}^{(r)}(\sheet{\alpha}x,I) \ovl{W}_{n-|I|+1}^{(h-r)}(\sheet{\alpha}x,J/I) +
\ovl{W}_{n+2}^{(h-1)}(\sheet{\alpha}x,\sheet{\alpha}x,J)  \cr
& &+ \sum_{j}
\partial_{x_j} \left( {{\ovl{W}_n^{(h)}(\sheet{\alpha}x,J/\{x_j\})\delta_{\alpha,\alpha_j}
-{\ovl{W}_n^{(h)}(\sheet{\alpha_j}{x}_j,J/\{x_j\})}} \over {(x-x_j)}}\right)
\eea
is a polynomial in the variable $x$, which is of degree at most $d-2$ and is independent on the choice of the sector $S_\alpha$.
\et

\proof{ in appendix \ref{approofthWngPng}}

\bt\label{thsym}
Each $W_n^{(h)}$ is a symmetric function of all its arguments.
\et
\proof{ in appendix \ref{approofthsym}, with the special case of $W_3^{(0)}$ in appendix \ref{approofthW3Krich}.}


\bt\label{thhomogeneity}
For $2-2h-n<0$, $W_n^{(h)}(\sheet{\alpha_1}{x}_1,\dots,\sheet{\alpha_n}{x}_n)$ is homogeneous of degree $2-2h-n$:
\bea
&&\left( \hbar\,{\partial\over \partial \hbar}+ \sum_{j=1}^{d+1} t_j\,{\partial\over \partial t_j}
+ \sum_{i=1}^\genus \epsilon_i\,{\partial\over \partial \epsilon_i} +t_0{\partial \over\partial t_0}
\right)W_n^{(h)}(\sheet{\alpha_1}{x}_1,\dots,\sheet{\alpha_n}{x}_n)\cr
&&\quad=(2-2h-n)\,W_n^{(h)}(\sheet{\alpha_1}{x}_1,\dots,\sheet{\alpha_n}{x}_n).
\eea
\et

\proof{
Under a change $t_k\to \l t_k$, $\hbar\to \l \hbar$, $\epsilon_i\to
\l\epsilon_i$, and $t_0\to \l t_0$ the Schr\"odinger equation remains unchanged, and
thus $\psi$ is unchanged. The kernel $K$ is changed to $K/\l$ and
nothing else is changed. By recursion, $W_n^{(h)}$ is then changed by
$\l^{2-2h-n}$.
}

\subsection{Diagrammatic representation}

The diagrammatic representation for the correlation functions structurally coincides with the one
for the correlation functions in one- and two-matrix models~\cite{Eyn1loop, ChekEynFg, CEO}.
We introduce the three kinds of propagators
$$
{\psset{unit=0.8}
\begin{pspicture}(-7,-1)(7,1)
\pcline[linewidth=1.5pt]{->>}(-7,0)(-5,0)
\rput[cb](-7,.3){$x$}
\rput[cb](-5,.3){$y$}
\rput[ct](-6,-0.6){$K(\sheet{\alpha}x,y)$}
\pcline[linewidth=1.5pt]{->}(-1,0)(1,0)
\rput[cb](-1,.3){$x$}
\rput[cb](1,.3){$y$}
\rput[ct](0,-0.6){$G(\sheet{\alpha}x,\sheet{\beta}y)$}
\pcline[linewidth=1.5pt](5,0)(7,0)
\rput[cb](5,.3){$x$}
\rput[cb](7,.3){$y$}
\rput[ct](6,-0.6){$B(\sheet{\alpha}x,\sheet{\beta}y)$}
\end{pspicture}
}
$$
and assume the partial ordering from ``infinity'' to ``$\acycle$-cycles'' to be from left to the right in graphical expressions. We
represent the terms $W_n^{(g)}(J)$ via the graphs with three-valent vertices; we assign its own variable
$\xi$ to every inner vertex and assume the integration w.r.t. this variable along the contour ${\mathcal C}_D$,
the order of integration depends on which vertex is closer to the ``$\acycle$-cycles'':
we begin with integrating at the innermost vertex. We also have $n$ outer legs (one-valent vertices) corresponding to
the points $\sheet{\alpha_i}x_i$, \ $i=1,\dots,n$. They are assumed to lie outside all the
inner integrations. For example, the term $W_3^{(0)}(x_1,x_2,x_3)$ then has the form
$$
{\psset{unit=0.8}
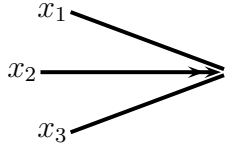
\begin{pspicture}(-2,-1.5)(2,1.5)
\pcline[linewidth=1.5pt]{->>}(-1,0)(2,0)
\pcline[linewidth=1.5pt](-0.5,1)(2.05,0.05)
\pcline[linewidth=1.5pt](-0.5,-1)(2.05,-0.05)
\rput[rc](-0.8,1){$x_1$}
\rput[rc](-1.3,0){$x_2$}
\rput[rc](-.8,-1){$x_3$}
\end{pspicture}
}
$$
whereas the recurrent relation (\ref{mainrecformula}) takes the form
$$
{\psset{unit=0.8}
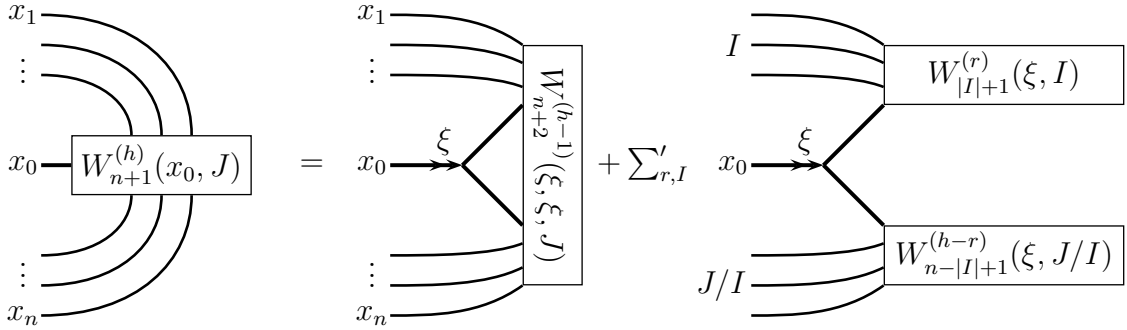
\begin{pspicture}(-8,-3)(8,3)
\psframe[linewidth=0.5pt,fillstyle=solid,fillcolor=white](-8,-0.5)(-5,0.5)
\rput[cc](-6.5,0){$W^{(h)}_{n+1}(x_0,J)$}
\pcline[linewidth=1.5pt](-8.5,0)(-8,0)
\psbezier[linewidth=1pt](-8.5,-1.5)(-7.5,-1.5)(-7,-1)(-7,-0.5)
\psbezier[linewidth=1pt](-8.5,-2)(-7.2,-2)(-6.5,-1.3)(-6.5,-0.5)
\psbezier[linewidth=1pt](-8.5,-2.5)(-6.9,-2.5)(-6,-1.6)(-6,-0.5)
\psbezier[linewidth=1pt](-8.5,1.5)(-7.5,1.5)(-7,1)(-7,0.5)
\psbezier[linewidth=1pt](-8.5,2)(-7.2,2)(-6.5,1.3)(-6.5,0.5)
\psbezier[linewidth=1pt](-8.5,2.5)(-6.9,2.5)(-6,1.6)(-6,0.5)
\rput[rc](-8.8,2.5){$x_1$}
\rput[rc](-8.8,1.7){$\vdots$}
\rput[rc](-8.8,0){$x_0$}
\rput[rc](-8.8,-1.7){$\vdots$}
\rput[rc](-8.8,-2.5){$x_n$}
\rput[cc](-4,0){$=$}
\pcline[linewidth=1.5pt]{->>}(-2.7,0)(-1.5,0)
\pcline[linewidth=1.5pt](-1.5,0)(-.5,1)
\pcline[linewidth=1.5pt](-1.5,0)(-.5,-1)
\psframe[linewidth=0.5pt,fillstyle=solid,fillcolor=white](-0.5,-2)(0.5,2)
\rput{-90}(0,0){{$W^{(h-1)}_{n+2}(\xi,\xi,J)$}}
\psbezier[linewidth=1pt](-2.7,-1.5)(-2,-1.5)(-1,-1.5)(-0.5,-1.3)
\psbezier[linewidth=1pt](-2.7,-2)(-2,-2)(-1.1,-2)(-0.5,-1.7)
\psbezier[linewidth=1pt](-2.7,-2.5)(-2,-2.5)(-1.2,-2.5)(-0.5,-2)
\psbezier[linewidth=1pt](-2.7,1.5)(-2,1.5)(-1,1.5)(-0.5,1.3)
\psbezier[linewidth=1pt](-2.7,2)(-2,2)(-1.1,2)(-0.5,1.7)
\psbezier[linewidth=1pt](-2.7,2.5)(-2,2.5)(-1.2,2.5)(-0.5,2)
\rput[cb](-1.8,0.4){$\xi$}
\rput[rc](-3,2.5){$x_1$}
\rput[rc](-3,1.7){$\vdots$}
\rput[rc](-3,0){$x_0$}
\rput[rc](-3,-1.7){$\vdots$}
\rput[rc](-3,-2.5){$x_n$}
\rput[cc](1.5,0){$+\sum'_{r,I}$}
\pcline[linewidth=1.5pt]{->>}(3.3,0)(4.5,0)
\pcline[linewidth=1.5pt](4.5,0)(5.5,1)
\pcline[linewidth=1.5pt](4.5,0)(5.5,-1)
\psframe[linewidth=0.5pt,fillstyle=solid,fillcolor=white](5.5,1)(9.5,2)
\psframe[linewidth=0.5pt,fillstyle=solid,fillcolor=white](5.5,-1)(9.5,-2)
\rput[cc](7.5,1.5){{$W^{(r)}_{|I|+1}(\xi,I)$}}
\rput[cc](7.5,-1.5){{$W^{(h-r)}_{n-|I|+1}(\xi,J/I)$}}
\psbezier[linewidth=1pt](3.3,-1.5)(4,-1.5)(5,-1.5)(5.5,-1.3)
\psbezier[linewidth=1pt](3.3,-2)(4,-2)(4.9,-2)(5.5,-1.7)
\psbezier[linewidth=1pt](3.3,-2.5)(4,-2.5)(4.8,-2.5)(5.5,-2)
\psbezier[linewidth=1pt](3.3,1.5)(4,1.5)(5,1.5)(5.5,1.3)
\psbezier[linewidth=1pt](3.3,2)(4,2)(4.9,2)(5.5,1.7)
\psbezier[linewidth=1pt](3.3,2.5)(4,2.5)(4.8,2.5)(5.5,2)
\rput[cb](4.2,0.4){$\xi$}
\rput[rc](3,2){$I$}
\rput[rc](3,0){$x_0$}
\rput[rc](2.8,-2){$J/I$}
%
%
\end{pspicture}
}
$$

We now formulate the diagrammatic technique for constructing the functions $W_n^{(h)}(J)$ for $n>0$ and $2g-2+n>0$.
Formally it is the same as the one in \cite{Eyn1loop,CEO}.

We comprise all the diagrams with the corresponding automorphism multipliers such that
\begin{itemize}
\item for $W_n^{(h)}(J)$ a diagram contains exactly $n$ external legs and $h$ loops;
\item we segregate one variable, say, $x_1$, and take all the maximum connected rooted subtrees starting
at the vertex $x_1$ and not going to any other external leg;
\item we associate the directed propagators $K(x,y)$ with all the edges of the rooted subtree; the direction
is always from the root to branches;
\item all other propagators that comprise exactly $h$ inner propagators and $n-1$ remaining external legs
are $B(x,y)$ if the vertices $x$ and $y$ are distinct and $\overline B(x,x)$ for the loop composed
from the single propagator;
\item each rooted subtree establishes the partial ordering on the set of three-valent vertices of the diagram;
we allow the inner propagators $B(x,y)$ to connect {\em only} the comparable vertices (a vertex is comparable
to itself).
\end{itemize}

\section{Deformations \label{variations}}

In this section, we consider the variations of correlation functions
$W_n^{(g)}$ under infinitesimal variations of the Schr\"odinger
potential $U(x)$ or $\hbar$. Infinitesimal variations of the
resolvent $\om(x)$ can be decomposed on the basis of ``meromorphic
forms'' $v_k(\sheet{\alpha}x)$, $k=1,\dots$; we set these forms
to be dual to special cycles with the duality kernel being
the Bergman kernel. It turns out that the classical $\hbar=0$ formulas
retain their form for $\hbar\neq0$.

\subsection{Variation of the resolvent}

We consider an infinitesimal polynomial variation:
$$
U\to U+\delta U, \qquad \hbar\to \hbar+\delta \hbar.
$$
Since $U=V'^2/4 - \hbar V''/2 -P$, we have
\beq
\delta U = {V'\over 2}\delta V' - {\hbar\over 2}\delta V'' - {\delta \hbar\over 2}\delta V'' - \delta P.
\eeq
We can consider variations of $V'(z)$ w.r.t. the higher times $t_k$, \ $k=1,\dots$, as well:
\beq
\delta V'(x) = \sum_{k=1} \delta t_k\,x^{k-1}.
\eeq
Then, for $k\le d+1$, $\delta P$ is of degree at most $d-1$ whereas for $k>d+1$ the polynomial $\delta P$
has the degree at most $k-2$.

Computing $\delta\left({\psi'_\alpha(x)}/{\psi_\alpha(x)}\right)$, we obtain
\beq
\delta\left({\psi'_\alpha(x)}/{\psi_\alpha(x)}\right) =
{1\over \hbar^2\,\psi_\alpha^2(x)}\,\int_{\infty_\alpha}^x\, \psi_\alpha^2(x') \left(\delta U(x')- 2\,{\delta \hbar\over \hbar}\,U(x')\right)\, dx',
\label{variationy}
\eeq
and for $\om(\sheet{\alpha}x) = V'(x)/2+\hbar \psi'_\alpha(x)/\psi_\alpha(x)$ we have
\beq \label{variationw(x)}
\delta \om(\sheet{\alpha}x) = {\delta V'(x)\over 2} + \delta \hbar \, {\psi_\alpha'(x)\over \psi_\alpha(x)}
+ {1\over \hbar\,\psi^2_\alpha(x)}\,\int_{\infty_\alpha}^x\, \psi^2_\alpha(x') \left(\delta U(x')- 2\,{\delta \hbar\over \hbar}\,U(x')\right)\, dx'.
\eeq

\subsection{Variations w.r.t. ``flat'' coordinates}

We choose a system of ``flat"
coordinates on our genus-$(d-1)$ manifold:
\beq
\epsilon_1,\dots,\epsilon_{d-1}, t_0,t_1,\dots\,.
\eeq

\subsubsection{Variations w.r.t. the filling fractions}

For the filling fraction $\delta \epsilon_\alpha$ we have $\delta V'=0$ and thus
\beq
\delta U(x) = -\delta P(x)
\eeq
where $\deg \delta P\leq d-2$, so we can decompose it in the basis of $h_\alpha$:
\beq
\delta P(x) = \sum_{\alpha'} c_{\alpha'}\,h_{\alpha'}.
\eeq
So, from (\ref{variationw(x)}), we have
\beq
\delta \om(x) = -\sum_{\alpha'} c_{\alpha'} \,v_{\alpha'}(x),
\eeq
and because $2i\pi \epsilon_{\alpha'} = \oint_{\acycle_{\alpha'}} \om$, we obtain
\beq
2i\pi \,\delta_{\alpha,\alpha'} = \oint_{\acycle_{\alpha'}} \delta \om = -\sum_{\alpha''} \oint_{\acycle_{\alpha'}} c_{\alpha''}\,v_{\alpha''}
= -c_{\alpha'}
\eeq
Therefore $\delta U(x)/\delta_{\epsilon_\alpha} = 2i\pi h_\alpha(x)$ and
\beq
\delta_{\epsilon_\alpha}\om(\sheet{\beta}x) = 2i\pi \, v_\alpha(\sheet{\beta}x) = \oint_{\bcycle_\alpha}\, B(\sheet{\beta}x,z)\,dz.
\eeq

The flat coordinate $\epsilon_\alpha$ is dual to the holomorphic form $v_\alpha$, which is itself dual to the cycle $\bcycle_\alpha$:
\beq
\epsilon_\alpha = {1\over 2i\pi}\,\oint_{\acycle_\alpha}\, \om,
\qquad
\delta_{\epsilon_\alpha} \om = 2i\pi\, v_\alpha = \oint_{\bcycle_\alpha}\, B.
\eeq

\subsubsection{Variation w.r.t. $t_0$}

We have
\beq
\delta U(x) = -\delta P(x) = -t_{d+1}\,x^{d-1} + Q(x),
\eeq
where $\deg Q\leq d-2$.
Using Eq. (\ref{variationw(x)}) we obtain
\beq
\delta \om(\sheet{\alpha}x) = {1\over \psi_\alpha^2(x)}\,\int_{\infty_\alpha}^x\, (-t_{d+1}x'^{d-1}+Q(x'))\,\psi_\alpha^2(x')\,dx',
\eeq
and the polynomial $Q$ must be chosen such that $\oint_{\acycle_i} \delta
\om=0$ We therefore have
\bea
\delta \om(\sheet{\alpha}x)
&=& -\,t_{d+1}\,K_{d-1}(\sheet{\alpha}x) \cr
&=& -t_{d+1}\,\Big( \widehat K_{d-1}(\sheet{\alpha}x) - \sum_{\beta=1}^{d-1}\, v_\beta(\sheet{\alpha}x)\,
\oint_{\acycle_\beta} \widehat K_{d-1}(x')\,dx'\Big) \cr
&=& v_d(\sheet{\alpha}x),
\eea
where
\beq
\widehat K_k(\sheet{\alpha}x) = {1\over \psi_\alpha^2(x)}\,\int^x_{\infty_\alpha} x'^k\,\psi_\alpha^2(x')\, dx',
\eeq
and $K_k(\sheet{\alpha}x)$ is the $k$th term in the large $z$ expansion of
$K(\sheet{\alpha}x,z)= -\sum_{k=0}^\infty {K_k(\sheet{\alpha}x,z)\over z^{k+1}}$ computed in
theorem \ref{thKlargez}. From theorem \ref{thGlargexz} we have
$G(x,\infty_\alpha) = \eta_\alpha t_{d+1}\,K_{d-1}(\sheet{\alpha}x) $. This shows that
\beq
\delta_{t_0}\om(\sheet{\alpha}x)= {1\over 2}\,(G(\sheet{\alpha}x,\infty_{\tilde d_+})-G(\sheet{\alpha}x,\infty_{\tilde d_-}))
= \int_{\infty_{\tilde d_-}}^{\infty_{\tilde d_+}} B(\sheet{\alpha}x,z)\,dz.
\eeq
The integral is taken here over the last, $d$th $\widetilde B$-cycle.

The flat coordinate $t_0$ is then dual to the 3rd kind
meromorphic form $-2G(\sheet{\alpha}x,\infty_\cdot)$, which is itself dual to the
cycle $[\infty_{\tilde d_-},\infty_{\tilde d_+}]$:
\beq
t_0=\oint_{{\mathcal C}_D}\om(z)dz, \qquad
\delta_{t_{0}} \om(\sheet{\alpha}x) = \int_{\infty_{\tilde d_-}}^{\infty_{\tilde d_+}}\, B(\sheet{\alpha}x,z)\,dz.
\eeq

\subsubsection{Variations w.r.t. $t_k, k=1\dots $. The two-point correlation function.}

Because
$$
t_k=\oint_{{\mathcal C}_D}\hbar \psi'(z)/\psi(z) z^{-k}dz,\quad k=0,1,\dots\,,
$$
the conditions $\partial t_k/\partial t_r=\delta_{k,r}$ and $\partial t_k/\partial \epsilon_\beta=0$ imply
\beq
\oint_{{\mathcal C}_D}\frac{\partial}{\partial t_r}\left(\hbar\frac{\psi'(z)}{\psi(z)}\right)z^{-k}dz=\delta_{k,r};
\qquad
\oint_{\acycle_\beta}\frac{\partial}{\partial t_r}\left(\hbar\frac{\psi'(z)}{\psi(z)}\right)dz=0,
\eeq
and from the general form (\ref{variationy}) of variation, we conclude that (cf. (\ref{residue}) and (\ref{norm}))
\beq
\frac{\partial}{\partial t_r}\left(\hbar\frac{\psi'_\alpha(x)}{\psi_\alpha(x)}\right)=v_{d+r}(\sheet{\alpha}x).
\eeq
In turn, we have the following lemma.

\bl
\beq
\label{vdr}
v_{d+r}(\sheet{\alpha}x)=\frac{1}{2i\pi}\oint_{{\mathcal C}_D>x} B(\sheet{\alpha}x,z)\frac{z^r}{r}dz,\quad r=1,\dots\,.
\eeq
\el

\proof{
That the expression in (\ref{vdr}) has the desired structure follows from the explicit form of the kernel $B$;
we need only to verify the normalization conditions. Vanishing of $\acycle$-cycle integrals is obvious; we must verify
\beq
\delta_{d,l}=\frac1{2i\pi}\oint_{{\mathcal C}_D}x^{-l}v_{d+r}(x)dx=\frac{1}{(2i\pi)^2}\oint_{{\mathcal C}_D>x}dz\oint_{{\mathcal C}_D}dx
\frac{z^r}{r}B(z,x)x^{-l}.
\eeq
Interchanging the order of integration contours and using that $x^{-l}B(z,x)\sim x^{-l-2}$ as $x\to\infty$, we observe that
the only nonzero contribution comes from the double pole at $x=z$, which gives
$$
\frac1{2i\pi}\oint_{{\mathcal C}_D}dx\, x^{-l}\left.\left(\frac{\partial}{\partial z}\frac{z^r}{r}\right)\right|_{z=x}
=\frac1{2i\pi}\oint_{{\mathcal C}_D}dx x^{r-l-1}=\delta_{r,l}.\quad\square
$$
}

We now define the {\em operator of the loop insertion}
$$
\frac{\partial}{\partial V(y)}:=\sum_{r=1}^\infty r y^{-r-1}\frac{\partial}{\partial t_r}
$$
applying which to $\hbar\psi'/\psi$, we obtain
\beq
\frac{\partial}{\partial V(y)} \left(\hbar\frac{\psi'_\alpha(x)}{\psi_\alpha(x)}\right)=\sum_{r=1}^\infty y^{-r-1}
\oint_{y>{\mathcal C}_D}B(\sheet{\alpha}x,z)z^r dz,
\eeq
and since $\oint_{{\mathcal C}_D}B(\sheet{\alpha}x,z) dz=0$, we add the term with $r=0$ into the sum obtaining
$$
\oint_{y>{\mathcal C}_D}B(\sheet{\alpha}x,z)\frac{1}{y-z} dz
$$
in the right-hand side. Note that the point $y$  lies between some infinity, say, $\infty_\beta$ and the integration
contour ${\mathcal C}_D$. Pulling the contour of integration through the point $y$ to infinity we obtain zero due to the
asymptotic conditions for $B(\sheet{\alpha}x,\sheet{\beta}z)$, so the only nonvanishing contribution comes from the residue at $y=z$,
which eventually gives
\beq
\frac{\partial}{\partial V(y)} \left(\hbar\frac{\psi'_\alpha(x)}{\psi_\alpha(x)}\right)=-\frac12 B(\sheet{\alpha}x,\sheet{\beta}y).
\label{two-point}
\eeq
Correspondingly, since $\frac{\partial}{\partial V(y)}V'(x)=\frac{1}{(y-x)^2}$, we obtain for the {\em two-point correlation function}
$W_2^{(0)}(x,y)$:
\beq
W_2^{(0)}(\sheet{\alpha}x,\sheet{\beta}y):=\frac{\partial}{\partial V(y)}\om(\sheet{\alpha}x)=-\frac12 B(\sheet{\alpha}x,\sheet{\beta}y)
+\frac{1}{2(y-x)^2}.
\eeq

\subsection{Variation of higher correlation functions}

Note that for all the above variations w.r.t. the flat coordinates,
we have a cycle $\delta\om^*$ and a (sector-independent) function $\Lambda_{\delta\om}^*$ such that
\beq
\delta\om(\sheet{\alpha}x) = \int_{\delta\om^*}\, B(\sheet{\alpha}x,z)\,\Lambda_{\delta\om}^*(z)\,dz.
\eeq

The theorem below allows computing infinitesimal variations of any $W_n^{(h)}$ under a variation of the Schr\"odinger equation.

\bt
Under an infinitesimal deformation $U\to U+\delta U$,
we have:
\beq
\delta W_n^{(h)}({\sheet{\alpha_1}x}_1,\dots,{\sheet{\alpha_n}x}_n) = \int_{\delta\om^*}\,
W_{n+1}^{(h)}({\sheet{\alpha_1}x}_1,\dots,{\sheet{\alpha_n}x}_n,x')\,\Lambda^*(x')\,dx'
\eeq
where $(\delta\om^*,\Lambda_{\delta\om}^*)$ is the dual cycle to the deformation of the resolvent $\om\to \om+\delta \om$.
\et

\proof{
We prove this theorem by induction. We begin with the loop equation  for $W_n^{(h)}(\sheet{\alpha}x,J)$:
\beq
\left(2\hbar\frac{\psi'_\alpha(x)}{\psi_\alpha(x)}+\hbar \partial_x\right)\,W_n^{(h)}(\sheet{\alpha}x;J)
+ U_n^{(h)}(\sheet{\alpha}x,\sheet{\alpha}x;J) = P_n^{(h)}(x,J).
\label{loopeq}
\eeq
Taking a variation $\delta$ w.r.t. any of the flat coordinate, we have
\bea
&&\left(2\hbar\frac{\psi'_\alpha(x)}{\psi_\alpha(x)}+\hbar \partial_x\right)\,\delta W_n^{(h)}(\sheet{\alpha}x,J)
+ \left(2\delta \hbar\frac{\psi'_\alpha(x)}{\psi_\alpha(x)}\right) W_n^{(h)}(\sheet{\alpha}x,J)
 + \delta U_n^{(h)}(\sheet{\alpha}x,\sheet{\alpha}x;J) \cr
 &&=\delta P_n^{(h)}(x,J),
\eea
where $\delta P_n^{(h)}(x,J)$ is a polynomial in $x$ of degree at most $d-2$. Here both
$\delta U_n^{(h)}(\sheet{\alpha}x,\sheet{\alpha}x;J)$ and
$2\delta \hbar\frac{\psi'_\alpha(x)}{\psi_\alpha(x)}=\int_{\delta\om^*}2B(\sheet{\alpha}x,x')\Lambda^*(x')dx'$
can be expressed by the induction assumption in the dual-cycle-integration form; meanwhile
\beq
\delta U_n^{(h)}(\sheet{\alpha}x,\sheet{\alpha}x;J)=\int_{\delta\om^*}U_{n+1}^{(h)}(\sheet{\alpha}x,\sheet{\alpha}x;J,x')\Lambda^*(x')dx'
-\int_{\delta\om^*}2B(\sheet{\alpha}x,x')\Lambda^*(x')dx'\cdot W_n^{(h)}(\sheet{\alpha}x,J),
\label{inter}
\eeq
because no term containing the two-point correlation function $W_2^{(0)}(\sheet{\alpha}x,x')$ enters
$\delta U_n^{(h)}(\sheet{\alpha}x,\sheet{\alpha}x;J)$.

Using the loop equation of form (\ref{loopeq}) relating $W_{n+1}^{(h)}(\sheet{\alpha}x;J,x')$ and
$U_{n+1}^{(h)}(\sheet{\alpha}x,\sheet{\alpha}x;J,x')$, we observe that the second term in the r.h.s. of (\ref{inter})
cancels the contribution of $2\delta \hbar\frac{\psi'_\alpha(x)}{\psi_\alpha(x)}$, and we obtain that
\bea
&&\left(2\hbar\frac{\psi'_\alpha(x)}{\psi_\alpha(x)}+\hbar \partial_x\right)
\left(\int_{\om^*} W_{n+1}^{(h)}(\sheet{\alpha}x,J,x')\Lambda^*(x')dx' - \delta W_n^{(h)}(\sheet{\alpha}x,J)\right) \cr
 &=& \delta P_n^{(h)}(x,J)-\int_{\om^*}P_{n+1}^{(h)}(x,J,x')\Lambda^*(x')dx' \cr
 &=& \sum_{i=1}^{d-1} \alpha_i(J)\, h_i(x)
\eea
with the r.h.s. being a polynomial in $x$ of degree not higher than $d-2$ expressed in the basis of the polynomials $h_i(x)$.
Using \ref{variationw(x)}, we have
\beq
\int_{\om^*} W_{n+1}^{(h)}(\sheet{\alpha}x,J,x')\Lambda^*(x')dx' - \delta W_n^{(h)}(\sheet{\alpha}x,J)
= \sum_{i=1}^{d-1} \alpha_i(J)\, v_i(x)
\eeq
but since both $W_n^{(h)}(x,J)$ and $W_{n+1}^{(h)}(x,J,x')$ have vanishing $\acycle$-cycle integrals, we have that $\alpha_i=0$, i.e.
\beq
\delta W_n^{(h)}(\sheet{\alpha}x,J) = \int_{\om^*} W_{n+1}^{(h)}(\sheet{\alpha}x,J,x')\Lambda^*(x')dx'
\eeq
}

\bc
\label{cor-epsilon}
$$
\frac{\partial W_n^{(h)}(J)}{\partial \epsilon_\alpha}=\oint_{{\mathcal B}_\alpha}W_{n+1}^{(h)}(J,x')dx'\quad \hbox{for any}\quad n\ge0,h\ge 0.
$$
\ec

\section{Classical and quantum geometry: summary}\label{summary}

In the table below we summarize the comparison between items in
classical algebraic geometry and their quantum counterparts.

\noindent
\begin{tabular}{|l
@{ $\,\, | \,\,$ } l
@{ $\,\, | \,\,$ }
p{220pt}|}
\hline
& {\bf classical $\quad \hbar=0$} & {\bf quantum} \\
\hline
 plane curve:  &
$E(x,y)=\sum_{i,j} E_{i,j} x^i\,y^j $ &
$E(x,y)=\sum_{i,j} E_{i,j} x^i\,y^j \, , \quad [y,x]=\hbar$ \\
& $E(x,y)=0$ &  $E(x,\hbar \partial_x)\psi=0$  \\
\hline
hyperelliptical  &  $y^2=U(x) $ & $y^2-U(x) \, , \quad [y,x]=\hbar$,
 \\
plane curve: &  $\deg U=2d$ &
$\hbar^2\psi'' = U\, \psi$  \\
\hline
Potential: &   \multicolumn{2}{c|} { $V'(x)=2(\sqrt{U(x)})_+$ } \\
\hline
resolvent: &
$\om(x) = {V'(x)/2} + y $.
&
$\om(\sheet{\alpha}x) = {V'(x)/2} +\hbar {\psi_\alpha'(x)\over \psi_\alpha(x)} $.\\
\hline
physical sheet(s): &  $y\sim_\infty - {1\over 2}V'(x) $, $\om \sim t_0/x$
&
$\hbar \psi'/\psi \sim_\infty - {1\over 2}V'(x) $, $\om \sim t_0/x$\\
&& sectors where $\psi\sim e^{-{V\over 2\hbar}}$\\
\hline
branchpoints: & simple zeroes of $U(x)$ & half-lines of accumulations \\
& $U(a_i)=0$, $U'(a_i)\neq 0$ & of zeroes of $\psi$ \\
& $i=1,\dots,2d+2$ &  $L_i$, $i=1,\dots,2d+2$ \\
\hline
double points: & double zeroes of $U(x)$ & half-lines without accumulations \\
& $U(\widehat a_i)=0$, $U'(\widehat a_i)= 0$ & of zeroes of $\psi$ (?) \\
\hline
genus $g=-1$ & degenerate surface & $\psi\,\ee{V/2\hbar} =$polynomial \\
\hline
\end{tabular}
\bigskip

\noindent
\begin{tabular}{|l
@{ $\,\, | \,\,$ } l
@{ $\,\, | \,\,$ }
p{220pt}|}
\hline
& classical $\quad \hbar=0$& quantum \\
\hline
$\acycle_\alpha$-cycles  & surround pairs of  & surround pairs of half-lines \\
$\alpha=1,\dots,g$  & branchpoints  & of accumulating zeroes \\
\hline
extra $\acycle_d$-cycle & surrounds last pair of  & surrounds last pair of half-lines  \\
$\alpha=d$ & branchpoints  &  of accumulating zeroes \\
\hline
$\bcycle$-cycles & \multicolumn{2}{c|} { $\acycle_i\cap \bcycle_j=\delta_{i,j}$  } \\
\hline
Holomorphic  & $v_i(x) = {-h_i(x) \over 2\sqrt{U(x)}}\,$ & $v_i(\sheet{\alpha}x)
= {1\over \hbar \psi_\alpha^2(x)}\int^x_{\infty_\alpha} \psi_\alpha^2(x')\,h_i(x')\,dx' $ \\
forms, 1st kind & \multicolumn{2}{c|} {$h_i=$polynomials, $\deg h_i\leq d-2$} \\
differentials & \multicolumn{2}{c|} {normalized: $\oint_{\acycle_\alpha} v_i(x)\,dx = \delta_{\alpha,i}, \, \alpha=1,\dots,d-1$ }\\
\hline
Period matrix & \multicolumn{2}{c|} {$\tau_{i,j} = \oint_{\bcycle_j} v_i \quad $, $i,j=1,\dots,\genus$ , $\qquad \tau_{i,j}=\tau_{j,i}$} \\
\hline
Filling fractions & \multicolumn{2}{c|}{ $2i\pi\,\epsilon_\alpha = \oint_{\acycle_\alpha} \om(x)dx\quad $, $\alpha=1,\dots,\genus$,
 $\qquad \epsilon_{d}=t_0-\sum_{\alpha=1}^\genus \epsilon_\alpha$} \\
\hline
3rd kind form &  \multicolumn{2}{c|} {$G(x,z) \sim_{x\to z} 1/(z-x)$} \\
&  \multicolumn{2}{c|} {  $G(\sheet{\alpha}x,\sheet{\beta}z) = (2\om(\sheet{\beta}z)-V'(z)-\hbar\partial_z)K(\sheet{\alpha}x,z)$  } \\
\hline
Recursion kernel &  \multicolumn{2}{c|} {$K(\sheet{\alpha}x,z)$} \\
&  \multicolumn{2}{c|} { $K(\sheet{\alpha}x,z) = \widehat K(\sheet{\alpha}x,z) - \sum_i v_i(\sheet{\alpha}x)\,C_i(z)$ } \\
&  \multicolumn{2}{c|} { $C_i(z) = \oint_{\acycle_i}  \widehat K(x',z) dx'$ } \\
& $\widehat K(x,z) = {1\over z-x}\,{1\over 2\sqrt{U(x)}}$ & $\widehat K(\sheet{\alpha}x,z)
= {1\over \hbar \psi_\alpha^2(x)}\int^x_{\infty_\alpha} {\psi^2_\alpha(x')dx'\over x'-z}$ \\
\hline
Bergman kernel & \multicolumn{2}{c|} {$B(\sheet{\alpha}x,\sheet{\beta}z) = -{1\over 2}\,\partial_z\, G(\sheet{\alpha}x,\sheet{\beta}z)$} \\
2nd kind differential& \multicolumn{2}{c|} { $B(\sheet{\alpha}x,\sheet{\alpha}z)\sim 1/2(x-z)^2$}\\
\hline
Symmetry: & \multicolumn{2}{c|} { $B(\sheet{\alpha}x,\sheet{\beta}z)=B(\sheet{\beta}z,\sheet{\alpha}x)$}\\
\hline
& \multicolumn{2}{c|} {$\oint_{\acycle_i} B(x,\sheet{\beta}z)dx=0$ } \\
& \multicolumn{2}{c|} {$\oint_{\bcycle_i} B(x,\sheet{\beta}z)dx=2i\pi\, v_i(\sheet{\beta}z)$ } \\
\hline
Meromorphic  & ${\cal R}(x)dx = {r(x)dx\over 2\sqrt{U(x)}}$ &
${\cal R}(\sheet{\alpha}x) = {1\over \hbar\psi_\alpha^2(x)}\,\int_{\infty_\alpha}^x r(x')\,\psi\alpha^2(x')\,dx'$ \\
forms & \multicolumn{2}{c|} {$r(x)=$rational with poles $r_i$, $r(x)=O(x^{d-2})$}\\
&  & $\Res_{r_i} r(x')\psi^2(x')=0$\\
\hline
Higher & \multicolumn{2}{c|} {$W_{n+1}^{(h)}(\sheet{\alpha}x,J) = \sum_i {1\over 2i\pi}\oint_{{\cal C}_i}
K(\sheet{\alpha}x,z)\, dz\,\Big( W_{n+2}^{(h-1)}(z,z,J)$ } \\
correlators & \multicolumn{2}{c|} {$\qquad \quad + \sum'_{s+s'=h,\, I\uplus I'=J} W_{1+|I|}^{(s)}(z,I) W_{1+|I'|}^{(s')}(z,I') \Big)$ } \\
& \multicolumn{2}{c|} { where ${\cal C}_i$ surrounds the branchpoint $L_i$,\ } $\cup_i {1\over 2i\pi}\oint_{{\cal C}_i}=\oint_{{\mathcal C}_D}$ \\
\hline
Symmetry & \multicolumn{2}{c|} {$W_{n}^{(g)}(x_1,x_2,\dots,x_n) =
W_{n}^{(g)}(x_{\sigma(1)},x_{\sigma(2)},\dots,x_{\sigma(n)})\, , \qquad\sigma\in S_n$} \\
\hline
Variations and & \multicolumn{2}{c|} {$U(x) \to U(x)+ \delta U(x)$}\\
dual cycle & \multicolumn{2}{c|} {  $\delta U^*$: $\delta \om(\sheet{\alpha}x) = \int_{\delta U^*} B(\sheet{\alpha}x,x')\,
\Lambda_{\delta U}(x')\,\,dx'$}\\
\hline
$\delta V'=\sum \delta t_{k}\, x^{k-1}$ & \multicolumn{2}{c|} {$\delta_{t_k} \om(\sheet{\alpha}x)
= \oint_{{\mathcal C}_D} B(\sheet{\alpha}x,x')\,{x'^k\over k}\,dx' $} \\
\hline
variation $\delta t_0$ & \multicolumn{2}{c|} {$\delta_{t_0} \om(\sheet{\alpha}x)
=\int_{\infty_{\tilde d_-}}^{\infty_{\tilde d_+}} B(\sheet{\alpha}x,x')\,dx' $} \\
\hline
variation $\delta \epsilon_i$ & \multicolumn{2}{c|} {$\delta_{\epsilon_i} \om(\sheet{\alpha}x) = \oint_{\bcycle_i} B(\sheet{\alpha}x,x')\,dx' $} \\
\hline
Variations of & \multicolumn{2}{c|} {$\delta W_{n}^{(h)}(x_1,\dots,x_n) = \int_{\delta U^*} W_{n+1}^{(h)}(x_1,\dots,x_n,x')
\,\, \Lambda_{\delta U}(x')\,\, dx'$}\\
higher correlators & \multicolumn{2}{c|} {} \\
%
%
%
\hline
\end{tabular}

\section{Application: Matrix models \label{secMM}}

The main reason for introducing $W_n^{(h)}$ is that they satisfy the loop equations for the random $\beta$-eigenvalue ensembles.
We can therefore identify them with the correlation functions (resolvents) of these ensembles.

Consider a (possibly formal) matrix integral:
\beq
Z = \int_{E_{N,\beta}}\, dM\,\, \ee{-{N\sqrt\beta\over t_0}\,\tr V(M)}
\eeq
where $V(x)$ is some polynomial, and where $E_{N,1}=H_N$ is the set
of hermitian matrices of size $N$, $E_{N,1/2}$ is the set of real
symmetric matrices of size $N$ and $E_{N,2}$ is the set of
quaternion self dual matrices of size $N$ (see \cite{Mehtabook}).

Alternatively, we can integrate over the angular part and get an integral over eigenvalues only \cite{Mehtabook}:
\beq
\label{betamodel}
Z= \int d\lambda_1\dots d\lambda_N\,\, |\Delta(\lambda)|^{2\beta}\,\, \prod_{i=1}^N\, \ee{-{N\sqrt\beta\over t_0}\,V(\lambda_i)},
\eeq
where $\Delta(\lambda)=\prod_{i<j}(\l_j-\l_i)$ is the Vandermonde determinant.

We can then generalize the matrix model to arbitrary values of $\beta$ taking the integral (\ref{betamodel}) as a
definition of the $\beta$-model integral. For this, we take
\beq
\hbar = {t_0\over N}\,\left(\sqrt\beta-{1\over \sqrt\beta}\right).
\eeq
Notice that $\hbar=0$ correspond to the Hermitian case $\beta=1$, and $\hbar\to -\hbar$ corresponds to $\beta\to 1/\beta$.

\subsection{Correlation functions and loop equations}

We define the correlation functions (the resolvents)
\beq
W_k(x_1,\dots,x_k) =  \beta^{k/2}\,\, \left\langle \sum_{i_1,\dots,i_k} {1\over x_1-\lambda_{i_1}} \dots {1\over x_k-\lambda_{i_k}} \right\rangle_c
\eeq
and
\beq
W_0 = {\mathcal F} = \log {\mathcal Z}.
\eeq
When considering variations in the potential $V(x)$, we again assume these resolvents to satisfy the asymptotic
conditions sectorwise, which means that they are defined also sectorwise.

And we assume (this is automatically true if we are considering
formal matrix integrals), that there is a large $N$ expansion of the
type (where we assume $\hbar=O(1)$):
\beq\label{WkgdevtopMM}
W_k(\sheet{\alpha_1}{x}_1,\dots,\sheet{\alpha_k}{x}_k)
= \sum_{h=0}^\infty (N/t_0)^{2-2h-k} W_k^{(h)}(\sheet{\alpha_1}{x}_1,\dots,\sheet{\alpha_k}{x}_k)
\eeq
\beq
W_0 = {\mathcal F} = \sum_{h=0}^\infty (N/t_0)^{2-2h} W_0^{(h)} \equiv \sum_{h=0}^\infty (N/t_0)^{2-2h} {\mathcal F}_h.
\eeq
The loop equations are obtained by integration by parts, for example:
\beq
0 = \sum_i \int d\l_1\dots d\l_N {\partial\over \partial \l_i}\left( {1\over x-\l_i}\, |\Delta(\l)|^{2\beta}\,
\prod_j \ee{-{N\sqrt\beta\over t_0}\,V(\l_j)}\right)
\eeq
gives:
\bea
0 &=& \sum_i \left<{1\over (x-\l_i)^2} + 2\beta\sum_{j\neq i}{1\over x-\l_i}{1\over \l_i-\l_j}
- {N\sqrt\beta\over t_0}\,{V'(\l_i)\over x-\l_i}\right> \cr
&=& \sum_i \left<{1\over (x-\l_i)^2} + \beta\sum_{j\neq i}{1\over x-\l_i}{1\over x-\l_j} - {N\sqrt\beta\over t_0}\,{V'(\l_i)\over x-\l_i}\right> \cr
&=& \sum_i \left<{1-\beta\over (x-\l_i)^2} + \beta\sum_{j}{1\over x-\l_i}{1\over x-\l_j} - {N\sqrt\beta\over t_0}\,{V'(\l_i)\over x-\l_i}\right> \cr
&=& (\beta-1){1\over \sqrt\beta} W'_1(x) + \beta ({1\over \beta}W_1^2(x) + {1\over \beta}W_2(x,x)) \cr
&& \qquad - {N\sqrt\beta\over t_0}\, \left({1\over\sqrt\beta} V'(x)W_1(x) - \sum_i \left< {V'(x)-V'(\l_i)\over x-\l_i}\right>\right) \cr
\eea
We define the polynomial
\beq
P_1(x) = {\sqrt\beta}\, \sum_i \left< {V'(x)-V'(\l_i)\over x-\l_i}\right> = (V' \, W_1)_+.
\eeq
We thus have the loop equation of~\cite{ChekEynbeta}
\beq
W_1^2(x) + \hbar{N\over t_0}\, W_1'(x) + W_2(x,x) = {N\over t_0}\, \left(V'(x)W_1(x)-P_1(x)\right)
\eeq

Using the expansion (\ref{WkgdevtopMM}) we come to the Ricatti equation
\beq
{W^{(0)}_1}(x)^2 + \hbar\,{\partial_x} W_1^{(0)}(x)  = V'(x)W^{(0)}_1(x)-P^{(0)}_1(x)
\eeq
satisfied by $\om(x)$:
\beq
W_1^{(0)}(x)=\om(x).
\eeq
The correlation functions of $\beta$-eigenvalue models obey therefore the topological recursion of definition \ref{defWng}.

\subsection{Variation w.r.t. $\hbar$}

In this subsection, we use the analogy with the $\beta$-eigenvalue ensemble to hint the possible form of the last remaining
building block of our construction, which is the variation w.r.t. $\hbar$, the exponent of the
Vandermonde determinant in (\ref{betamodel}). Up to irrelevant multipliers, we can consider $\beta(\partial/\partial\beta)$
instead of $\hbar(\partial/\partial\hbar)$, for which we have
\beq
\beta\frac{\partial}{\partial\beta}\log{\cal Z}\sim \frac{2\beta}{\cal Z}
\int d\lambda_1\dots d\lambda_N\,\, \Delta(\lambda)^{2\beta}\log|\Delta(\lambda)|\prod_{i=1}^N\, e^{-{N\sqrt\beta\over t_0}\,V(\lambda_i)},
\label{1}
\eeq
so the logarithm of the Vandermonde determinant appears.

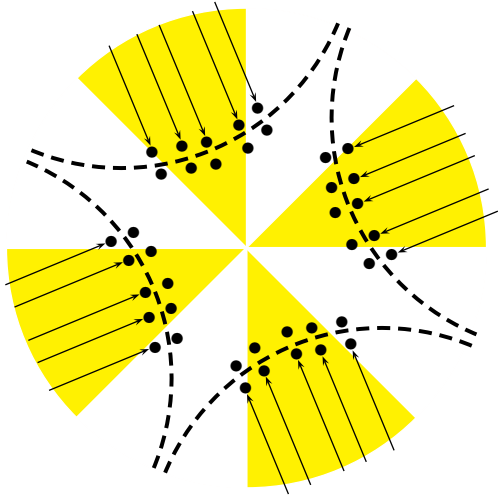
\begin{figure}[h]
\begin{center}
{\psset{unit=0.8}
\begin{pspicture}(-8,-5)(8,5)
\newcommand{\PATTERN}[1]{%
{\psset{unit=#1}
\pscircle[linecolor=white,linewidth=0.5pt,fillstyle=solid,fillcolor=yellow](0,0){4.0}
\pswedge[linecolor=white,linewidth=0.5pt,fillstyle=solid,fillcolor=white](0,0){4.}{45}{90}
\rput{90}(0,0){\pswedge[linecolor=white,fillstyle=solid,fillcolor=white](0,0){4.}{45}{90}}
\rput{180}(0,0){\pswedge[linecolor=white,fillstyle=solid,fillcolor=white](0,0){4.}{45}{90}}
\rput{270}(0,0){\pswedge[linecolor=white,fillstyle=solid,fillcolor=white](0,0){4.}{45}{90}}
}
}
\newcommand{\POLES}[1]{%
{\psset{unit=#1}
\rput(1.6,0){\makebox(0,0){{\small$\bullet$}}}
\rput(2.0,0){\makebox(0,0){{\small$\bullet$}}}
\pcline[linewidth=0.5pt]{->}(4.1,0)(2.1,0)
\rput(1.7,0.4){\makebox(0,0){{\small$\bullet$}}}
\rput(2.1,0.4){\makebox(0,0){{\small$\bullet$}}}
\pcline[linewidth=0.5pt]{->}(4.1,0.4)(2.2,0.4)
\rput(1.8,0.9){\makebox(0,0){{\small$\bullet$}}}
\rput(2.2,0.9){\makebox(0,0){{\small$\bullet$}}}
\pcline[linewidth=0.5pt]{->}(4.1,0.9)(2.3,0.9)
\rput(1.65,-0.6){\makebox(0,0){{\small$\bullet$}}}
\rput(2.05,-0.6){\makebox(0,0){{\small$\bullet$}}}
\pcline[linewidth=0.5pt]{->}(4.1,-0.6)(2.15,-0.6)
\rput(1.8,-1){\makebox(0,0){{\small$\bullet$}}}
\rput(2.2,-1){\makebox(0,0){{\small$\bullet$}}}
\pcline[linewidth=0.5pt]{->}(4.1,-1)(2.3,-1)
}
}
\rput(0,0){\PATTERN{1}}
\rput{22.5}(0,0){\POLES{1}}
\rput{112.5}(0,0){\POLES{1}}
\rput{202.5}(0,0){\POLES{1}}
\rput{292.5}(0,0){\POLES{1}}
\rput{22.5}(0,0){\psarc[linestyle=dashed, linewidth=1.5pt](5.66,0){3.9}{135}{225}}
\rput{112.5}(0,0){\psarc[linestyle=dashed, linewidth=1.5pt](5.66,0){3.9}{135}{225}}
\rput{202.5}(0,0){\psarc[linestyle=dashed, linewidth=1.5pt](5.66,0){3.9}{135}{225}}
\rput{292.5}(0,0){\psarc[linestyle=dashed, linewidth=1.5pt](5.66,0){3.9}{135}{225}}
\end{pspicture}
}
\caption{The origin of the integration contour ${\mathcal C}_D$ in the matrix-model concept.
The inner dots are $\lambda_i$ and the outer dots are $\lambda_i+\delta_\gamma$ ($\gamma=0,2,4,6$);
thin arrowed lines are the logarithmic cuts.}
\end{center}
\label{fig:CD-matrix}
\end{figure}

It seems impossible to construct such a term
from $W_1(\sheet{\alpha}x)$, but we can use the {\em two}-point correlation function $W_2(\sheet{\alpha}x,\sheet{\gamma}y)$ instead. Adopting
a $\beta$-model inspired definition of $W_2(\sheet{\alpha}x,\sheet{\beta}y)$ as a two-resolvent correlation function (not necessarily connected),
$$
W_2(x,y)=\frac1{\mathcal Z}\int D_{N}\lambda \sum_{i=1}^{N}\frac{1}{x-\lambda_i}
\sum_{j=1}^{N}\frac{1}{y-\lambda_j} |\Delta(\lambda)|^{\hbar N}e^{-{N\sqrt\beta\over t_0}\,V(\lambda)}
$$
we then introduce the regularization (both IR and UV, if speaking in the physical terms). At this point, we also split all the
eigenvalues $\lambda_i$ into clusters each of which corresponds to some sector $S_\gamma$; for each term $1/(y-\lambda_i)$
we then integrate w.r.t. $y$ from $\Lambda_\gamma$ to $x+\delta_\gamma$ along the straight lines all of which are parallel;
the regularization parameters depend only on the sector number $\gamma$, and the limit of
removed regularization corresponds to $\Lambda_\gamma\to \infty_\gamma$ and $\delta_\gamma\to 0$. We then obtain
\bea
&{}&\frac{2\beta}{\cal Z}\int_{\Lambda_\gamma}^{x+\delta_\gamma}d\xi W_2(\sheet{\alpha}x,\sheet{\gamma}\xi)\sim \cr
&&\qquad\sim \int d\lambda_1\dots d\lambda_N\,\, \Delta(\lambda)^{2\beta}
\sum_{i=1}^{N}\frac{1}{x-\lambda_i}
\sum_\gamma\sum_{j_\gamma=1}^{\epsilon_\gamma}\int_{\Lambda_\gamma}^{x+\delta_\gamma}\frac{d\xi}{\xi-\lambda_{j_\gamma}}
\prod_{i=1}^N\, e^{-{N\sqrt\beta\over t_0}\,V(\lambda_i)}\cr
&&\qquad= \int d\lambda_1\dots d\lambda_N\,\, \Delta(\lambda)^{2\beta}
\sum_{i=1}^{N}\frac{1}{x-\lambda_i}\times\cr
&&\qquad\quad\times
\sum_\gamma\sum_{j_\gamma=1}^{\epsilon_\gamma}
\Bigl[\log\left|x+\delta_\gamma-\lambda_{j_\gamma}\right|-\log|\Lambda_\gamma|+O(1/\Lambda_\gamma)\Bigr]
\prod_{i=1}^N\, e^{-{N\sqrt\beta\over t_0}\,V(\lambda_i)}.
\label{mamo}
\eea
We now want to perform the integration over the variable $x$ to obtain the expression (\ref{1}). Obviously,
we need to choose the integration contour in a rather specific way: we want it to encircle all the poles
in $\lambda_i$ in the variable $x$ leaving outside all the logarithmic cuts from $\infty_\gamma$ to $\lambda_i+\delta_\gamma$
in the corresponding sector (see Fig.~\ref{fig:CD-matrix}). Given such a contour, we can then perform the integration
w.r.t. $x$ by residues at the points $\lambda_i$ (recall that in the eigenvalue model pattern
we do not have boundaries between sectors inside the complex plane as yet; they appear because
of the collective effect of taking into account the $\lambda$ poles and due to sectorwise regularization chosen).
Evaluating the integral over $x$ in (\ref{mamo}) by residues at $\lambda_i$, we obtain
\bea
&&\frac{2\beta}{\cal Z}\int d\lambda_1\dots d\lambda_N\,\, \Delta(\lambda)^{2\beta}
\Bigl[\sum_{i=1}^{N}\sum_\gamma\sum_{j_\gamma=1}^{\epsilon_\gamma}
\log\left|\lambda_i+\delta_\gamma-\lambda_{j_\gamma}\right|-\cr
&&\qquad-N\sum_\gamma\epsilon_\gamma\log|\Lambda_\gamma|+O(1/\Lambda_\gamma)\Bigr]
\prod_{i=1}^N\, e^{-{N\sqrt\beta\over t_0}\,V(\lambda_i)}\cr
&=&\frac{2\beta}{\cal Z}\int d\lambda_1\dots d\lambda_N\,\, \Delta(\lambda)^{2\beta}
\left|\log\prod_{i\ne j}\bigl(\lambda_i-\lambda_{j}+\delta_\gamma\bigr)\right|-\cr
&&\qquad\quad-2\beta N\sum_\gamma\epsilon_\gamma\log|\Lambda_\gamma|+2\beta \sum_\gamma\epsilon_\gamma\log|\delta_\gamma|,
\label{mamo-1}
\eea
where the first term in the r.h.s. gives the desired integral (\ref{1}) as $\delta_\gamma\to 0$, and
the last two terms are divergent in the limit of regularization removed. These two terms depend however
only on the occupation numbers therefore contributing to potential-independent part of ${\mathcal F}_0$ only and
we can remove them by the proper normalization.

\section{The free energy}\label{secFree}

We use the variations and theorem \ref{thhomogeneity} to define the ${\mathcal F}_h$.

\subsection{The operator $\widehat H$}

Theorem \ref{thhomogeneity} gives:
\beq
(2-2h-n-\hbar\,\partial_{\hbar})\,W_n^{(h)} = \left(t_0\,\partial_{t_0}
+\sum_{k=1}^{d+1}\,t_k\,\partial_{t_k}+\sum_{i=1}^g \epsilon_i \partial_{\epsilon_i}\right)\,W_n^{(h)}
\eeq
In section~\ref{variations}, we expressed
the derivatives of $W_n^{(h)}$ as integrals of $W_{n+1}^{(h)}$ up to the action of $\hbar\frac{\partial}{\partial \hbar}$,
\beq
(2-2h-n-\hbar\,\partial_{\hbar})\,W_n^{(h)} = \widehat H.\,W_{n+1}^{(h)}= \widehat H.\frac{\partial}{\partial V}\,W_{n}^{(h)}
\label{homogeneity}
\eeq
where $\widehat H$ is the linear operator acting as follows:
\beq
\widehat H.f(x) = t_0\,\oint_{\td{\bcycle}_d} f +\sum_{j=1}^{d+1} \int_{{\mathcal C}_D} {t_{j}\,x^j\over j}\,f
+ \sum_{i=1}^\genus \epsilon_i\,\oint_{\bcycle_i} f.
\eeq

We define $W_0^{(h)}={\mathcal F}_h$ for $n=0$ and $h\geq 2$ as

\bd
The free energy ${\mathcal F}_h$ for $h\geq 2$ is the functions for which
\beq
(2-2h-\hbar\,\partial_{\hbar})\,{\mathcal F}_h = \widehat H.\,W_{1}^{(h)}
\eeq
\ed





\subsection{The derivative $\hbar\frac{\partial}{\partial \hbar}$}

The matrix-model considerations in preceding section imply that constructing the derivative in $\hbar$
of the correlation function $W^{(h)}_n(J)$ would involve resolvents of order $n+2$. That is,
\bea
\hbar\frac{\partial}{\partial \hbar}W_n^{(h)}(J)&=&\int_{{\mathcal C}_{D_\xi}}\,d\xi\,\left[
\int_{\infty}^\xi W_{n+2}^{(h-1)}(\overline{\xi}',\xi,J)d\xi'+\right.\cr
&&+\left.\sum_{r=0}^h\sum_{I\subseteq J}\int_{\infty}^\xi W_{|I|+1}^{(r)}(\overline{\xi}',I)d\xi'
\cdot W_{n-|I|+1}^{(h-r)}(\xi,J/I)\right],
\label{derHbar}
\eea
where $\overline\xi$ must be taken to be an ``innermost'' variable in the sense that taking into account that
$\int^\xi B(\xi',y)=G(\xi,y)$, we replace all the appearances
$$
{\psset{unit=0.8}
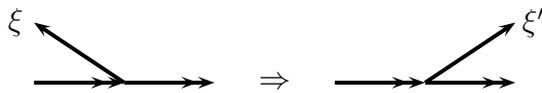
\begin{pspicture}(-4,-1.5)(4,1.5)
\pcline[linewidth=1.5pt]{->>}(-4,0)(-2.5,0)
\pcline[linewidth=1.5pt]{->>}(-2.5,0)(-1,0)
\pcline[linewidth=1.5pt]{->}(-2.5,0)(-4,1)
\rput[cc](0,0){$\Rightarrow$}
\pcline[linewidth=1.5pt]{->>}(1,0)(2.5,0)
\pcline[linewidth=1.5pt]{->>}(2.5,0)(4,0)
\pcline[linewidth=1.5pt]{->}(2.5,0)(4,1)
\rput[rc](-4.3,1){$\xi$}
\rput[lc](4.3,1){$\xi'$}
\end{pspicture}
}
$$
and
$$
{\psset{unit=0.8}
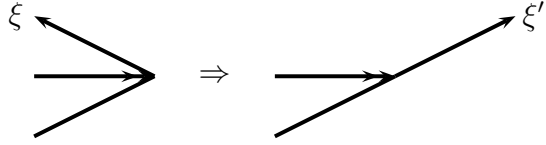
\begin{pspicture}(-3,-1.5)(3,1.5)
\pcline[linewidth=1.5pt]{->>}(-3,0)(-1,0)
\pcline[linewidth=1.5pt](-3,-1)(-1,0)
\pcline[linewidth=1.5pt]{->}(-1,0)(-3,1)
\rput[cc](0,0){$\Rightarrow$}
\pcline[linewidth=1.5pt]{->>}(1,0)(3,0)
\pcline[linewidth=1.5pt](1,-1)(3,0)
\pcline[linewidth=1.5pt]{->}(3,0)(5,1)
\rput[rc](-3.3,1){$\xi$}
\rput[lc](5.3,1){$\xi'$}
\end{pspicture}
}
$$
with {\em no additional factors}.

Note that the sum in (\ref{derHbar}) ranges all cases, not necessarily stable ones, so we begin with
studying nonstable contributions to stable cases ($2h-2+n>0$).
Note that all these contributions then come from the second term in (\ref{derHbar}).

\subsubsection{Case $r=0,\,I=\emptyset$ and $r=h,\,I=J$}

We consider the situation when $n\ge1$. We can then fix $x_1$  to be the root of all the subtrees composed
from the $K$-propagators and $\xi$ can then be the variable of any of external $B$-legs. Then, the contribution in
$W^{(h)}_n(J)$ comprises all the insertions
$$
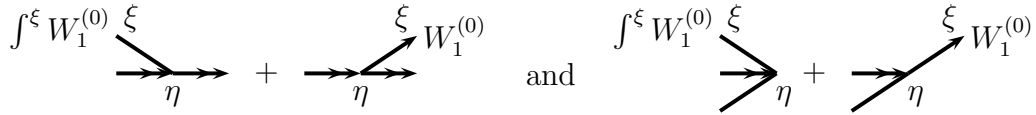
\hbox{
{\psset{unit=0.5}
\begin{pspicture}(-7,-2)(7,2)
\pcline[linewidth=1.5pt]{->>}(-4,0)(-2.5,0)
\pcline[linewidth=1.5pt]{->>}(-2.5,0)(-1,0)
\pcline[linewidth=1.5pt](-2.5,0)(-4,1)
\rput[ct](-2.5,-0.6){$\eta$}
\rput[lb](-3.8,1.1){$\xi$}
\rput[rc](-5.5,1.2){$\int^\xi W_1^{(0)}$}
\rput[cc](0,0){$+$}
\pcline[linewidth=1.5pt]{->>}(1,0)(2.5,0)
\pcline[linewidth=1.5pt]{->>}(2.5,0)(4,0)
\pcline[linewidth=1.5pt]{->}(2.5,0)(4,1)
\rput[ct](2.5,-0.6){$\eta$}
\rput[rb](3.8,1.1){$\xi$}
\rput[lc](5,1){$W_1^{(0)}$}
\end{pspicture}
}
{\psset{unit=0.5}
\begin{pspicture}(-7,-2)(7,2)
\rput[cc](-7,0){and}
\pcline[linewidth=1.5pt]{->>}(-2.5,0)(-1,0)
\pcline[linewidth=1.5pt](-2.5,-1)(-1,0)
\pcline[linewidth=1.5pt](-1,0)(-2.5,1)
\rput[ct](-0.8,-0.6){$\eta$}
\rput[lb](-2.3,1.1){$\xi$}
\rput[rc](-4,1.2){$\int^\xi W_1^{(0)}$}
\rput[cc](0,0){$+$}
\pcline[linewidth=1.5pt]{->>}(1,0)(2.5,0)
\pcline[linewidth=1.5pt]{->}(2.5,0)(4,1)
\pcline[linewidth=1.5pt](1,-1)(2.5,0)
\rput[ct](2.7,-0.6){$\eta$}
\rput[rb](3.8,1.1){$\xi$}
\rput[lc](5,1){$W_1^{(0)}$}
\end{pspicture}
}
}
$$
Let us consider the first case; the second one can be treated analogously. We push the integration
over $\xi$ through the one over $\eta$ in the second diagram and as the result we obtain
$$
{\psset{unit=0.5}
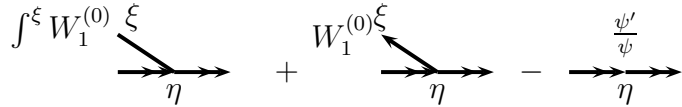
\begin{pspicture}(-7,-2)(7,2)
\pcline[linewidth=1.5pt]{->>}(-8,0)(-6.5,0)
\pcline[linewidth=1.5pt]{->>}(-6.5,0)(-5,0)
\pcline[linewidth=1.5pt](-6.5,0)(-8,1)
\rput[ct](-6.5,-0.6){$\eta$}
\rput[lb](-7.8,1.1){$\xi$}
\rput[rc](-9.5,1.2){$\int^\xi W_1^{(0)}$}
\rput[cc](-3.5,0){$+$}
\pcline[linewidth=1.5pt]{->>}(-1,0)(.5,0)
\pcline[linewidth=1.5pt]{->>}(.5,0)(2,0)
\pcline[linewidth=1.5pt]{->}(.5,0)(-1,1)
\rput[ct](.5,-0.6){$\eta$}
\rput[rb](-.8,1.1){$\xi$}
\rput[lc](-2,1){$W_1^{(0)}$}
\rput[cc](3,0){$-$}
\pcline[linewidth=1.5pt]{->>}(4,0)(5.5,0)
\pcline[linewidth=1.5pt]{->>}(5.5,0)(7,0)
\rput[ct](5.5,-0.6){$\eta$}
\rput[cb](5.5,1){$\frac{\psi'}{\psi}$}
\end{pspicture}
}
$$
Here the sum of the first two terms contains the integral of the total derivative of the function
$\int^\xi W_1^{(0)}(\xi')d\xi' G(\eta,\xi)$, and since $G(\eta,\xi)\sim O(\xi^{-1})$ and
$\int^\xi W_1^{(0)}(\xi')d\xi'\sim \int^\xi t_0 d\xi'/\xi'\sim t_0\log |\xi|$, this contribution vanishes.
Only the third contribution coming from the residue at $\xi=\eta$ survives, and this contribution is
nothing but minus the action of the $\widehat H$ operator on the external leg $B(\eta,\xi)$, so
$$
{\psset{unit=0.5}
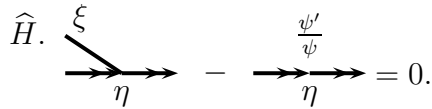
\begin{pspicture}(-7,-2)(7,2)
\pcline[linewidth=1.5pt]{->>}(-4,0)(-2.5,0)
\pcline[linewidth=1.5pt]{->>}(-2.5,0)(-1,0)
\pcline[linewidth=1.5pt](-2.5,0)(-4,1)
\rput[ct](-2.5,-0.6){$\eta$}
\rput[lb](-3.8,1.1){$\xi$}
\rput[rc](-5,1.2){$\widehat H.$}
\rput[cc](0,0){$-$}
\pcline[linewidth=1.5pt]{->>}(1,0)(2.5,0)
\pcline[linewidth=1.5pt]{->>}(2.5,0)(4,0)
\rput[ct](2.5,-0.6){$\eta$}
\rput[cb](2.5,1){$\frac{\psi'}{\psi}$}
\rput[cc](5,0){$=0.$}
\end{pspicture}
}
$$
So, the total contribution of the two cases $r=0,\,I=\emptyset$ and $r=h,\,I=J$ {\em exactly cancels the
action of the $\widehat H$ operator}.

\subsubsection{Case $r=0,\,I=\{x_1\}$}

We begin with the identity
\beq
\int_{x>{\mathcal C}_{D_\xi}>y}G(\sheet{\alpha}x,\xi)K(\xi,y)=-K(\sheet{\alpha}x,y).
\label{convolutionK}
\eeq
(Here and hereafter inequalities of the type $x>{\mathcal C}_{D_\xi}>y$ indicates the mutual positions of points and integration contours.)
Indeed, representing $G(\sheet{\alpha}x,\sheet{\gamma}\xi)=\psi^2_\gamma(\xi) \partial_\xi {K(\sheet{\alpha}x,\xi)\over \psi^2_\gamma(\xi)}$
and integrating by parts, we obtain
$$
\sum_\gamma \Bigl.K(\sheet{\alpha}x,\xi)K(\sheet{\gamma}\xi,y)\Bigr|_{\infty_{\td\gamma_-}}^{\infty_{\td\gamma_+}}-
\int_{x>{\mathcal C}_{D_\xi}>y}K(\sheet{\alpha}x,\xi)\left[\frac{1}{\xi-y}+\sum_j h_j(\xi)C_j(y)\right],
$$
the substitution apparently gives zero, and in the second term only the residue at $\xi=y$ contributes thus
producing (\ref{convolutionK}). An obvious corollary is the second convolution formula
\beq
\int_{x>{\mathcal C}_{D_\xi}>y}G(\sheet{\alpha}x,\xi)B(\xi,\sheet{\beta}y)=-B(\sheet{\alpha}x,\sheet{\beta}y).
\label{convolutionB}
\eeq

In the case $r=0,\,I=\{x_1\}$, we have the diagram
$$
{\psset{unit=0.8}
\begin{pspicture}(-3,-1)(5,1)
\pcline[linewidth=1.5pt]{->}(-3,0)(-1.5,0)
\pcline[linewidth=1.5pt](-1.5,0)(-1,0)
\rput[rc](-3.3,0){$x_1$}
\rput[rc](-1.5,-0.6){$\xi$}
\psframe(-1,-0.6)(3,0.6)
\rput[cc](1,0){$W_n^{(h)}(\xi,J/\{x_1\})$}
\rput[lc](5,0){$=-W_n^{(h)}(J)$}
\end{pspicture}
}
$$

\subsubsection{Case $r=h,\,I=J/\{x_n\}$}

Here, we need another identity
\beq
\int_{x,y>{\mathcal C}_{D_\xi}}G(\sheet{\alpha}x,\xi)B(\xi,\sheet{\beta}y)=0.
\label{convolutionzero}
\eeq
to obtain it, we represent the functions $G$ and $B$ through the kernel $K$, that is, we have
\bea
&&\int_{x,y>{\mathcal C}_{D_\xi}}G(\sheet{\alpha}x,\xi)B(\xi,\sheet{\beta}y)=\cr
&&=\int_{x,y>{\mathcal C}_{D_\xi}}\left(\partial_\xi-2\frac{\psi'_\gamma(\xi)}{\psi_\gamma(\xi)}\right)K(\sheet{\alpha}x,\xi)
\partial_\xi\left(\partial_\xi-2\frac{\psi'_\gamma(\xi)}{\psi_\gamma(\xi)}\right)K(\sheet{\beta}y,\xi)\cr
&&=\Bigl.K(\sheet{\alpha}x,\xi)B(\sheet{\beta}y,\sheet{\gamma}\xi)\Bigr|_{\infty_-}^{\infty_+}-
\int_{x,y>{\mathcal C}_{D_\xi}}K(\sheet{\alpha}x,\xi)
\left(\partial_\xi+2\frac{\psi'_\gamma(\xi)}{\psi_\gamma(\xi)}\right)
\partial_\xi\left(\partial_\xi-2\frac{\psi'_\gamma(\xi)}{\psi_\gamma(\xi)}\right)K(\sheet{\beta}y,\xi).
\nonumber
\eea
Here the substitution gives zero, and the third-order differential operator acting on the kernel $K(\sheet{\beta}y,\xi)$
is again the Gelfand--Dikii operator (\ref{KdVoperator}) that
does not depend on the sector $\gamma$. The integrand is also obviously regular at all zeros of $\psi$-functions, so
the total integration over $\td\acycle$-cycles just gives zero.

Therefore, the contribution of the case $r=h,\,I=J/\{x_n\}$ is zero, and the total contribution of all the unstable cases
together with the action of the $\widehat H$-operator just gives minus one times the original contribution $W_n^{(h)}(J)$.

\subsection{Examples of application of $\hbar\frac{\partial}{\partial\hbar}$}

\subsubsection{Reconstructing $W_n^{(0)}(J)$}

We now apply formula (\ref{derHbar}) to reconstruct the correlation function $W_n^{(0)}(J)$. In the zero genus case, we
must take into account only the nonconnected contributions (the second term in (\ref{derHbar})) into account; we choose
the root of the first term, $\int_{\infty}^\xi W_{|I|+1}^{(r)}(\overline{\xi}',I)d\xi'$, to be $x_1$, the point $\overline\xi$
is then the end of some other (nonrooted) leg $G(\eta,\overline\xi)$ of the first diagram; for the second diagram we choose the root to be
at the end $\xi$ of leg with the corresponding propagator $K(\xi,\rho)$.
As the result of the integration over $\xi$, we obtain using (\ref{convolutionK}) that these two diagrams are sewed along the
propagator $K(\eta,\rho)$ thus producing the {\em connected} diagram with the maximum subtree of propagators $K$ rooted at the external point
$x_1$. We may now ask the question {\em how many times} the given diagram can be obtained as a composition of two diagrams in
the formula (\ref{derHbar})? We obtain this diagram by first breaking it into two parts by
cutting some of internal arrowed lines (including also
the external line $K(x_1,\kappa)$ if we take into account the nonstable contributions) and then sewing again along the same line;
obviously, we obtain this diagram as many times as the total number of arrowed lines (with the minus sign from (\ref{convolutionK})),
which is $2-n$ for $W_n^{(0)}(J)$. So, we see that adopting the definition (\ref{derHbar}) for the action of the operator
$\hbar\frac{\partial}{\partial \hbar}$, we obtain
\bea
\left(\hbar\frac{\partial}{\partial \hbar}+\widehat H.\frac{\partial}{\partial V}\right)W_n^{(0)}(J)=(2-n)W_n^{(0)}(J),
\eea
which is a particular case of formula (\ref{homogeneity}).

\subsubsection{Acting on ${\overline W}^{(0)}_2(x_1,x_2)$}

Here, we consider the action on a nonstable correlation function
${\overline W}^{(0)}_2(x_1,x_2)=B(\sheet{\alpha_1}{x}_1,\sheet{\alpha_2}{x}_2)-\frac{\delta_{\alpha_1,\alpha_2}}{(x_1-x_2)^2}$.
Excluding the terms that compensate the action of $\widehat H$, we have that the action of $\hbar\frac{\partial}{\partial\hbar}$ gives
\bea
&&\int_{x_1>{\mathcal C}_{D_\xi}>x_2}d\xi\, G(\sheet{\alpha_1}{x}_1,\xi)\left(B(\xi,\sheet{\alpha_2}{x}_2)-\frac1{(\xi-x_2)^2}\right)
-\int_{x_1,x_2>{\mathcal C}_{D_\xi}}d\xi\, \frac{1}{x_1-\xi}B(\sheet{\alpha_2}{x}_2,\xi)+\hbox{$x_1\leftrightarrow x_2$}\cr
&&=-2B(\sheet{\alpha_1}{x}_1,\sheet{\alpha_2}{x}_2)+B(\sheet{\alpha_1}{x}_1,\sheet{\alpha_2}{x}_2)+B(\sheet{\alpha_2}{x}_2,\sheet{\alpha_1}{x}_1)=0,
\eea
which again is in accordance with formula (\ref{homogeneity}).

\subsubsection{Acting on ${\mathcal F}_1$}

Here we demonstrate the first case of ``reconstructing'' the free energy term (although this is not true, to obtain the genuine
term ${\mathcal F}_1$ we need other methods, which are still missing; we must however demonstrate that applying formula
(\ref{homogeneity}) we get zero perhaps up to some irrelevant regularizing factors). In this case, no nonstable terms contribute;
the only contribution comes from the first term in (\ref{derHbar}), which gives
\bea
&&\left(\hbar\frac{\partial}{\partial\hbar}+\widehat H.\frac{\partial}{\partial V}\right){\mathcal F}_1=
\int_{{\mathcal C}_{D_\xi}}d\xi\,\left(G(\xi,\overline\xi)-\frac{1}{\xi-\overline\xi}\right)\cr
&&=\int_{{\mathcal C}_{D_\xi}}d\xi\,\left[\int_{\infty_\alpha}^{\xi+\delta_\alpha}d\xi'\,\frac{\partial}{\partial \xi}
\frac{\psi^2_\alpha(\xi')/\psi^2_\alpha(\xi)-1}{\xi'-\xi}
+\sum_j \int_{\infty_\alpha}^{\xi+\delta_\alpha}d\xi'\,h_j(\xi')\psi^2_\alpha(\xi')\left(\frac{C_j(\xi)}{\psi^2_\alpha(\xi)}\right)'.
\right]
\nonumber
\eea
Integrating by parts in the second term, we obtain (up to terms of order $O(\delta_\alpha)$)
$\int_{{\mathcal C}_{D_\xi}}d\xi\,h_j(\xi)C_j(\xi)$, and the integrand is sector-independent and nonsingular at zeros of
$\psi_\alpha$ thus giving zero upon integration. In the first term, integrating by parts in the variable $\xi'$
the term with $1/(\xi-\xi')^2)$ and taking into account that
$\lim_{\xi'\to\xi}\frac{1}{\xi'-\xi}\left(\frac{\psi^2_\alpha(\xi')}{\psi^2_\alpha(\xi)}\right)
=2\frac{\psi'_\alpha(\xi)}{\psi_\alpha(\xi)}$, we obtain
\bea
&&\int_{{\mathcal C}_{D_\xi}}d\xi\,\left[-2\frac{\psi'_\alpha(\xi)}{\psi_\alpha(\xi)}+\int_{\infty_\alpha}^{x+\delta_\alpha}d\xi'\,
\frac{2\frac{\psi'_\alpha(\xi')\psi_\alpha(\xi')}{\psi^2_\alpha(\xi)}-2\frac{\psi'_\alpha(\xi)\psi^2_\alpha(\xi')}{\psi^3_\alpha(\xi)}}{\xi'-\xi}
\right]\cr
&&=\int_{{\mathcal C}_{D_\xi}}d\xi\,\left[-2\frac{\psi'_\alpha(\xi)}{\psi_\alpha(\xi)}+\int_{\infty_\alpha}^{x+\delta_\alpha}d\xi'\,
\left(\frac2{\xi'-\xi}\frac{\psi'_\alpha(\xi')\psi_\alpha(\xi')}{\psi^2_\alpha(\xi)}+
\frac{\psi^2_\alpha(\xi')}{\xi'-\xi}\left(\frac1{\psi^2_\alpha(\xi)}\right)'\right)\right]\cr
&&=\int_{{\mathcal C}_{D_\xi}}d\xi\,\left[-2\frac{\psi'_\alpha(\xi)}{\psi_\alpha(\xi)}+\int_{\infty_\alpha}^{x+\delta_\alpha}d\xi'\,
\left(\frac2{\xi'-\xi}\frac{\psi'_\alpha(\xi')\psi_\alpha(\xi')}{\psi^2_\alpha(\xi)}-
\frac{\psi^2_\alpha(\xi')}{\psi^2_\alpha(\xi)}\frac1{(\xi'-\xi)^2}\right)\right]\cr
&&=\int_{{\mathcal C}_{D_\xi}}d\xi\,\left[-2\frac{\psi'_\alpha(\xi)}{\psi_\alpha(\xi)}+\int_{\infty_\alpha}^{x+\delta_\alpha}
\frac2{\xi'-\xi}\frac{\psi'_\alpha(\xi')\psi_\alpha(\xi')}{\psi^2_\alpha(\xi)}d\xi'+
\frac{\psi^2_\alpha(\xi')}{\psi^2_\alpha(\xi)}d\frac{1}{\xi'-\xi}\right]\cr
&&=\int_{{\mathcal C}_{D_\xi}}d\xi\,\left[-2\frac{\psi'_\alpha(\xi)}{\psi_\alpha(\xi)}+\int_{\infty_\alpha}^{x+\delta_\alpha}
\frac1{\xi'-\xi}\frac{\partial}{\partial \xi'}\frac{\psi^2_\alpha(\xi')}{\psi^2_\alpha(\xi)}d\xi'+
\frac{\psi^2_\alpha(\xi')}{\psi^2_\alpha(\xi)}d\frac{1}{\xi'-\xi}\right]\cr
&&=\int_{{\mathcal C}_{D_\xi}}d\xi\,\left[-2\frac{\psi'_\alpha(\xi)}{\psi_\alpha(\xi)}+
\left.\frac1{\xi'-\xi}\frac{\psi^2_\alpha(\xi')}{\psi^2_\alpha(\xi)}\right|_{\infty_\alpha}^{x+\delta_\alpha}
\right]\cr
&&=\int_{{\mathcal C}_{D_\xi}}d\xi\,\left(\frac{1}{\delta_\alpha}+O(\delta_\alpha)\right),
\nonumber
\eea
and the result is a constant, which is divergent in the limit of regularization removed but
is otherwise independent on all the variables (the same phenomenon occurs when calculating the corresponding action of the
$\widehat H$ operator on ${\mathcal F}_1$ in the standard matrix models, see \cite{ChekEynFg,ChekEynbeta}).

\subsubsection{Acting on $W^{(1)}_1(x)$}

In this case, we have two possible contributions: the one from nonstable graphs gives $W^{(1)}_1(x)$ with the (desired) factor
$-1$ whereas the second one would come from the first term in (\ref{derHbar}) originated from $W_3^{(0)}$ term, that is,
\beq
\int_{{\mathcal C}_{D_\xi}}d\xi\,\int_{{\mathcal C}_{D_\eta}}d\eta\,K(\sheet{\alpha}x,\eta)G(\eta,\overline\xi)B(\eta,\xi),
\eeq
where the contour of integration over $\eta$ goes between the points $\overline\xi$ and $\xi$. We set integrals of this type
to be zero, which provides the last required prescription for the diagrammatic technique describing the free energy terms
${\mathcal F}_h$.

\subsection{The term ${\mathcal F}_h$}

For the stable cases ($h\ne 0,1$), we can now formulate the diagrammatic technique for the term ${\mathcal F}_h$.
We need the diagrams describing the stable terms $W_1^{(r)}(\overline\xi)$ and $W_1^{(h-r)}(\xi)$
with $1\le r \le h-1$ and $W_2^{(h-1)}(\xi,\overline\xi)$.
$$
{\psset{unit=0.6}
\begin{pspicture}(-10,-2)(10,2)
\rput[rc](-9,0){$(2h-2){\mathcal F}_h=$}
\psframe(-6,-1)(-2,1)
\rput[cc](-4,0){$W_2^{(h-1)}$}
\rput[cc](-2.5,0.7){\small$\bullet$}
\rput[cc](-5.6,0){\small$\bullet$}
\rput[ct](-5.6,-0.5){$\rho$}
\rput[ct](-2.5,0.3){$\eta$}
\psbezier[linewidth=1.5pt]{->>}(-2.5,0.7)(-6.5,3.7)(-7.7,0)(-5.6,0)
\rput(-1,0){$-\sum\limits_{r=1}^{h-1}$}
\psframe(1,-1)(4,1)
\rput[cc](2.7,0){$W_1^{(r)}$}
\rput[cc](1.4,0){\small$\bullet$}
\rput[ct](1.4,-0.5){$\rho$}
\psframe(6,-1)(9,1)
\rput[cc](7.7,0){$W_1^{(h-r)}$}
\rput[cc](6.4,0){\small$\bullet$}
\rput[ct](6.4,-0.5){$\eta$}
\psbezier[linewidth=1.5pt]{->>}(3,2)(-1,2)(0,0)(1.4,0)
\rput[cc](3.3,2.2){$\int^\xi$}
\pcline[linewidth=1.5pt](3.5,1.85)(3.5,2.15)
\psbezier[linewidth=1.5pt]{->>}(3.5,2)(5,2)(5,0)(6.4,0)
\end{pspicture}
}
$$
Here in the first term the sum ranges all the diagrams contributing to $W_2^{(h-1)}$ that have distinct vertices to which the
external legs are attached; we then amputate both these legs and joint the vertices $\eta$ and $\rho$ (the $\rho$ vertex
is always the first three-valent vertex in the rooted tree) by the propagator $K(\eta,\rho)$. We took into account already the
integration over $\xi$ thus obtaining an extra minus sign. We cannot perform this integration that easily in the second term
in which the integration over $\xi$ is such that $\rho>{\mathcal C}_{D_\xi}>\eta$ and the symbol $\int^\xi$ indicates that
we must insert the integration
$$
\int_{\rho>{\mathcal C}_{D_\xi}>\eta}d\xi\,\int_{\infty_\alpha}^{\xi+\delta_\alpha}d\xi' K(\sheet{\alpha}{\xi'},\rho)K(\sheet{\alpha}{\xi},\eta)
$$
between the integrations over the variables $\rho$ and $\eta$.

\section{Conclusion}\label{conclusion}

We have defined quantum version of algebraic geometry notions, which allows us
to solve the loop equations in the arbitrary $\beta$-ensemble case.

The notion of branchpoints become
``blurred", a branchpoint is no longer a point, but an asymptotic
accumulation line along which we integrate instead of taking the residue at the branch point.

Another surprising property pertains to the cohomology
theory, which makes sense only if the cycle integral of any form
depends only on the homology class of the cycle, i.e., we need all
forms to have vanishing residues at the zeros $s_i$. This ``no-monodromy"
condition is automatically satisfied for our forms coming from the Schr\"odinger equation and it
is equivalent to the set of Bethe ansatz equations satisfied by $s_i$,
similar to what takes place in the Gaudin model \cite{BabBetGaudin}.

\medskip

In contrast to paper~I, there is no explicit dependence on the chosen
sector; however, even the total number of ${\acycle}$-cycles and rank of the
period matrix may vary depending on the choice of cuts in the complex plane.
This might pertain to that we do not have actual finite-genus (classical) Riemann
surface: analytical continuation may never result in sewing the corresponding
solutions to the Schr\"odinger equation, and we therefore deal with different
(finite-genus) sections of an ambient infinite-genus surface. So, indeed,
the genus is no longer deterministic.

\medskip

Using the sectorwise approach, we can define the symplectic invariants; in Appendix~\ref{F0-Gaussian}
below we present the first nontrivial calculation of this sort: the dependence of the leading
term on the occupation numbers.

\medskip

In this paper, we restricted ourselves to the case of hyperelliptic curves,
i.e. second order differential equations, that corresponds to 1-matrix model.
The first straightforward generalization pertains to including the logarithmic potentials
into consideration, which would produce the Nekrasov functions nonperturbatively in the parameter $\epsilon_2/\epsilon_1$.
A more challenging problem is to generalize this approach to
linear differential equations of any order, which would correspond to
a 2-matrix $\beta$-ensemble model. In this case, we are also presumably able to define
the notions of sheets, branchpoints, forms, and correlation functions. We expect
also the preservation of the Bethe ansatz property
ensuring a no-monodromy condition claiming that all cycle integrals depend
only on the homology classes of cycles. The difference between the hyperelliptic case and the
general case is comparable to the difference between the patterns of papers \cite{Eyn1loop}
and \cite{CEO}, i.e., the definition of the kernel $K$ must be more involved
and less explicit, but we postpone this discussion for further publications.

\medskip

It would be also interesting to see whether the quantities ${\mathcal F}_h$ possess a
symplectic invariance, or more precisely a ``canonical
invariance", i.e., whether they are invariant under any change $(x,y)\to (\td
x,\td y)$ such that $[\td y,\td x]=[y,x]=\hbar$.

\section*{Acknowledgments}
We would like to thank O. Babelon, M. Berg\`ere, G. Borrot, P. Di Francesco, V. Pasquier, A. Prats-Ferrer, A. Voros
for useful and fruitful discussions on this subject.
The work of L.Ch. is supported by the Russian Foundation for Basic Research grants 09-02-93105-CNRS$\_$a and 09-01-12150-ofi$\_$m,
by the Grant for Supporting Leading Scientific Schools NSh-795.2008.1, and by the Program Mathematical Methods for Nonlinear Dynamics.
The work of B.E. and O.M. is partly supported by the Enigma European network MRT-CT-2004-5652, by the ANR project
G\'eom\'etrie et int\'egrabilit\'e en physique math\'ematique ANR-05-BLAN-0029-01,
by the European Science Foundation through the Misgam program,
by the Quebec government with the FQRNT.


\setcounter{section}{0}

\appendix{Proof of theorem \ref{thWngPng}}
\label{approofthWngPng}

We now prove theorem \ref{thWngPng}, that all $W_n^{(h)}$'s satisfy
the loop equation, i.e.,
\bea
P_{n+1}^{(h)}(x;\sheet{\alpha_1}{x}_1,\dots,\sheet{\alpha_n}{x}_n)
 &=&
\hbar\left(2\frac{\psi_\alpha'(x)}{\psi_\alpha(x)}+\partial_{x}\right)
\overline{W}_{n+1}^{(h)}(\sheet{\alpha}{x},\sheet{\alpha_1}{x}_1,\dots,\sheet{\alpha_n}{x}_n)\cr
&& + \sum_{r=0}^h\sum'_{I\subset J} \ovl{W}_{|I|+1}^{(r)}(\sheet{\alpha}x,I) \ovl{W}_{n-|I|+1}^{(h-r)}(\sheet{\alpha}x,J/I) +
\ovl{W}_{n+2}^{(h-1)}(\sheet{\alpha}x,\sheet{\alpha}x,J)  \cr
& &+ \sum_{j}
\partial_{x_j} \left( {{\ovl{W}_n^{(h)}(\sheet{\alpha}x,J/\{x_j\})\delta_{\alpha,\alpha_j}
-{\ovl{W}_n^{(h)}(\sheet{\alpha_j}{x}_j,J/\{x_j\})}} \over {(x-x_j)}}\right)
\eea
is a polynomial in $x$ of degree at most $d-2$.

From the definition, we have (with $U$ from (\ref{U}))
\beq
W_{n+1}^{(g)}(\sheet{\alpha}x,J)=   {1\over 2i\pi} \oint_{{\cal C}}\, dz\,  K(\sheet{\alpha}x,z)\left(
 U_{n+2}^{(g-1)}(z,z,J)+\sum_j B(\sheet{\alpha_j}{x_j},z)W_n^{(g)}(z,J/\{x_j\})\right).
\eeq
Acting by $\hbar\left(2\frac{\psi'_\alpha(x)}{\psi_\alpha(x)}+\partial_x\right)$ on $K(\sheet{\alpha}x,z)$
gives $1/(x-z)+\sum_{j=1}^gh_j(x)C_j(z)$, and the second part is obviously polynomial satisfying assertions of the
theorem. Pulling the contour of
integration w.r.t. $z$ to infinity (with $x$ originally outside the integration contour) and
taking into account that the integral at infinity vanishes thanks to the asymptotic conditions, we find that only the
residue at $z=x$ and the residue at $z=x_j$ in the second term in the brackets give nonzero contributions; the result of
integration reads
$$
U_{n+2}^{(g-1)}(\sheet{\alpha}x,\sheet{\alpha}x,J)+\sum_j B(\sheet{\alpha_j}{x_j},\sheet{\alpha}x)W_n^{(g)}(\sheet{\alpha}x,J/\{x_j\})
+\sum_j \frac{\partial}{\partial x_j}\frac{W_n^{(g)}(J)}{x-x_j},
$$
so taking into account (\ref{normalization-W}), we obtain the assertion of the theorem. $\quad\square$

\appendix{The symmetricity of $W_3^{(0)}$}
\label{approofthW3Krich}

\bt \label{thW3Krich}
The three-point function $W_3^{(0)}$ is symmetric
\et

\proof{
Introducing $Y:=-2\hbar \psi'/\psi$, \ $W_3^{(0)}$ is by definition
\bea
&& W_3^{(0)}\bigr(\sheet{\alpha_0}{x_0},\sheet{\alpha_1}{x_1},\sheet{\alpha_2}{x_2}\bigl)\cr
&=&  {1\over i\pi}\oint_{{\cal C}_D} dx\, K(\sheet{\alpha_0}{x_0},x)B(\sheet{\alpha_1}{x_1},x)B(\sheet{\alpha_2}{x_2},x) \cr
 &=&  {1\over 4i\pi} \oint_{{\cal C}_D} dx\,  K_0 \, G_1^{'} \, G'_2 \cr
 &=&  {1\over 4i\pi}\oint_{{\cal C}_D} dx\, K_0 \left( (\hbar K''_1 +  Y K'_1 +  Y'
 K_1)(\hbar K''_2 + YK'_2 +Y' K_2) \right) \cr
 &=&  {1\over 4i\pi} \oint_{{\cal C}_D} dx\, K_0 \, (\, \hbar^2 K''_1 K''_2 +  \hbar Y (K'_1
 K''_2+K''_1 K'_2) +  \hbar Y' (K''_1 K_2 +K''_2 K_1) \cr
 && +  Y^2 K'_1 K'_2+  Y Y' (K_1 K'_2+K'_1 K_2)+
{Y'}^2 K_1 K_2 \,) \cr
 \eea
where we have introduced a shorthand notation $K_p = K(x_p,x)$, $G_p=G(x_p,x)$, all the
derivatives are w.r.t. $x$, and we omit indices indicating the sectors.

The combinations $K_0 K_1 K_2 f(x)$, where $f(x)$ is sector-independent ($f=1,U,U',\dots$), vanish upon integration w.r.t. $x$ because
each of $K_i$ is also sector-independent w.r.t. $x$. We can then use
the Ricatti equation $Y_i^2 = 2 \hbar Y_i' + 4U$ to replace $Y_i^2$ by $2 \hbar Y_i'$ and $ Y_i Y_i'$ by $\hbar Y_i''$, which gives
\bea
& & W_3^{(0)}(x_0,x_1,x_2)\cr
 &=&  {1\over 4i\pi}\oint_{{\cal C}_D} dx\, K_0 (  \hbar Y (K'_1
 K''_2+K''_1 K'_2) +  \hbar Y' (K''_1 K_2 +K''_2 K_1) \cr
 && + 2 \hbar Y' K'_1 K'_2+ \hbar Y'' (K_1 K'_2+K'_1 K_2)+
{Y'}^2 K_1 K_2 \,)\ \cr
 &=&  {1\over 4i\pi}\oint_{{\cal C}_D} dx\, K_0 \, (  \hbar Y (K'_1 K'_2)' +  \hbar Y' (K_1 K_2)'' + \hbar Y'' (K_1 K_2)'+  {Y'}^2 K_1 K_2 \,)\ \cr
 &=&   {1\over 4i\pi}\oint_{{\cal C}_D} dx\, {Y'}^2 K_0 K_1 K_2 +
 \hbar \big( Y'' K_0 (K_1 K_2)' - (Y K_0)' K'_1 K'_2 - (Y' K_0)' (K_1 K_2)' \big) \cr
 &=&  {1\over 4i\pi}\oint_{{\cal C}_D} dx\, {Y'}^2 K_0 K_1 K_2 -
 \hbar \big(  (Y K_0)' K'_1 K'_2 + Y' K_0' (K_1 K_2)' \big) \cr
&=&  {1\over 4i\pi}\oint_{{\cal C}_D}dx\, {Y'}^2 K_0 K_1 K_2 - \hbar Y K'_0 K'_1 K'_2 - \hbar Y' (K_0 K'_1 K'_2+K'_0 K_1 K'_2+K'_0 K'_1 K_2) \cr
\eea
This expression is readily symmetric in $x_0, x_1, x_2$ as claimed in theorem \ref{thsym}.
}

\appendix{Proof of theorem \ref{thsym}}
\label{approofthsym}

{\bf Theorem \ref{thsym}}
{\it
Each $W_n^{(g)}$ is a symmetric function of all its arguments.
}
\bigskip

\proof{
The special case of $W_3^{(0)}$ was proved in appendix \ref{approofthW3Krich} above. The symmetricity of
the two-point correlation function $\overline{W}_2^{(0)}$ was proved in Theorem \ref{thBergmanSymmetry}.

For technical reason, it is easier to proceed with the proof for {\em nonconnected} correlation functions. We introduce
two types of them:
\begin{itemize}
\item the correlation function
\beq
\label{stable-sum}
\widehat{W_n^{(h)}}(I)=\sum_{\hbox{partitions} \atop \hbox{$\{I_1,\dots, I_k\}$ of\ $I$}}\prod_{j=1}^k W_{n_j}^{(h_j)}(I_j)
\eeq
that comprises partitions of only stable ($2h_j+n_j-2>0$) type with $I_j\ne \emptyset$;
\item the correlation function
\beq
\label{nonstable-sum}
\widetilde{W_n^{(h)}}(I)=\sum_{\hbox{partitions} \atop \hbox{$\{I_1,\dots, I_k\}$ of\ $I$}}\prod_{j=1}^k W_{n_j}^{(h_j)}(I_j)
\eeq
that admits also two-point correlation functions $\overline{W}_2^{(0)}$ in the sums,
$2h_j+n_j-2\ge0$, with $I_j\ne \emptyset$;
\end{itemize}

The symmetricity of all
$\widehat{W_s^{(h')}}(I)$ with $s+2h'\le n+2h$
obviously implies the symmetricity of $\widetilde{W_n^{(h)}}(I)$

It is obvious from the definition that $\widehat{W_{n+1}^{(h)}}\bigr(\sheet{\alpha_0}{x_0},\sheet{\alpha_1}{x_1},\dots,\sheet{\alpha_n}{x_n}\bigl)$ is
symmetric in $x_1,x_2,\dots,x_n$, and therefore we need to show that (for $n\geq 1$):
\beq
\widehat{W_{n+1}^{(h)}}(\sheet{\alpha_0}{x_0},\sheet{\alpha_1}{x_1},J)-
\widehat{W_{n+1}^{(h)}}(\sheet{\alpha_1}{x_1},\sheet{\alpha_0}{x_0},J)=0,
\eeq
where $J=\{ \sheet{\alpha_2}{x_2},\dots,\sheet{\alpha_n}{x_n}\}$.

The proof is by recursion on $-\chi=2h-2+n$.

Assume that all $\widehat{W_k^{(h')}}$ and $\widetilde{W_k^{(h')}}$
with $2h'+k-2\leq 2h+n$ are symmetric.
We have:
\bea
\label{first-equation}
&& \widehat{W_{n+1}^{(h)}}(\sheet{\alpha_0}{x_0},\sheet{\alpha_1}{x_1},J) \cr
&=&\frac{1}{2\pi i} \oint_{\mathcal{C}_{\mathcal D_x}>y} dx\, K(\sheet{\alpha_0}{x_0},x) \Big(
\widetilde{W_{n+2}^{(h-1)}}(x,x,x_1,J) +  2B(\sheet{\alpha_1}{x_1},x)K(x,y) \widetilde{W_{n+1}^{(h-1)}}(y,J)\Big).
\eea
We first consider the product of functions $KBK$ in the second term: recalling that
$B(\sheet{\alpha_1}{x_1},\sheet{\beta}{x})=\partial_x\bigl(\partial_x-2\frac{\psi'_\beta(x)}{\psi_\beta(x)} \bigr)K(\sheet{\alpha_1}{x_1},x)$
and integrating by parts, we obtain
\bea
&& \frac{1}{2\pi i} \oint_{\mathcal{C}_{\mathcal D_x}>y} dx\, K(\sheet{\alpha_0}{x_0},x)B(\sheet{\alpha_1}{x_1},x)K(x,y)\cr
&=&-\frac{1}{2\pi i} \oint_{\mathcal{C}_{\mathcal D_x}>y} dx\,K'_x(\sheet{\alpha_0}{x_0},x)K'_x(\sheet{\alpha_1}{x_1},x)K(x,y)\cr
&&+\frac{1}{2\pi i} \oint_{\mathcal{C}_{\mathcal D_x}>y} dx\,K'_x(\sheet{\alpha_0}{x_0},x)K(\sheet{\alpha_1}{x_1},x)
2\frac{\psi'(x)}{\psi(x)}K(x,y)\cr
&&-\frac{1}{2\pi i} \oint_{\mathcal{C}_{\mathcal D_x}>y} dx\,K(\sheet{\alpha_0}{x_0},x)K'_x(\sheet{\alpha_1}{x_1},x)K'_x(x,y)\cr
&&+\frac{1}{2\pi i} \oint_{\mathcal{C}_{\mathcal D_x}>y} dx\,K(\sheet{\alpha_0}{x_0},x)K(\sheet{\alpha_1}{x_1},x)
2\frac{\psi'(x)}{\psi(x)}K'_x(x,y).
\eea
The first and the last terms in the right-hand side are already symmetric w.r.t. the replacement $x_0\leftrightarrow x_1$ and
we disregard them. Integrating by parts in the third term in the right-hand side we obtain one more symmetric term
with $K(\sheet{\alpha_0}{x_0},x)K(\sheet{\alpha_1}{x_1},x)K''_{xx}(x,y)$ (which we can disregard as well) plus the term with
$K'_x(\sheet{\alpha_0}{x_0},x)K(\sheet{\alpha_1}{x_1},x)K'_x(x,y)$. Combining the result with the second term, we obtain
\bea
&&\frac{1}{2\pi i} \oint_{\mathcal{C}_{\mathcal D_x}>y} dx\,K'_x(\sheet{\alpha_0}{x_0},x)K(\sheet{\alpha_1}{x_1},x)
\left(\partial_x+2\frac{\psi'(x)}{\psi(x)}\right)K(x,y)\cr
&=&\frac{1}{2\pi i} \oint_{\mathcal{C}_{\mathcal D_x}>y} dx\,K'_x(\sheet{\alpha_0}{x_0},x)K(\sheet{\alpha_1}{x_1},x)
\left(\frac{1}{x-y}+\sum_\beta h_\beta(x)C_\beta(y)\right),
\eea
where the integrand is sector-independent w.r.t. the variable $x$, so only the residue at $x=y$ (with the minus sign) contributes
in the second term in (\ref{first-equation}), which therefore becomes
\beq
\label{second-equation}
-\frac{1}{2\pi i} \oint_{\mathcal{C}_{\mathcal D_y}} dy\,2K'_y(\sheet{\alpha_0}{x_0},y)K(\sheet{\alpha_1}{x_1},y)\widetilde{W_{n+1}^{(h-1)}}(y,y,J).
\eeq

For the first term in (\ref{first-equation}), we use the induction assumption writing it in the form
\bea
&&\frac{1}{2\pi i} \oint_{\mathcal{C}_{\mathcal D_x}>y} dx\,
\frac{1}{2\pi i} \oint_{\mathcal{C}_{\mathcal D_y}} dy\,
 K(\sheet{\alpha_0}{x_0},x) K(\sheet{\alpha_1}{x_1},y)\times\cr
&&\quad\times
\Big(2B(x,y)^2\delta_{h=1,n=2}+4B(x,y)\widetilde{W_{n+1}^{(h-1)}}(x,y,J)'+ \widetilde{W_{n+3}^{(h-2)}}(x,x,y,y,J)'\Big),
\eea
where the prime indicates that no propagators of $B(x,y)$ type enter the expression, and no singularity
occurs in the corresponding terms upon interchanging the
order of contour integration w.r.t. $x$ and $y$. The last term is again obviously symmetric w.r.t. the
replacement $x_0\leftrightarrow x_1$.

The skew-symmetric part in the middle term is one-half of the residue coming from the
double-pole $-1/(x-y)^2$ in the expression for $B(x,y)$ (it comes again with the minus sign due to the choice of contour
ordering), so we obtain
\beq
\frac{1}{2\pi i} \oint_{\mathcal{C}_{\mathcal D_y}} dy\, 2K'_y(\sheet{\alpha_0}{x_0},y) K(\sheet{\alpha_1}{x_1},y)
\widetilde{W_{n+1}^{(h-1)}}(y,y,J)'+\hbox{symmetric term},
\eeq
which exactly cancel the term in (\ref{second-equation}) except the only case $g=1,\ n=2$ in which we use that
$B(x,y)=-1/(x-y)^2+{\overline W}_2^{(0)}(x,y)$ as $x\to y$, so
$$
2B(x,y)^2=2(x-y)^{-4}-4(x-y)^{-2}{\overline W}_2^{(0)}(x,y)+\hbox{regular},
$$
the most singular first term results in the integrand $K'''K$ that is sector-independent and vanishes, whereas the
second term produces
$$
\frac{1}{2\pi i} \oint_{\mathcal{C}_{\mathcal D_y}} dy\, 2K'_y(\sheet{\alpha_0}{x_0},y) K(\sheet{\alpha_1}{x_1},y)
\overline{W}_{2}^{(0)}(y,y)+\hbox{symmetric term},
$$
which kills the last remaining possible term in the expression (\ref{second-equation}). The theorem is proved.
}

\appendix{Calculating $\frac{\partial^3{\mathcal F}_0}{\partial t_0^3}$ in the Gaussian case}
\label{F0-Gaussian}

In this appendix, we calculate the singular part of the third derivative of ${\mathcal F}_0$ and integrate the
answer to obtain the singular part of ${\mathcal F}_0$ itself. Although we explicitly calculate only the
Gaussian model case, we propose the answer for the general model free energy singular part.

In the Gaussian model case with the potential $V(x)=x^2$, we have four sectors of solutions with the
asymptotic directions $\pm\infty,\ \pm i\infty$. As the basic solutions we take $\psi_+(x)$ and $\psi_-(x)$ that
decrease at the corresponding {\em imaginary} infinities $+i\infty$ and $-i\infty$. The real axis then plays the
role of the $\td\acycle$-cycle whereas the imaginary axis is the $\td\bcycle$-cycle.

We are interested in evaluating the singular part of the third-order derivative $\frac{\partial^3{\mathcal F}_0}{\partial t_0^3}$.
Let us first find out about the origin of this singular behavior. Apparently, local singularities at finite $t_0$ appear
when the solutions $\psi_+$ and $\psi_-$ coincide, which happens when $\psi_\pm:=\psi_n=H_n(ix)e^{x^2/2\hbar}$, where $H_n$ are the
Hermite polynomials, and
$$
\hbar^2\partial^2_x\psi_n(x)=x^2\psi_n(x)+(2n+1)\hbar \psi_n(x).
$$
From Corollary~\ref{cor-epsilon} we have
\beq
\label{formula-uno}
\frac{\partial^3{\mathcal F}_0}{\partial t_0^3}=\frac{1}{(2\pi i)^3}\oint_{\mathcal B}\!\oint_{\mathcal B}\!\oint_{\mathcal B}\,
dz_1\,dz_2\,dz_3\,W_3^{(0)}(z_1,z_2,z_3),
\eeq
and using that $W_3^{(0)}(\sheet{\alpha_1}{z_1},\sheet{\alpha_2}{z_2},\sheet{\alpha_3}{z_3})
=\oint_{{\mathcal C}_{\mathcal D}}d\xi K(\sheet{\alpha_1}{z_1},\xi)B(\sheet{\alpha_2}{z_2},\xi)B(\sheet{\alpha_3}{z_3},\xi)$
and that no singularities appear when integrating over $z_2$ and $z_3$ we obtain that each integral gives, by Theorem~\ref{thBergmanABcycles},
just $v_0(\sheet{\alpha}{\xi})$ with $\alpha=\pm$ and
\beq
\label{v0}
v_0(\sheet{\pm}{\xi})=C_0\frac{1}{\psi_\pm^2(\xi)}\int_{\pm i\infty}^\xi \psi_\pm^2(\rho)d\rho
\eeq
with the normalization constant $C_0$ such that
$$
\int_{-\infty}^{+\infty}\bigl[v_0(\sheet{-}{\xi})-v_0(\sheet{+}{\xi})\bigr]=1.
$$
Note that even in the case where $\psi_+=\psi_-$ the functions $v_0(\sheet{+}{\xi})$ and $v_0(\sheet{-}{\xi})$ differ because of
different lower limits of integrations, their difference is just $C_0\frac{1}{\psi^2(\xi)}\int_{-i\infty}^{+i\infty}\psi^2(\rho)d\rho$,
and the normalization constant $C_0$ at $\psi_+=\psi_-=\psi_n$ is
\beq
\label{C0}
C_0=\left(\int_{-\infty}^{+\infty}\frac{1}{\psi_n^2(\xi)}d\xi\right)^{-1} \left(\int_{-i\infty}^{+i\infty}\psi_n^2(\rho)d\rho\right)^{-1}.
\eeq
The remaining integral w.r.t. $z_1$ in (\ref{formula-uno}) develops singularity when $\psi_\pm\to \psi_n$ because the function
$K(\sheet{\pm}{z_1},\xi)$ develops a logarithmic cut on the $\td{\mathcal B}$-cycle, and using the explicit form (\ref{hatK}) for the
$K$-kernel (in this simplest case, $K=\widehat K$), we obtain
\beq
\label{formula-due}
\frac{\partial^3{\mathcal F}_0}{\partial t_0^3}=
\sum_\pm\int_{{\mathcal C}^\pm_\xi}d\xi \frac{1}{2\pi i}\int_{\pm i\infty}^{\mp i\infty}dz
\frac{1}{\hbar}\frac{1}{\psi^2_{\pm}(z)}\int_{\pm i\infty}^z d\rho \frac{\psi_\pm^2(\rho)}{\rho-\xi}v_0^2(\sheet{\pm}{\xi}),
\eeq
where the contour ${\mathcal C}^\pm_\xi$ goes between $\pm \infty$ and $\mp infty$ encircling the point $\rho$. The singularity
occurs when $\rho$ (and, correspondingly, $z$) tends to $-i\infty$ for $\psi_+$ and to $+i\infty$ for $\psi_-$; this singular part
comes from the residue at $\xi=\rho$, and we obtain that the expression in (\ref{formula-due}) is
$$
\sum_\pm\int_{0}^{\mp i\infty}dz\frac{1}{\hbar}
\frac{1}{\psi^2_{\pm}(z)}\int_{0}^z d\rho \psi_\pm^2(\rho)v_0^2(\sheet{\pm}{\rho})+
\hbox{regular}
$$
and using the explicit expressions (\ref{v0}) and (\ref{C0}) for $v_0$, we obtain
\bea
&&\hbox{sing.}\left(\frac{\partial^3{\mathcal F}_0}{\partial t_0^3}\right)\cr
&&\quad=\sum_\pm\int_{0}^{\mp i\infty}dz\frac{1}{\hbar}
\frac{1}{\psi^2_{\pm}(z)}\int_{0}^z d\rho \frac{1}{\psi_\pm^2(\rho)}\left[\int_{\pm i\infty}^\rho \psi_\pm^2(s)ds\right]^2C_0^2,
\eea
where the singularity occurs at the upper integration limit for $z$ and $\rho$ when $\psi_+,\psi_-\to \psi_n$, and
the term in the square brackets is nonsingular in this limit, so we can replace it by its limiting value, which cancel
exactly the corresponding term in the normalization constant $C_0$ (see (\ref{C0}); the integral w.r.t. $z$ and $\rho$
can be separated, and we obtain the final expression
\beq
\label{sing}
\hbox{sing.}\left(\frac{\partial^3{\mathcal F}_0}{\partial t_0^3}\right)
=\sum_\pm \frac{1}{2\hbar} \left[\int_{0}^{\mp i\infty}dz
\frac{1}{\psi^2_{\pm}(z)}\right]^2 \left[\int_{-\infty}^{+\infty}\frac{1}{\psi_\pm^2(x)}dx\right]^{-2}.
\eeq

We now calculate $t_0$ when $\psi_+,\psi_-\to \psi_n$. Choosing $\psi_-(x)=\psi_+(x)\int_{-\infty}^x\frac{d\xi}{\psi^2_+(\xi)}$
and taking into account that the number of poles of solutions outside the $\td{\mathcal A}$-cycle is $n$, we have
\bea
t_0&=&-\hbar n+\hbar\int_{-\infty}^{+\infty}\left(\frac{\psi'_-}{\psi_-}-\frac{\psi'_+}{\psi_+}\right)
=-\hbar n +\hbar\int_{-\infty}^{+\infty}\frac{1}{\psi_+\psi_-}\cr
&=&-\hbar n+\hbar\int_{-\infty}^{+\infty}\frac{dz}{\psi^2_+(z)}
\frac{1}{\int_{-i\infty}^0\frac{d\xi}{\psi_+^2(\xi)}+\int_{0}^z\frac{d\xi}{\psi_+^2(\xi)}}.
\eea
The first integral in the denominator diverges as $\psi_+\to\psi_n$ and denoting this integral as $\Lambda$,
we have
\beq
\label{asymptotics}
\Bigl.t_0\Bigr|_{\psi_+\to\psi_n}=-\hbar n+\hbar\frac{\int_{-\infty}^{+\infty}\frac{dz}{\psi^2_+(z)}}
{\int_{-i\infty}^0\frac{d\xi}{\psi_+^2(\xi)}}+O(\Lambda^{-2}).
\eeq
Comparing this expression with (\ref{sing}), we obtain
\beq
\hbox{sing.}\left(\frac{\partial^3{\mathcal F}_0}{\partial t_0^3}\right)=\frac{1}{\hbar(n+t_0/\hbar)^2}, \quad n\in{\mathbb Z}_{+,0},
\eeq
that is, this derivative has double poles with the coefficient $1/\hbar$ at all points $t_0=-\hbar n$, $n=0,1,\dots$. A function that
exhibits such a behavior is obviously $\Gamma$-function, so we have that, up to an entire function,
\beq
\hbox{sing.}\left(\frac{\partial^3{\mathcal F}_0}{\partial t_0^3}\right)\simeq \frac{1}{\hbar}\bigl[\log \Gamma\bigr]''(t_0/\hbar),
\eeq
and, in turn,
\beq
\label{sing-F0}
\hbox{sing.}{\mathcal F}_0\simeq \hbar^2 \bigl[\hbox{$\int$}\log \Gamma\bigr](t_0/\hbar).
\eeq
Turning to the asymptotic behavior of $\int \log \Gamma(x)$ at large positive $x$ we observe that the
leading term is $\frac12 x^2\log x$, which is exactly what we might expect from the matrix-model-like arguments:
we must be able to apply semiclassical approximation at large positive $t_0/\hbar$, and in this regime we have
the leading asymptotic behavior of Gaussian matrix model, i.e.,
$\hbox{sing.}{\mathcal F}_0\simeq \frac12 t^2_0\log(t_0)$ modulo polynomial terms (of order not higher than two).

We may therefore put forward the following conjecture.

\begin{conjecture}
The singular part of ${\mathcal F}_0$ for any potential $V_{d+1}(x)$ has the form $\hbar^2\sum_{i=1}^d \frac12 [\int \log \Gamma](\td\epsilon_i/\hbar)$
where $\td\epsilon_i$ are the occupation numbers on the cycles $\td\acycle_i$.
\end{conjecture}

\end{document}